\documentclass[iop,numberedappendix]{emulateapj}
\slugcomment{Published in \apj} 
\usepackage{natbib}
\usepackage{amsmath}
\usepackage{booktabs}
\usepackage{longtable}
\usepackage{color}
\usepackage{graphicx, subfigure}
\usepackage{hyperref}
\newcommand{\di}{\mathrm{d}} 
\newcommand{\mr}{\mathrm} 
\newcommand{\Ht}{\mathrm{H_2}} 
\newcommand{\Ho}{\mathrm{H}} 
\newcommand{\Hplus}{\mathrm{H^+}}
\newcommand{\He}{\mathrm{He}} 
\newcommand{\Heplus}{\mathrm{He^+}}
\newcommand{\HCOplus}{\mathrm{HCO^+}}
\newcommand{\CO}{\mathrm{CO}} 
\newcommand{\CI}{\mathrm{C}}
\newcommand{\OI}{\mathrm{O}}
\newcommand{\CII}{\mathrm{C^+}} 
\newcommand{\CHx}{\mathrm{CH_x}} 
\newcommand{\CH}{\mathrm{CH}} 
\newcommand{\OHx}{\mathrm{OH_x}} 
\newcommand{\Oplus}{\mathrm{O^+}} 
\newcommand{\Si}{\mathrm{Si}} 
\newcommand{\Siplus}{\mathrm{Si^+}}

\defcitealias{NL1999}{NL99}
\defcitealias{BFJ2010}{BFJ2010}
\defcitealias{CF2004}{CF2004}

\begin{document}
\title{A simple and accurate network for hydrogen and carbon chemistry in the
interstellar medium}
\author{Munan Gong\altaffilmark{1}, Eve C. Ostriker\altaffilmark{1}
and Mark G. Wolfire\altaffilmark{2}}
\altaffiltext{1}{Department of Astrophysical Sciences, Princeton University,
Princeton, New Jersey 08544, USA; 
munan@princeton.edu, eco@astro.princeton.edu}
\altaffiltext{2}{Department of Astronomy, University of Maryland, College Park, 
MD 20742-2421, USA; mwolfire@astro.umd.edu}

\begin{abstract}
Chemistry plays an important role in the interstellar medium (ISM), regulating
the heating
and cooling of the gas and determining abundances of molecular species
that trace gas properties in observations. Although solving the time-dependent
equations is necessary for accurate abundances and temperature in the 
dynamic ISM, a full chemical network is too computationally 
expensive to incorporate in numerical simulations.
In this paper, we propose a new simplified chemical network for hydrogen and carbon
chemistry in the atomic and molecular ISM. 
We compare results from our chemical network in detail with results from a full 
photodissociation region (PDR) code, and also with the 
\citet{NL1999} (NL99) network previously adopted in the simulation literature.
We show that our chemical network gives similar results to the PDR code in
the equilibrium abundances of all species over a
wide range of densities, temperature, and metallicities,
whereas the NL99 network shows significant disagreement.
Applying our network in 1D models, we find that the $\CO$-dominated regime
delimits the coldest gas and that the corresponding temperature 
tracks the cosmic-ray ionization rate in molecular clouds. We provide
a simple fit for the locus of $\CO$-dominated regions as a function of gas density
and column. 
We also compare with observations of diffuse and
translucent clouds. We find that the $\CO$, $\CHx$ and $\OHx$ abundances are 
consistent with equilibrium predictions for
densities $n=100-1000~\mr{cm^{-3}}$, but the predicted equilibrium $\CI$
abundance is higher than observations, signaling the potential importance of
non-equilibrium/dynamical effects.
\end{abstract}

\section{Introduction}
Chemistry plays a key role in the interstellar medium (ISM) and star formation. 
In the ISM, a range of chemical species acts as
agents for heating and cooling of the gas, regulating the gas
temperature. Specific atomic and molecular species provide observational tracers 
for the gas physical properties, such as density and temperature.
One of the most important molecules in the ISM is carbon monoxide ($\CO$). $\CO$ is 
a main coolant for most of the molecular ISM via its rotational transitions
\citep{GL1978, NK1993, NLM1995, Omukai2010}, and also widely used as an observational 
tracer for molecular hydrogen ($\Ht$) and star forming regions in both our
Milky Way and other galaxies 
\citep[e.g.,][]{Bolatto2008, Leroy2008, Roman-Duval2010, Saintonge2011, KE2012,
Tacconi2013, Bolatto2013, Dobbs2014, HD2015}. Cooling by $\CO$ enables gas to
reach low temperatures at which it collapses gravitationally to form stars. In
return, feedback from star formation affects the chemical state of
the ISM, as far-ultraviolet (FUV) radiation from massive stars
dissociates and ionizes the gas. In particular, photodissociation of $\CO$
produces $\CI$, and ionization of $\CI$ produces $\CII$, with their
fine structure lines among the most important coolants of atomic gas
\citep{Wolfire1995, Wolfire2003}.
The turbulence generated by supernova
explosions leads to mixing of the gas composition
and also changes the gas density distribution.

Traditionally, steady-state 1D photodissociation region (PDR) models with
complex chemical networks have been used intensively 
to understand the detailed chemical structure of the
ISM \citep[e.g.,][]{Kaufman1999, Meudon2006, Rollig2007, Wolfire2008, WHM2010,
Hollenbach2012}.
Although much progress has been made using PDR models, 
solving the time-dependent equations is necessary for accurate abundances and
temperature in the dynamic ISM. Typical
star forming giant molecular clouds (GMCs) are
not in chemical equilibrium \citep{LHH1984, Papadopoulos2004}, and are pervaded by 
supersonic turbulence that creates complex internal structure on dynamical
timescales comparable to chemical timescales \citep{MK2004,Glover2010}.
In recent years, efforts have therefore been increasing to incorporate
 time-dependent chemistry within both local \citep{Walch2015, Girichidis2016}
 and global \citep{RS2016a, RS2016b} galactic hydrodynamic
and magnetohydrodynamic numerical simulations of the ISM.
However, the full chemical network used in PDR models
is too computationally expensive
to incorporate in large numerical simulations.
Modeling the chemical evolution of a group of $N_\mr{spec}$ species requires
solving coupled ordinary differential equations (ODEs) of $N_\mr{spec}$
dimensions. Because the chemical timescales in the reaction set have extreme
variation, the set of ODEs is usually very stiff and has to be
solved implicitly, which results in a computational cost scaling as
$N_\mr{spec}^3$.

For this reason, various authors have proposed approximations to simplify
the chemical network for gas in star-forming regions 
\citep{NL1997, NL1999, KC2008, Glover2010, Richings2014}.
The general approach is to reduce the number of species, retaining only hydrogen
and the most important species for cooling and observation, such
as $\CII$, $\CI$, and $\CO$. The number of reactions can also be reduced
by capturing just the most important formation and destruction pathways for these
species. \citet{GC2012} compared a number of simplified networks by
looking at the $\CII$, $\CI$, $\CO$, and temperature distributions 
in 3D simulations of turbulent molecular clouds. They found that all of the
models they explored give similar results for density and temperature distributions,
suggesting the gas thermodynamics is not very sensitive to the detailed chemistry.
The $\CO$ distribution, however, is very different in various networks. 
They concluded that the simplified network of 
\citet[][hereafter \citetalias{NL1999}]{NL1999}
reproduces the $\CO$ abundance in more complicated models of
\citet{Glover2010}, and therefore recommend the \citetalias{NL1999} network for
simulating $\CO$ formation. 

However, there are several limitations to the
work of \citet{GC2012}. They did not carry out a detailed comparison between the
different networks and more complex PDR models. Moreover, the comparisons they
made are only under limited physical conditions. Notably, they used 
a cosmic-ray ionization rate $\xi_\Ho = 10^{-17}~\mr{s^{-1} H^{-1}}$, 
which is an order of magnitude lower than the
recent estimates from observations of Milky Way molecular clouds, 
$\xi_\Ho \approx 2\times 10^{-16}~\mr{s^{-1} H^{-1}}$ 
\citep{Indriolo2007, Hollenbach2012}. 
We show that when using the realistic cosmic-ray ionization rate, 
the \citetalias{NL1999} network significantly
underproduces $\CO$ compared to observations (see Section
\ref{section:code_test}).

In this paper, we propose a new lightweight and accurate
chemical network for following carbon and hydrogen chemistry 
in simulations of 
the atomic and molecular ISM. In Section \ref{section:chemical_model}, 
we describe our chemical network (which is an extension of \citetalias{NL1999}),
related heating and cooling processes, and
numerical methods. Then we carry out detailed comparison with both a PDR code and the
original \citetalias{NL1999} network in Section \ref{section:code_test}.
Finally,
in Section \ref{section:application}, we apply our network to simple 1D cloud models,
and show the results from these applications in comparison to observations.

\section{The Chemical Model}\label{section:chemical_model}
\subsection{Chemical Reaction Network}
\subsubsection{General Framework}
The chemical network in this paper is listed in 
in Tables \ref{table:chem1} and \ref{table:chem2}, and described in detail in
Appendix \ref{section:chem_desp}. 
Our chemical network is based on the \citetalias{NL1999} network in \citet{NL1999}
and \citet{GC2012}, with significant modification and extension. 
Eighteen species are considered: $\Ho$, $\Ht$, $\Hplus$, $\mr{H_2^+}$, $\mr{H_3^+}$,
$\He$, $\mr{He^+}$, $\OI$, $\Oplus$, $\CI$, $\CII$, $\CO$, 
$\mr{HCO^+}$, $\Si$, $\Siplus$, $\mr{e}$, $\CHx$ and $\OHx$.
Following \citetalias{NL1999}, $\CHx$ (including $\mr{CH}$,  $\mr{CH_2}$,
$\mr{CH^+}$,  $\mr{CH_2^+}$,  $\mr{CH_3^+}$) and $\OHx$ (including $\mr{OH}$,  
$\mr{H_2O}$, $\mr{OH^+}$, $\mr{H_2O^+}$ $\mr{H_3O^+}$) are pseudo-species, 
and their treatment is described in Appendix 
\ref{section:CHx} and \ref{section:OHx}.
Among the 18 species in our network, 12 of them are
independently calculated, and 6 of them, $\Ho$, $\He$, $\CI$, $\OI$, $\Si$ and
$\mr{e}$, are derived from conservation of total hydrogen, helium, carbon,
oxygen, and silicon nuclei, and the conservation of total charge.
\footnote {We have compared the approach of using conservation laws
        with directly calculating the abundance of all species from rate equations, 
    and found the results are the same. When calculating the
abundance of  $\Ho$, $\CI$, $\OI$, and $\mr{e}$ from conservation laws, we
assume all $\mathrm{CH_x}$ and $\mathrm{OH_x}$ are in the form of
$\mathrm{CH}$ and $\mathrm{OH}$.
Because the abundance of $\mathrm{CH_x}$ and $\mathrm{OH_x}$ is usually very small
compared to the hydrogen in $\Ho$ or $\Ht$, carbon in $\CI$, $\CII$ or $\CO$,
oxygen in $\OI$, and electrons provided by $\Hplus$ or $\CII$,
this assumption has no significant effect on the chemical network.}
The gas-phase abundances of total helium nuclei are
$x_\mr{He, tot}=n_\mr{He, tot}/n=0.1$. The gas-phase abundance of  
total carbon, oxygen, and silicon nuclei are assumed to be proportional to the gas
metallicity relative to the solar neighborhood $Z_\mr{g}$, and we adopt the
values $x_\mr{C, tot}=1.6\times 10^{-4} Z_\mr{g}$ \citep{Sofia2004}, 
$x_\mr{O, tot}=3.2\times 10^{-4} Z_\mr{g}$ \citep{SS1996}, and 
$x_\mr{Si, tot}=1.7\times 10^{-6} Z_\mr{g}$ \citep{Cardelli1994}. 
We do not explicitly follow the evolution of dust grains, but instead
assume that the grain-assisted reaction rates scale with the dust abundance
relative to the solar neighborhood $Z_d$. We assume the gas metallicity 
$Z_g$ and dust abundance $Z_d$ vary
simultaneously, and use a single parameter for metallicity 
relative to the solar neighborhood,
\begin{equation}
  Z = Z_\mr{g} = Z_\mr{d}.
\end{equation}

We have updated all reaction rates in the original \citetalias{NL1999} network,
according to the most recent values in the
UMIST \citep{McElroy2013} and KIDA \citep{KIDA2010} catalogs, and other
references in Tables \ref{table:chem1} and \ref{table:chem2}.
Notably, we have a
higher rate of $\mr{C^+ + OH \rightarrow CO^+ + H}$, 
and a higher rate of $\mr{C^+ + e \rightarrow C}$ by
including both radiative and dielectronic recombination. These higher
rates lower the electron abundance 
and aid the formation of $\mr{OH_x}$ and $\mr{HCO^+}$, resulting in more
efficient $\CO$ formation.

We have also modified and extended the \citetalias{NL1999} network, as described in detail in
Appendix \ref{section:NL99_extension}. One important extension is the addition
of grain-assisted recombination of $\CII$ and $\mr{He^+}$. Although
\citet{GC2012} concluded these grain reactions do not affect $\CO$ formation
at the low cosmic-ray ionization rate $\xi_\Ho=10^{-17} \mr{s^{-1} H^{-1}}$,
we found that with the updated cosmic-ray ionization rate
$\xi_\Ho=2\times 10^{-16} \mr{s^{-1} H^{-1}}$ \citep{Indriolo2007, Indriolo2015},
these reactions are critical for $\CO$ formation at moderate densities 
$n \lesssim 1000~\mr{cm^{-3}}$. This is because
electrons formed by cosmic-ray ionization inhibit $\CO$ formation,
while $\mr{He^+}$ ions formed by cosmic rays destroy $\CO$ \citep{Bisbas2015}.  
Grain-assisted recombination reduces the abundance of $\mr{e}$ and $\mr{He^+}$.
With these extensions, we have 50 reactions in total, including
collisional reactions, grain-assisted recombination and $\Ht$ formation,
photodissociation by FUV radiation, and cosmic-ray ionization.

\subsubsection{Photochemistry\label{section:photo-chemstry}}
The photodissociation reactions induced by FUV depend on the radiation field
strength. We assume that the incident radiation field scales with the standard
interstellar radiation field determined by \citet{Draine1978}, 
using the parameter $\chi$ as the field strength relative to 
$J_\mr{FUV} = 2.7 \times 10^{-3} \mr{erg~cm^{-2} s^{-1}}$
($G_0=1.7$ in \citet{Habing1968} units). 
The photochemistry reaction rates appropriate for
this radiation field are listed in Table \ref{table:chem2}.

In optically thick regions, the radiation is attenuated by dust
and by molecular line shielding. In a
plane-parallel slab geometry with beamed incident radiation field from one
direction, the photodissociation rates of optically thick regions $R_\mr{thick}$
can be related to the rates in optically thin regions $R_\mr{thin}$ by 
\begin{equation}\label{eq:R_thick}
    R_\mr{thick} = \chi R_\mr{thin} f_\mr{shield} = \chi R_\mr{thin} f_\mr{dust}
    f_\mr{s},
\end{equation}
where $f_\mr{dust} = \exp(-\gamma A_V)$ is the dust-shielding factor and
$f_\mr{s}$ is the self-shielding of $\CI$, $\CO$ and $\Ht$ (see text below). The
values of the parameter $\gamma$ appropriate for the different reactions
are listed in Table
\ref{table:chem2}. The visual extinction $A_V$ is calculated as 
\begin{equation}
    A_V = \frac{N Z_\di}{1.87\times 10^{21} \mr{cm^{-2}}}~,
\end{equation}
where $N = N_\mr{H} + 2N_\Ht$ is the total 
hydrogen column density along the line of sight.
This corresponds to $R_V\equiv A_V/E(B-V)=3.1$
and $N/E(B-V) = 5.8\times 10^{21}\mr{cm^{-2}}$, 
appropriate for the diffuse ISM \citep{BSD1978}. In clouds with slab geometry
and isotropic radiation field impinging from one side (Section
\ref{section:application}), $f_\mr{shield}$ is
an average value calculated from different incident angles. 
For more general, non-slab geometry, photo rates at 
a given location would be calculated by angle averages of 
Equation (\ref{eq:R_thick}), as both $\chi$ and $f_\mr{shield}$ vary with direction.

For photoelectric
heating on dust (see Section \ref{section:hc}), the heating
rate depend on the radiation field strength between $6-13.6~\mr{eV}$ which
affects the charge states of grains. For radiation in the FUV important
for the photoelectric effect, we use a constant dust cross-section
$\sigma_\mr{d, PE} = 10^{-21} Z_\di \mr{cm^2 H^{-1}} = 1.87
A_V/N$ to calculate the attenuation by dust,
\begin{equation}
    \chi_\mr{PE}(N) = \chi f_\mr{dust, PE} = \chi \exp(-\sigma_\mr{d, PE}
    N);
\end{equation}
our adopted cross-section is close to the value $1.8 A_V/N$ 
used in previous PDR models.

In addition to dust shielding, we also include the $\Ht$ self-shielding, 
$\CO$ self-shielding, $\CI$ self-shielding, and shielding by $\Ht$ of $\CO$ and
$\CI$. The photodissociation rate of $\Ht$ can be written as 
\begin{equation}
    R_\mr{thick,H_2} = \chi R_\mr{thin,H_2}  f_\mr{dust} f_\mr{s, H_2}(N_\mathrm{H_2}).
\end{equation}
We use the results in \citet{DB1996} for $\mathrm{H_2}$ self-shielding:
\begin{align}
    &f_{s, \mathrm{H_2}}(N_\mathrm{H_2}) = \frac{0.965}{(1+x/b_5)^2}\\
    & + \frac{0.035}{(1+x)^{0.5}} \exp[-8.5\times 10^{-4}(1+x)^{0.5}], 
\end{align}
where $x \equiv N_\mathrm{H_2}/(5\times 10^{14} \mathrm{cm^{-2}})$ and 
$b_5 = b/(\mathrm{km/s})$ for $b$ the velocity dispersion. In regions where
$2x(\mathrm{H_2}) = 2n(\mathrm{H_2})/n \gtrsim 0.1$, the
self-shielding of $\mathrm{H_2}$ is in the wings of the line profile, and the 
$\mathrm{H_2}$ fraction is not sensitive to the choice of $b_5$. 
Here we use a constant value $b_5=3$.

Similarly, the photodissociation rate of $\CO$ can be written as 
\begin{equation}
    R_\mr{thick,CO} = \chi R_\mr{thin,CO} f_\mr{dust} 
                        f_\mr{s, CO}(N_\mathrm{CO}, N_\mathrm{H_2}).
\end{equation}
The shielding factor $f_\mr{s, CO}(N_\mathrm{CO}, N_\mathrm{H_2})$
is interpolated from Table 5 in \citet{Visser2009} for a
given column density of $\mathrm{CO}$ and $\mathrm{H_2}$. This
accounts for both the 
$\mathrm{CO}$ self-shielding and the shielding of $\CO$ by $\mathrm{H_2}$. 
In \citet{Visser2009} Table 5, 
the velocity dispersion of $\mathrm{CO}$ is assumed to be 
$b(\mathrm{CO})=0.3\mathrm{km/s}$ \footnote{We have also tried to use the
$b(\CO)=3\mathrm{km/s}$ results in \citet{Visser2009}, and found that 
$f_\mr{s, CO}$ changes within $\sim 50\%$, and the $\CO$ fraction is not
sensitive to $b(\CO)$. }, the excitation temperature
$T_\mathrm{ex}(\mathrm{CO}) = 5\mathrm{K}$ and
$N(^{12}\mathrm{CO})/N(^{13}\mathrm{CO})=69$.

Lastly, the photodissociation rate of $\CI$ is
\begin{equation}
    R_\mr{thick,C} = \chi R_\mr{thin,C} f_\mr{dust} 
                        f_\mr{s, C}(N_\mathrm{C}, N_\mathrm{H_2}).
\end{equation}
We adopt the treatment in \citet{TH1985}. The shielding factor 
$f_\mr{s, C}(N_\mathrm{C}, N_\mathrm{H_2}) = \exp(-\tau_\CI)f_\mr{s,C}(\Ht)$,
where $\tau_\CI = 1.6\times 10^{-17}N_\CI/\mr{cm^{-2}}$, $f_\mr{s,C}(\Ht) =
\exp(-r_\Ht)/(1+r_\Ht)$, and $r_\Ht = 2.8\times 10^{-22} N_\Ht/\mr{cm^{-2}}$.

\subsection{Heating and Cooling Processes\label{section:hc}}
The heating and cooling processes included in our models are listed in Table
\ref{table:thermo}, and described in detail in Appendix
\ref{section:heating_cooling}. Here we describe the general processes we
considered. The time rate of change of the gas thermal energy per
$\Ho$ nucleus $e_\mr{g, sp}$ is given by
\begin{equation}\label{eq:e_g}
    \frac{\di e_\mr{g, sp}}{\di t} = \Gamma_\mr{tot} - \Lambda_\mr{tot}.
\end{equation}

The total heating rate per $\Ho$ nucleus is
\begin{equation}
    \Gamma_\mr{tot} = \Gamma_\mr{PE} + \Gamma_\mr{CR}
      + \Gamma_\mr{H2gr} + \Gamma_\mr{H2pump} + \Gamma_\mr{H2diss},
\end{equation}
where $\Gamma_\mr{PE}$ is the heating from the photoelectric effect on dust,
$\Gamma_\mr{CR}$ the cosmic-ray heating, $\Gamma_\mr{H2gr}$ the heating from
$\Ht$ formation on dust grains, $\Gamma_\mr{H2pump}$ the heating from $\Ht$ UV pumping, and 
$\Gamma_\mr{H2diss}$ the heating from $\Ht$ photodissociation. For typical
environments in the cold atomic and molecular ISM, the photoelectric heating
dominates at $A_V \lesssim 1$ and the cosmic-ray heating dominates at 
$A_V \gtrsim 1$.

The total cooling rate per $\Ho$ nucleus is
\begin{equation}
    \Lambda_\mr{tot} = \Lambda_\mr{line} + \Lambda_\mr{dust} + \Lambda_\mr{rec} 
      + \Lambda_\mr{H2coll} + \Lambda_\mr{Hion},
\end{equation}
where $\Lambda_\mr{line}$ is the line cooling by atomic and molecular species,
$\Lambda_\mr{dust}$ the cooling from gas--grain collisions,
$\Lambda_\mr{rec}$ the cooling by electron
recombination on dust grain, $\Lambda_\mr{H2coll}$ the cooling by collisional
dissociation of $\Ht$, and $\Lambda_\mr{Hion}$ the cooling by collisional
ionization of $\mr{H}$. The line cooling dominates in typical atomic and
diffuse
molecular ISM where the gas densities are not high enough for dust cooling to
be important ($n \lesssim 10^6~\mr{cm^{-3}}$).
We include line cooling of $\OI$, 
$\CI$, $\CII$ fine structure lines,
the Ly$\alpha$ line of $\Ho$, $\CO$ rotational lines, and the $\Ht$
vibration and rotational lines.

The gas temperature is related to the gas energy by \citep{DESPOTIC}
\begin{equation}
    T = \frac{e_\mr{g, sp}}{c_{v, \Ho}}.
\end{equation}
The specific heat at constant volume $c_{v, \Ho} = \frac{1}{2}k_B f$, where 
$k_B=1.381\times 10^{-16}~\mr{erg~K^{-1}}$ is the Boltzmann constant and
$f=\sum f_s x_s$ is the degree of freedom in all species per $\Ho$ nucleus.
For $\mr{H}$, $\He$, $\mr{He^+}$, $\mr{H^+}$, and $e$, $f_s=3$ from
translational degrees of freedom. For
$\Ht$, the excitation temperature for rotational and vibrational levels are
$\theta_\mr{rot}=170.6 \mr{K}$ and $\theta_\mr{vib}=5984 \mr{K}$
\citep{Tomida2013}. 
Assuming a fixed ortho-to-para ratio of $\Ht$, the rotational
and vibrational heat capacities of $\Ht$ are very small at typical molecular cloud
temperatures $T\lesssim 50 \mr{K}$. Therefore, we ignore the contribution of
rotational and vibrational levels of $\Ht$ to the specific
heat, and simply use
\begin{align}
    c_{v, \Ho} &= \frac{3}{2} k_B (x_\mr{H} + x_\mr{H_2}+ x_\mr{H^+} 
        + x_\mr{He} + x_\mr{He^+} + x_\mr{e})\\
        &= \frac{3}{2} k_B [(1-x_\Ht) + x_\mr{He, tot} + x_\mr{e}].
\end{align}

\subsection{Numerical Method\label{section:numerical_method}}
We consider a simple one-dimensional slab model with uniform density $n$, and
FUV radiation incident from one side of the slab, similar to typical 1D
PDR (photodissociation region) models. The incident radiation field is
expected to be relatively isotropic over $2\pi$ steradians. In Section
\ref{section:code_test}, we use the approach of \citet{WHM2010} in order
to compare with their PDR code: the isotropic radiation field is approximated
by assuming a unidirectional flux incident at an angle
of $60^\circ$ to the normal of the slab surface. Thus, with the incident radiation
field strength $\chi$, the field strength at the perpendicular 
column density $N$ from the
slab surface will be $\chi_\mr{eff}=(\chi/2) f_\mr{shield}(2N)$, where the factor
$1/2$ in $\chi$ comes from the one-sided slab and the factor $2$ in $N$
comes from the angle of $60^\circ$. In Section \ref{section:application}, we
directly calculate the isotropic radiation field strength by averaging over the
incident angels.

In each model, we divide a slab into $10^3$ logarithmically spaced grid zones in
the range of $N=10^{17}/Z-10^{22}/Z~\mr{cm^{-2}}$. In each zone, 
walking inward from the lowest $N$ to the highest $N$ in order, 
we calculate the equilibrium chemistry and temperature. 
This is because the chemical states of all exterior zones are needed to
calculate the radiation field shielding factor for the next zone.

The evolution of abundances of chemical species $x_s = n_s/n$ is determined
by a set of chemical reactions:
\begin{equation}\label{eq:chem_react}
\begin{split}
    \frac{\di x_s}{\di t} = &\sum_{i,j} (\pm k_\mr{2body,s}^{ij}n x_{i}
    x_{j}) + \sum_i (\pm k_\mr{gr,s}^i n x_{i})\\
    &+ \sum_i (\pm k_\mr{cr,s}^i x_{i}) + 
    \sum_i (\pm k_{\gamma, \mr{s}}^i x_{i}).
\end{split}
\end{equation}
Here $k_\mr{2body,s}^{ij}$, $k_\mr{gr,s}^i$, $k_\mr{cr,s}^i$,
and $k_{\gamma, \mr{s}}^i$ are the rates for the
two-body, grain surface, cosmic-ray and photodissociation reactions listed in
Tables \ref{table:chem1} and \ref{table:chem2} that have species $s$ either as
a reactant or product. $x_i$ and $x_j$ are the abundance of the
corresponding reactant species, and the sign is negative if $x_s$ is a
reactant of the chemical reaction and positive if $x_s$ is a product.
The chemical reactions in Equation (\ref{eq:chem_react}) and
the energy evolution in Equation (\ref{eq:e_g}) give a coupled ODE (ordinary
differential equation) system of 13 variables (12 independently calculated 
species and 1 energy equation).
The derived species ($\Ho$, $\He$, $\CI$, $\OI$, $\Si$, and $\mr{e}$) 
are not directly calculated from Equation \ref{eq:chem_react}, 
but from the conservation of atomic nuclei and charge. 
To efficiently solve this set of stiff ODEs, we use the open-source
CVODE package \citep{CVODE}, which adopts a method of 
implicit backward differentiation formulas.

\section{Code Test and Comparison}\label{section:code_test}
In this section, we test the agreement between the greatly simplified chemical
network in this paper and a PDR code with much more sophisticated chemistry and 
radiation transfer. We also compare with the widely used original \citetalias{NL1999} network 
and point out the importance of our modifications.

The PDR code used here is derived from the \citet{TH1985}
PDR code, which has been in continuous use, and updated and maintained
since its original inception. The model has been used extensively to analyze
observations in a diverse set of environments including the diffuse ISM
\citep[e.g.,][]{Wolfire2003, Wolfire2008, Sonnentrucker2015},
low-mass molecular clouds \citep{LeeStanimirovi2014, Burton2015}, 
Galactic GMCs \citep{WHM2010}, intense Galactic PDRs
\citep{Sheffer2011}, and extragalactic PDRs 
\citep[e.g.,][]{Kaufman2006, Stacey2010}.
Recent updates to the chemistry, heating, and cooling rates are given in
\citet{WHM2010}, \citet{Hollenbach2012} and \citet{NW2016}.
The code calculates the thermal balance temperature and steady-state
abundances of 74 species using 322 reactions. 
The dominant gas-phase species are $\Ho$, $\mr{He}$, $\CI$, $\mr{O}$, $\Ht$,
$\mr{O_2}$, $\mr{OH}$, $\CO$, $\mr{H_2O}$, $\mr{e^-}$, $\mr{H^+}$, $\mr{He^+}$,
$\CII$, $\mr{O^+}$, $\mr{OH^+}$, $\mr{CO^+}$, $\mr{H_2O^+}$, $\mr{HCO^+}$,
$\mr{H_3O^+}$, $\mr{H_2^+}$, $\mr{H_3^+}$, $\mr{CH^+}$, $\mr{CH_2^+}$,
$\mr{CH_3^+}$,
$\mr{CH}$, $\mr{CH_2}$, $\mr{CH_3}$, $\mr{CH_4}$, $\mr{Mg}$, $\mr{Mg^+}$,
$\mr{Si}$, $\mr{Si^+}$, $\mr{SiH_2^+}$, $\mr{SiH}$, $\mr{SiO}$, $\mr{Fe}$,
$\mr{Fe^+}$, $\mr{S}$, $\mr{S^+}$, $\mr{SiO^+}$,
$\mr{SO^+}$, $\mr{HOSi^+ }$,$\mr{H_2^*}$, $\mr{H^-}$, $\mr{PAH^-}$, $\mr{PAH}$,
and $\mr{PAH^+}$. Also included are fluorine and chlorine
chemistry as in \citet{NW2009}. Freeze-out of atoms and molecules
on grains (such as $\CO$ and $\mr{H_2O}$) and grain surface reactions such as
$\mr{OH}$ and $\mr{H_2O}$ formation are included as in 
\citet{Hollenbach2009, Hollenbach2012} but
are turned off for most of the comparisons in this paper. The effects of
grain surface reactions on chemistry are discussed in Appendix \ref{section:SR}.
For comparison to the results presented in this paper, we use
a modified version of the code in which
the geometry is plane parallel and the density is constant.
In addition, we
use the photoionization, photodissociation, and cosmic-ray
induced photo rates from from \citet{Heays2017}, consistent with Tables
\ref{table:chem1} and \ref{table:chem2} in this paper.
We use the approximation in \citet{WHM2010} for  an isotropically incident
radiation field assuming a single ray incident at 60 degrees to the normal
and use the $\Ht$ self-shielding treatment as in \citet{TH1985}.

The \citetalias{NL1999} network we compare with is similar to that used in \citet{GC2012}.
It is based on two main parts: the $\CO$ chemistry in \citetalias{NL1999}, and
the hydrogen chemistry in \citet{GM2007}. In order to make a meaningful comparison
of the difference between the $\CO$ chemistry in this paper and the \citetalias{NL1999} network,
we updated all the reaction rates in the \citetalias{NL1999} network to the values in
this paper (see Tables \ref{table:chem1} and \ref{table:chem2}), and adopted the
same gas-phase atomic abundances of carbon and oxygen. We have also added the
$\mr{H_3^+}$ destruction channel $\mathrm{H_3^+ + e \rightarrow 3 H}$ in
addition to the reaction $\mathrm{H_3^+ + e \rightarrow H_2 + H}$ in
\citet{NL1999}, to have the hydrogen chemistry consistent with
that used in \citet{Glover2010} and this work.
\footnote{Caution needs to be taken here: if the $\mr{H_3^+}$ destruction
channel $\mathrm{H_3^+ + e \rightarrow 3 H}$ is not included, the artificially
elevated $\mr{H_3^+}$ abundance would cause $\CO$ to form much more efficiently 
then it should.}

We run slab models described in Section \ref{section:numerical_method} with
densities $n=50-1000~\mr{cm^{-3}}$, and ambient radiation field
$\chi=1$. We compare the results based on the 
\citetalias{NL1999} network (as described above)
with those from our chemical network.
We similarly run the PDR code, and compare results,
using the same configurations of density and incident radiation field.
Here we focus on the comparison of the chemistry network and fix the temperature at
$T=20~\mr{K}$ and metallicity $Z=1$. We have also made comparisons with the
PDR code at $Z=0.1$, and by solving the chemistry and temperature
simultaneously. The detailed results are shown in Appendix \ref{section:Z0p1}
and \ref{section:TTB}.
The treatment of grain-assisted recombinations in the PDR code is based on the
results in \citet{Wolfire2008}, and is slightly different from the rate in
\citet{WD2001b} that we adopted. Thus, for the purpose of comparison,
we multiply the rates of
reactions 2-5 in Table \ref{table:chem2} by a factor of 0.6 in our models 
to match the rates in the PDR code. For further discussions of the grain-assisted 
recombination rates, see Appendix \ref{section:GR1}.
We use the recent measurement
of the primary cosmic-ray ionization rate per hydrogen
$\xi_\mr{H}=2\times 10^{-16}\mr{s^{-1}H^{-1}}$
\citep{Indriolo2007} if not specified otherwise.
We also run some fixed temperature models with the low cosmic-ray
ionization rate $\xi_\mr{H}=10^{-17}\mr{s^{-1}H^{-1}}$, in order to compare
with results from the
previous literature that adopted this rate \citep[e.g.][]{GC2012}. 

\begin{figure*}[htbp]
\centering
     \begin{center}
      \subfigure[$\xi_\mr{H}=2\times 10^{-16}\mr{s^{-1}H^{-1}}$]{%
            \includegraphics[width=0.49\textwidth]{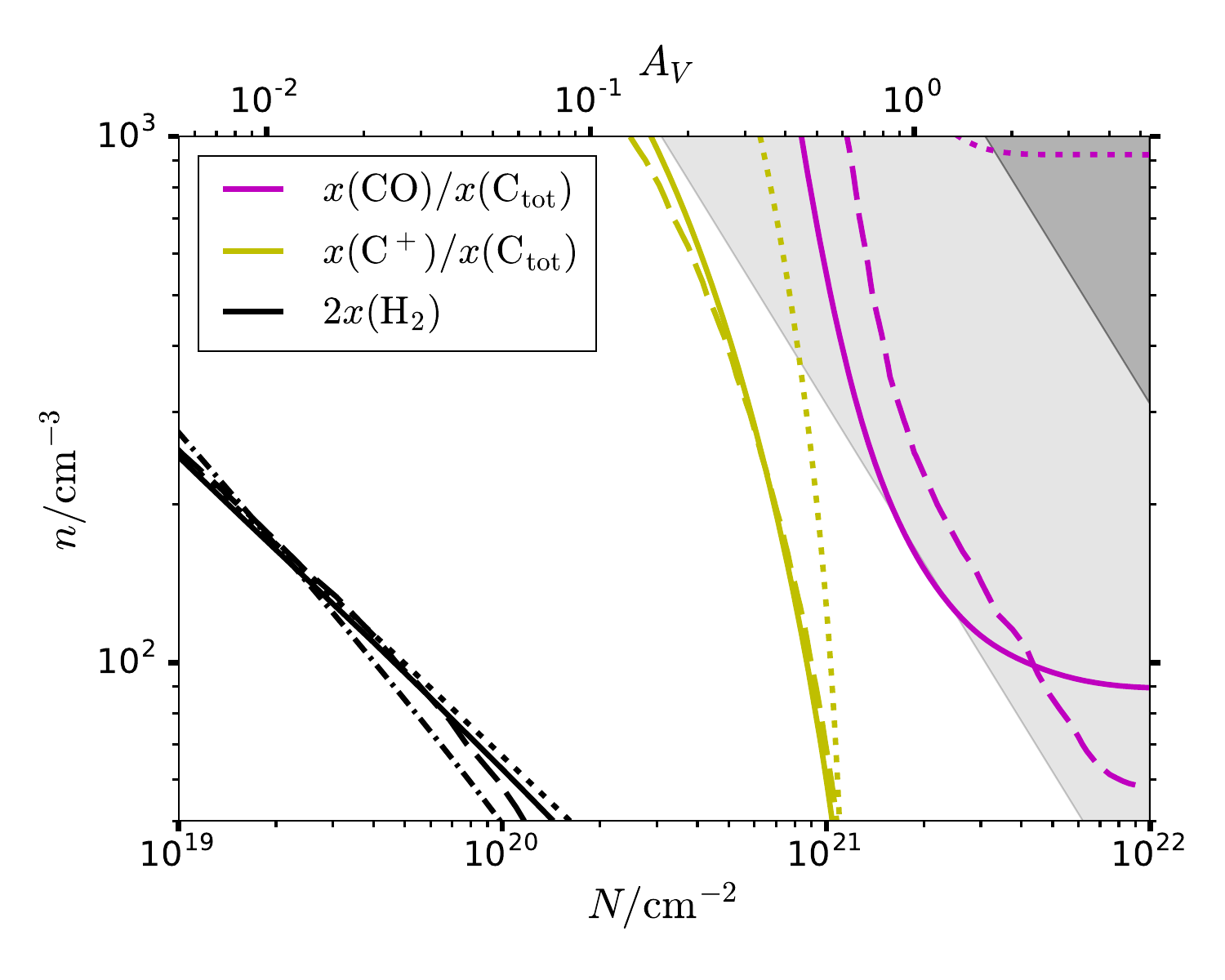}
        }
      \subfigure[$\xi_\mr{H}=10^{-17}\mr{s^{-1}H^{-1}}$]{%
           \includegraphics[width=0.49\textwidth]{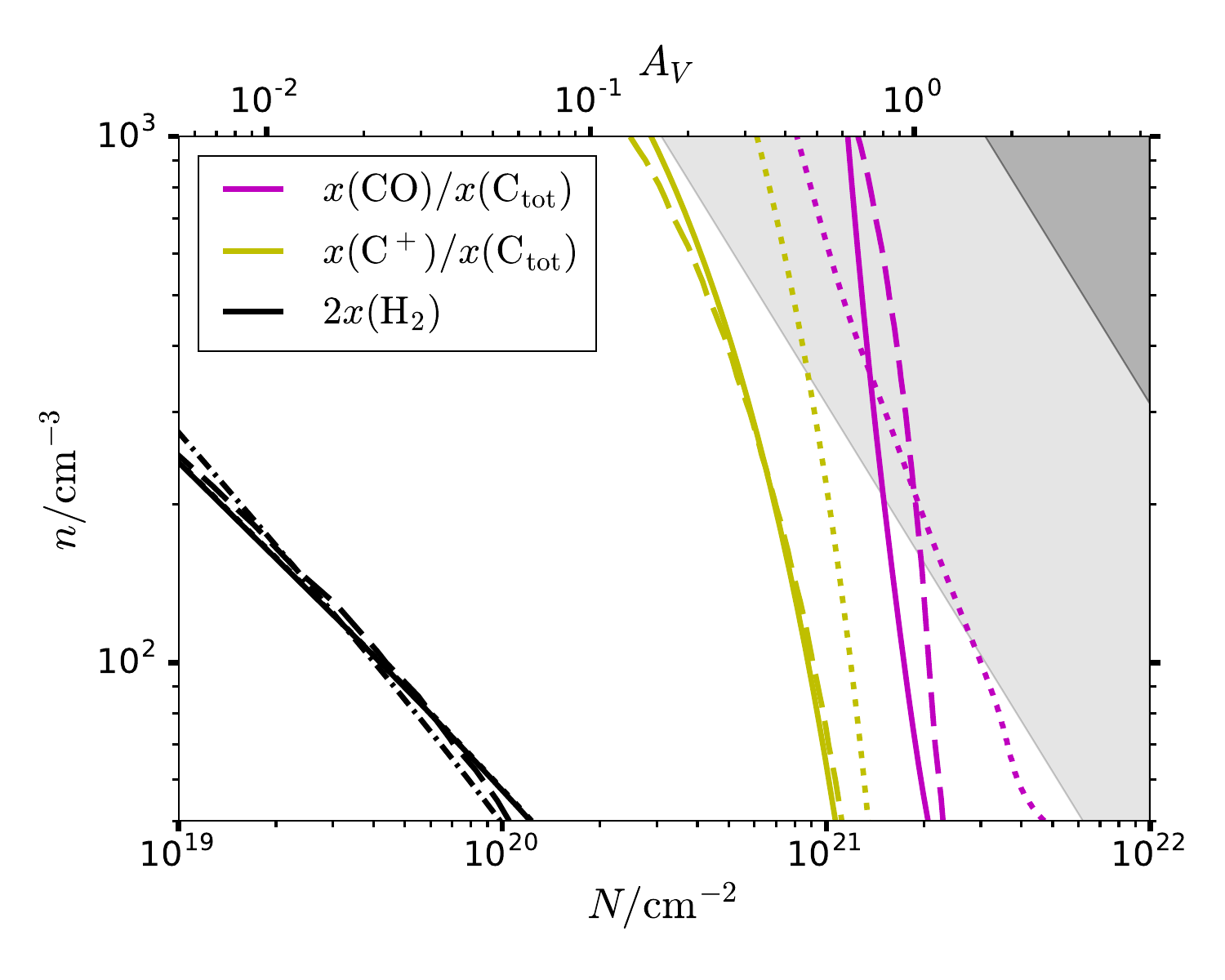}
        } 

    \end{center}
    \caption{Contours showing locations of $\Ho$ to $\Ht$, $\CII$ to $\CI$, 
        and $\CI$ to $\CO$ transitions in the $N$--$n$ plane.
        The magenta, yellow, and black lines are
        contours where $x_\CO/x_\mr{C_{tot}}$, $x_\CII/x_\mr{C_{tot}}$, and
        $2x_\Ht$ are respectively equal to 0.5. 
        The solid, dashed and dotted lines are the results
        of the chemistry network in this paper, the PDR code, and the \citetalias{NL1999}
        network, respectively. The black dashed-dotted line plots the analytic formula in
        \citet{BS2016} for the $\mathrm{H}$ to $\mathrm{H_2}$ transition.
        The left panel (a) is using cosmic-ray ionization rate
        $\xi_\mr{H}=2\times 10^{-16}\mr{s^{-1}H^{-1}}$
        \citep{Indriolo2007}, and the right panel (b) is using the old low
        cosmic-ray ionization rate $\xi_\mr{H}= 10^{-17}\mr{s^{-1}H^{-1}}$.
        The gas temperature is fixed at $T=20\mr{K}$. The regions where the
        $\Ht$ formation timescale is comparable to or shorter than the
        typical turbulence crossing timescale are shaded dark gray
        for $t_\Ht/t_\mr{dyn}<1$ and light gray for $t_\Ht/t_\mr{dyn}<10$ 
        (see Equation (\ref{eq:n_N_H2}) ).
        \label{fig:contour_nH}
       }
\end{figure*}

In our models, from the edge of the slab to the inside, 
the FUV radiation field is attenuated by
both dust and molecular lines. As $N$ and $A_V$ increases, the abundance of
molecular hydrogen increases due to $\Ht$ self-shielding, 
while the atomic hydrogen abundance decreases.
The $\CII$ abundance decreases due to dust
shielding, with most carbon in $\CI$ at intermediate $A_V$. Above $A_V \sim 1$,
cosmic-ray ionization of $\Ht$ creates $\mr{H_3^+}$, which reacts with $\CI$
and $\OI$ to form $\CHx$ and $\OHx$ molecules. $\CHx$ and $\OHx$ molecules
further mediate the formation of $\CO$.

\begin{figure*}[htbp]
     \begin{center}
        \subfigure[$n=100~\mr{cm^{-3}}$]{%
            \includegraphics[width=0.49\textwidth]{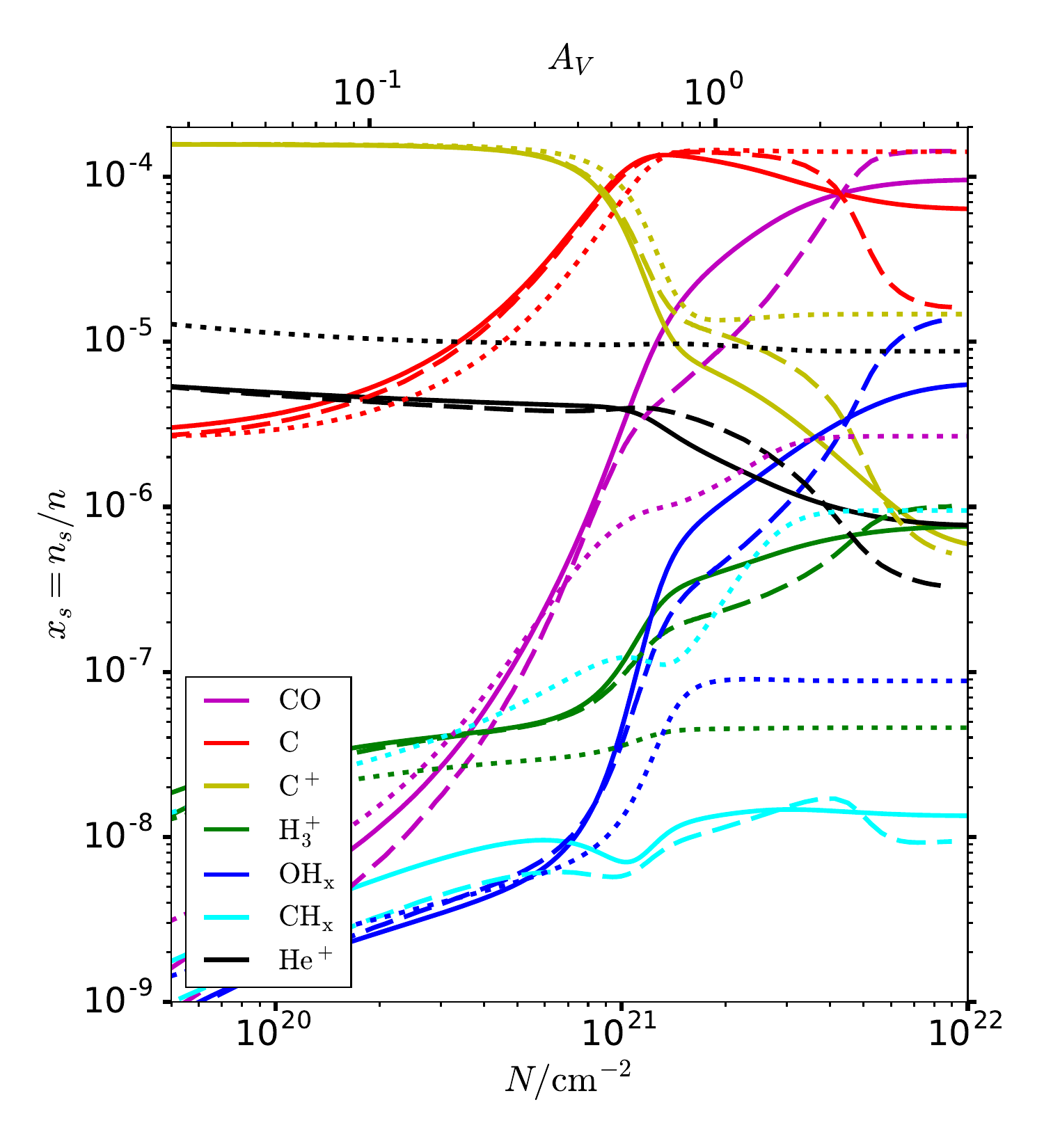}
        }%
        \subfigure[$n=1000~\mr{cm^{-3}}$]{%
           \includegraphics[width=0.49\textwidth]{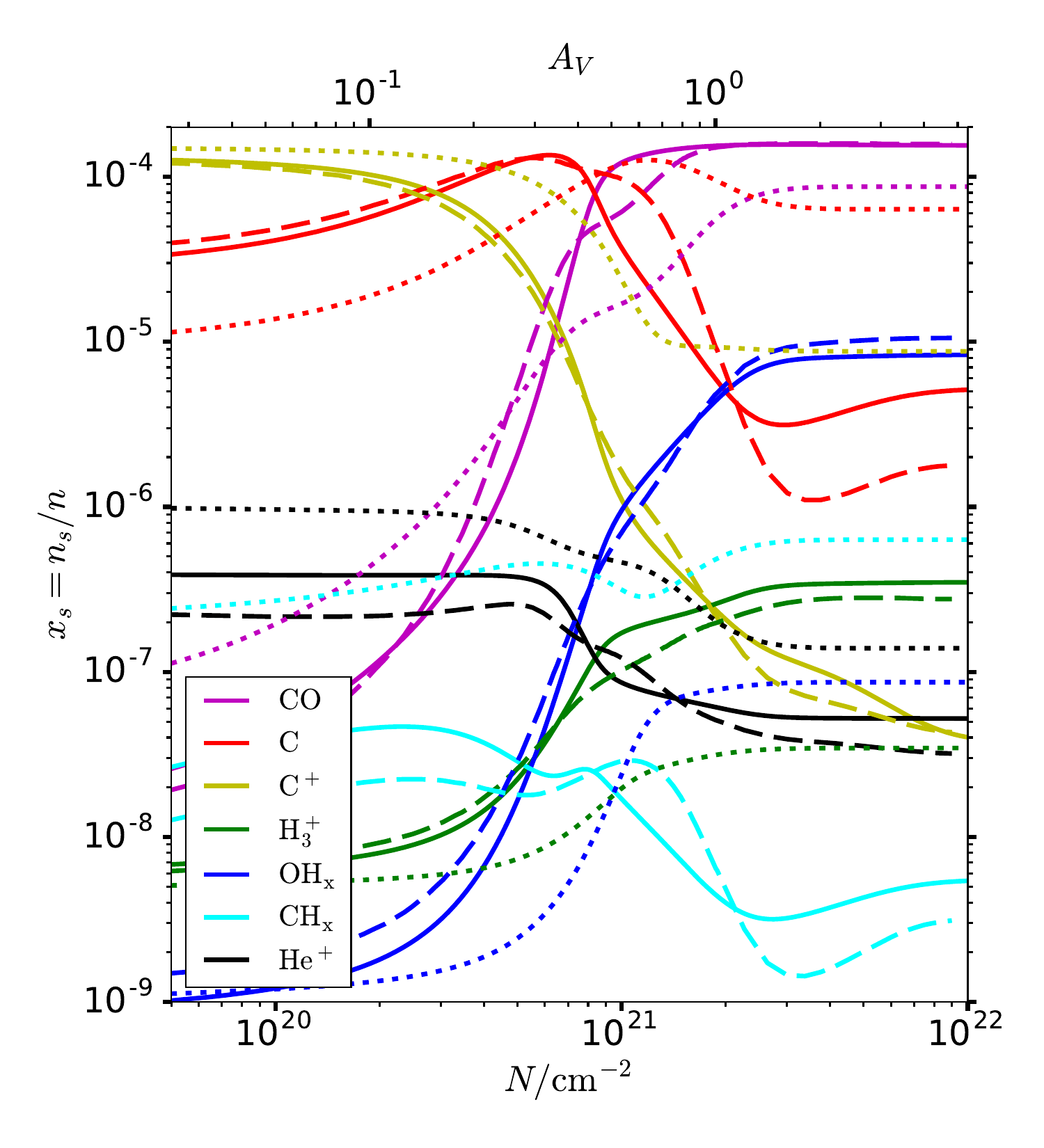}
        } 

    \end{center}
    \caption{Abundances of different species as a function of $A_V$ and $N$ at
        densities
        $n=100\mr{cm^{-3}}$ in panel (a) and $n=1000\mr{cm^{-3}}$ 
        in panel (b). The cosmic-ray ionization rate here is $\xi_\mr{H}=2\times
        10^{-16}\mr{s^{-1}H^{-1}}$, corresponding to panel (a) of Figure
        \ref{fig:contour_nH}. Gas temperature is fixed at
        $T=20~\mr{K}$. See also Figures
        \ref{fig:species_all_nH50-200} and \ref{fig:species_all_nH500-1000} in 
        Appendix \ref{section:add_plots} for abundances of 
        all species at densities between $n=50-1000\mr{cm^{-3}}$. 
        The abundances of different species $x_i=n_i/n$ are
        plotted in different colors: $\CO$ (magenta), $\CI$ (red), $\CII$
        (yellow), $\mr{H_3^+}$ (green), $\OHx$ (blue), $\CHx$ (cyan), and
        $\Heplus$ (black). As in Figure \ref{fig:contour_nH}, the solid,
        dashed and dotted lines respectively represent results
        of the chemistry network in this paper, the PDR code, and the \citetalias{NL1999}
        network. Note that the \citetalias{NL1999} network substantially
        underestimates the $\CO$ abundance at $n=100\mr{cm^{-3}}$,
        and overestimates the $\CII$ abundance  
        (estimates of other species also fail in parts of parameter
        space). Overall, our network shows good agreement with the PDR code
        results.
        \label{fig:species_nH}
    }
\end{figure*}

Figure \ref{fig:contour_nH} illustrates the locations of the 
$\Ho$ to $\Ht$, $\CII$ to $\CI$, and $\CI$ to $\CO$ transitions in 
the $N$--$n$ plane. In panel (a) of Figure \ref{fig:contour_nH}, 
we show that there is a very good agreement on the location of the $\CII$ to $\CI$ 
transition between our chemical network and the PDR code (yellow solid and dashed lines).
There are some differences in the location of $\CI$ to $\CO$ transitions 
(magenta solid and dashed lines),
but the detailed abundance and column of $\CO$ agrees within a factor of $\sim 2$ at a
given $A_V$ (see Figures \ref{fig:species_nH} and \ref{fig:Ni_nH}, and discussions below).
The \citetalias{NL1999} network, however, yields both $\CII$ to $\CI$ and
$\CI$ to $\CO$ transitions at a significant higher $A_V$ (yellow and magenta dotted
lines). Especially at $n \lesssim 500\mr{cm^{-3}}$, the $\CO$ abundance in the
\citetalias{NL1999} network remains very
low up to $A_V=5$. This is inconsistent with observations of $\CO$ in diffuse
molecular clouds with $n \lesssim 200 \mr{cm^{-3}}$, and $A_V \lesssim 1$
\citep{MBM1985, Burgh2007, Sheffer2008, Goldsmith2013}. 
The failure of the \citetalias{NL1999} network here is mainly due to
two reasons. Firstly and most importantly, the \citetalias{NL1999} network does not include $\CII$
recombination on dust grains. Because the rate of $\CII$ recombination on dust
grains is higher than the direct recombination with free electrons, the
$\CII$ to $\CI$ transition is pushed to higher $A_V$, and the $\CII$ abundance at 
$A_V = 1-5$ regions is elevated. This impedes $\CO$ formation: the electrons
from $\CII$ destroy $\mr{H_3^+}$, which is an important reactant for the first
step of $\CO$ formation. Second, the \citetalias{NL1999} network does not include $\Heplus$
recombination on dust grains, which is the main destruction channel for
$\Heplus$ at solar metallicity.
In well-shielded regions, the reaction $\mathrm{He^+ + CO \rightarrow C^+ + O + He}$
is the main channel for $\CO$ destruction. Therefore, the elevated $\Heplus$
abundance also suppresses $\CO$ in the \citetalias{NL1999} network 
(see also Figure \ref{fig:species_nH} for $\mr{H_3^+}$ and $\Heplus$ abundances). 
Without recombination of $\mr{C^+}$ and $\mr{He^+}$ on grains, the transition of
$\CI$ to $\CO$ in the \citetalias{NL1999} network occurs at at a quite high
density, $n>500~\mathrm{cm^{-3}}$.

The $\Ho$ to $\Ht$ transition is the same in our network and the 
\citetalias{NL1999} network, and shows very good agreement with the PDR
code. The same $\Ho$ to $\Ht$ transition in our network and the
\citetalias{NL1999} network is simply because the hydrogen chemistry in both networks are identical,
and $\Ht$ formation is largely unaffected by other species. Our results also
show very good agreement with the analytic formula for the $\Ho$ to $\Ht$
transition in \citet{BS2016} (see their Equation (39)).

\begin{figure*}[htbp]
     \begin{center}
        \subfigure[$n=100~\mr{cm^{-3}}$]{%
           \includegraphics[width=0.49\textwidth]{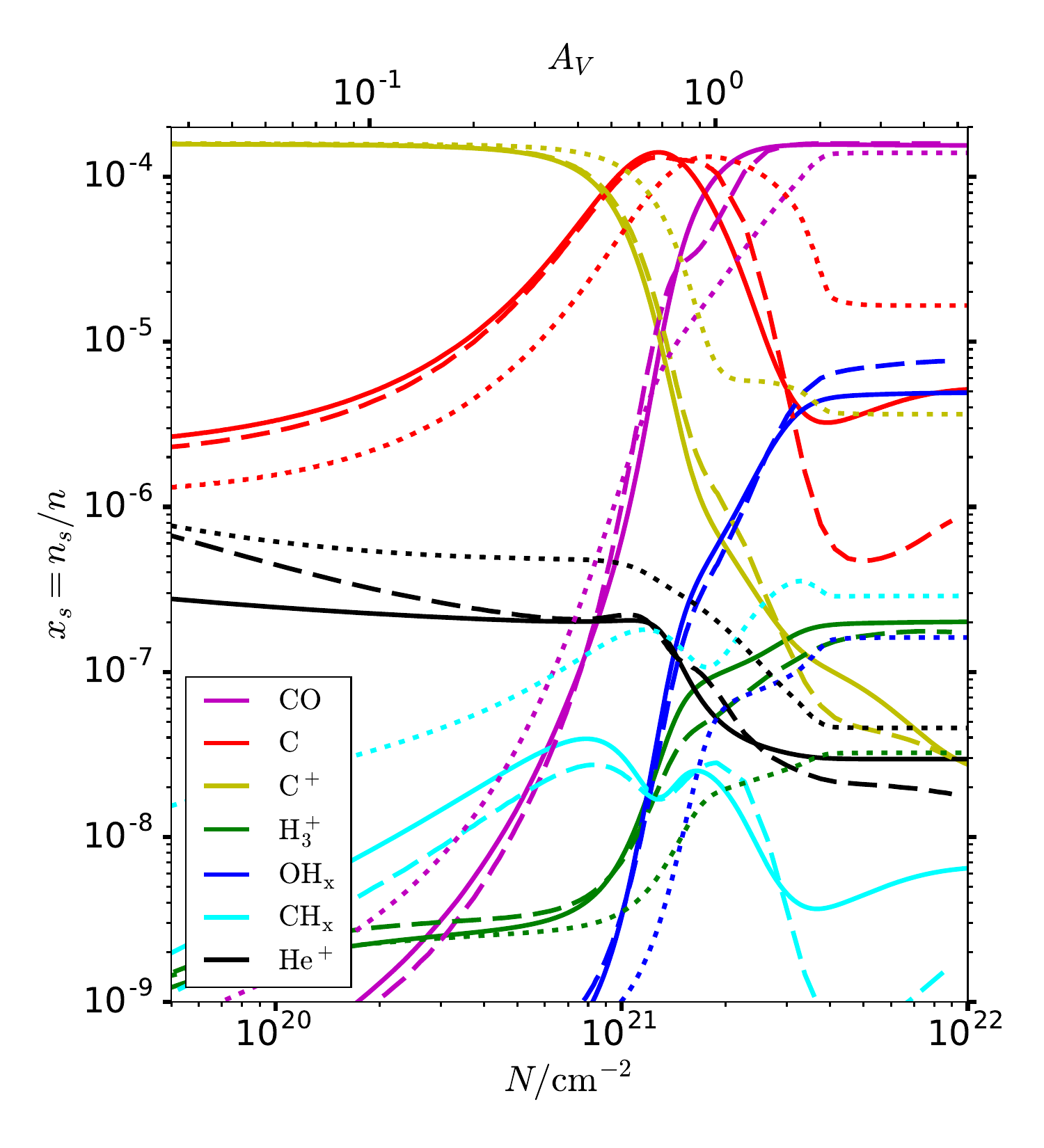}
        }%
        \subfigure[$n=1000~\mr{cm^{-3}}$]{%
           \includegraphics[width=0.49\textwidth]{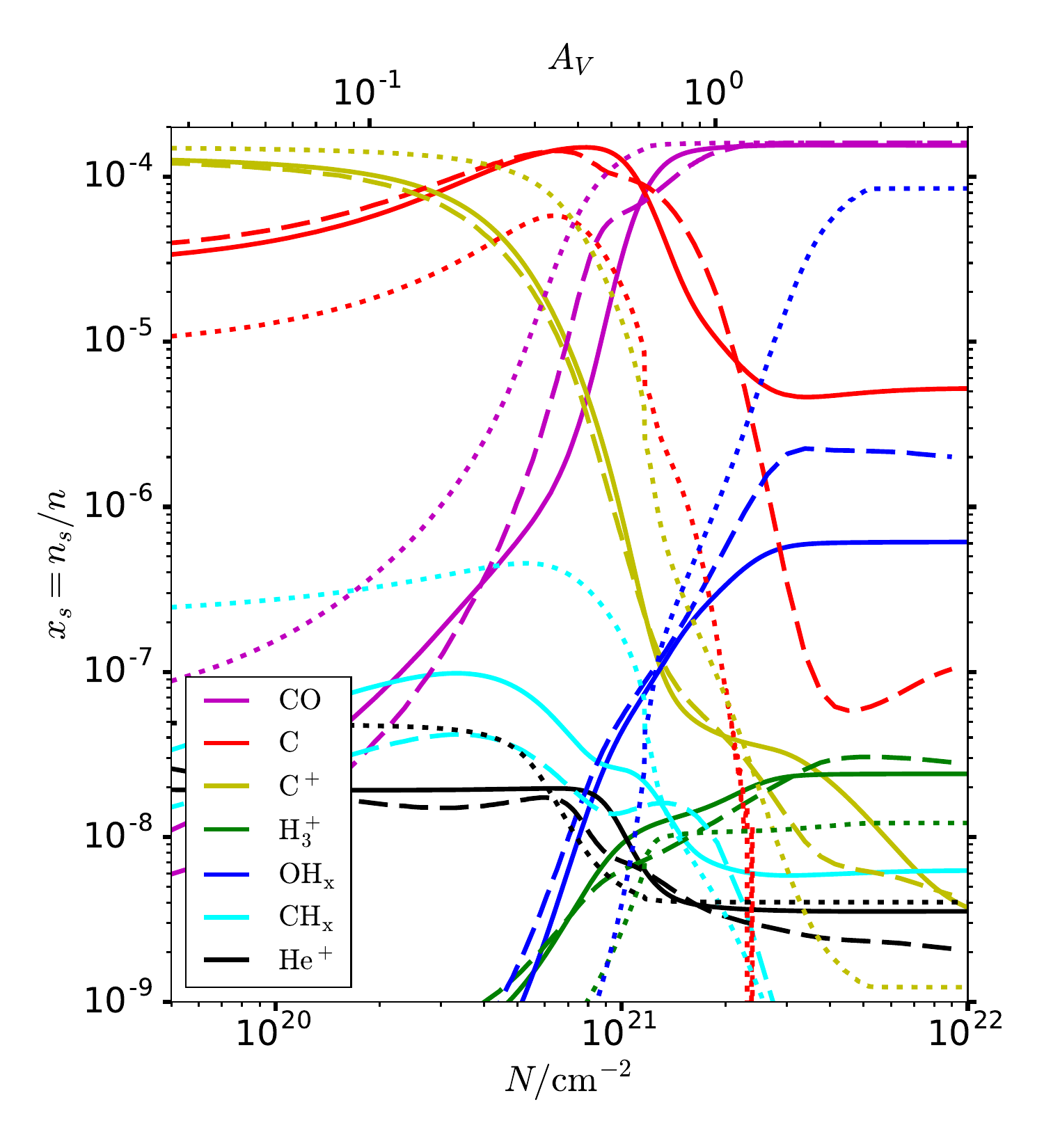}
        } 

    \end{center}
    \caption{Similar to Figure \ref{fig:species_nH}, except with the low cosmic-ray
    ionization rate $\xi_\mr{H}=10^{-17}\mr{s^{-1}H^{-1}}$, as in panel
    (b) of Figure \ref{fig:contour_nH}. At a low cosmic-ray rate, the \citetalias{NL1999}
    network results are more similar to ours, although in certain parts of parameter
    space there are still large discrepancies.
    \label{fig:species_CR1m17_nH}
}
\end{figure*}

\begin{figure*}[htbp]
     \begin{center}
        \subfigure[$n=100~\mr{cm^{-3}}$]{%
           \includegraphics[width=0.49\textwidth]{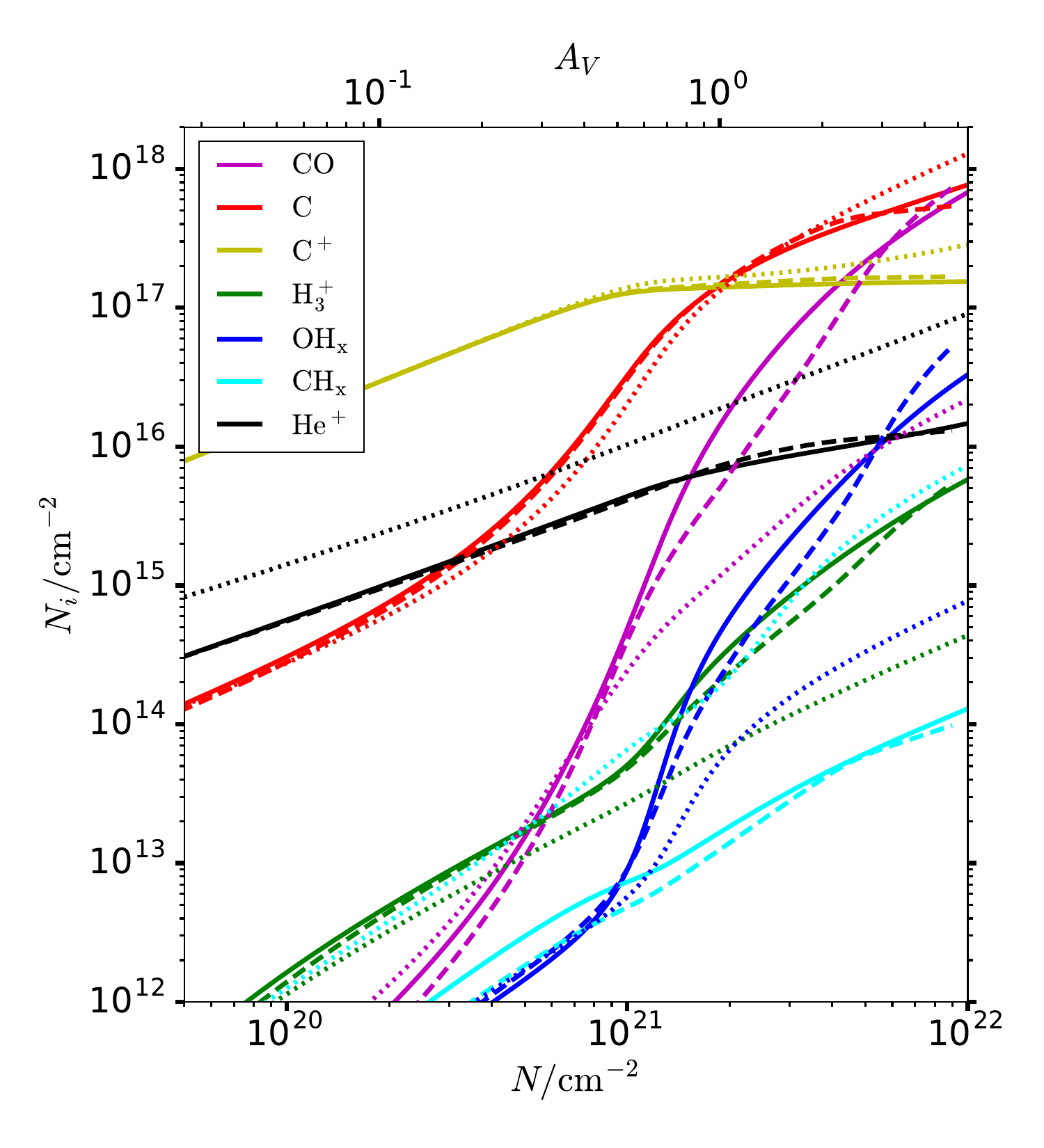}
        }%
        \subfigure[$n=1000~\mr{cm^{-3}}$]{%
           \includegraphics[width=0.49\textwidth]{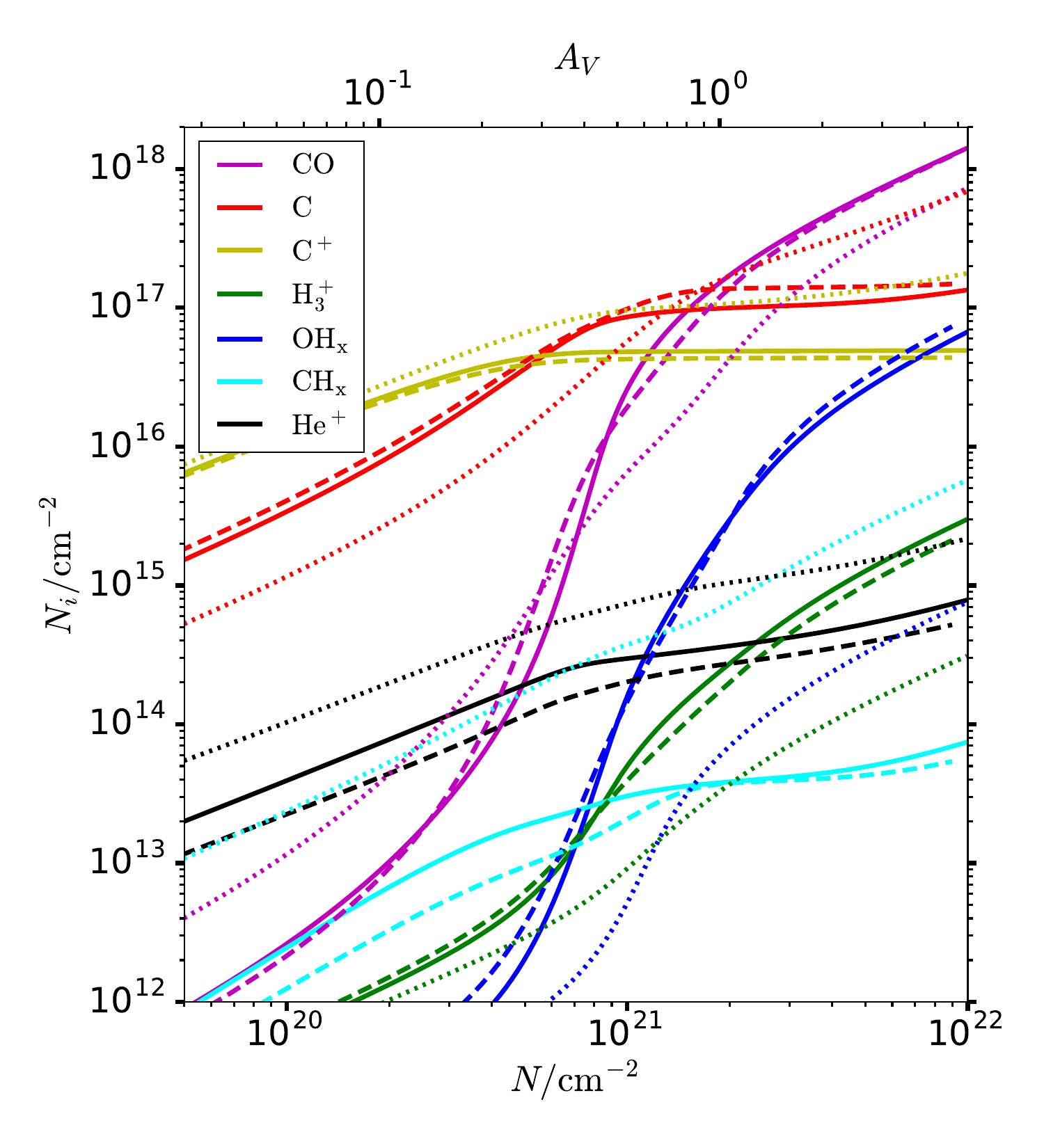}
        } 

    \end{center}
    \caption{Integrated column density of different species $N_i$ as a function
    of $N_H$ and $A_V$, for our network, the PDR code, and the \citetalias{NL1999} network. The
colors and symbols are the same as in Figures \ref{fig:species_nH}. The
cosmic-ray ionization rate here is $\xi_\mr{H}=2\times 10^{-16}\mr{s^{-1}H^{-1}}$.
See also Figures \ref{fig:Ni_all_nH50-200} and \ref{fig:Ni_all_nH500-1000} in
Appendix \ref{section:add_plots} for all species at densities 
$n=50-1000\mr{cm^{-3}}$.
    \label{fig:Ni_nH}
}
\end{figure*}

Panel (b) in Figure \ref{fig:contour_nH} shows an interesting comparison using
the low cosmic-ray ionization rate $\xi_\mr{H}=10^{-17}\mr{s^{-1}H^{-1}}$. With
a low cosmic-ray rate, the
location of the $\CI$ to $\CO$ transition is similar in the \citetalias{NL1999} network and our
network. This is because with the low cosmic-ray rate, $\CO$
destruction by $\Heplus$ is less efficient due to the lower abundance of $\Heplus$
created by cosmic rays, and the $\CO$ abundance is now mainly limited by
photodissociation. The dependence of the photodissociation rate on dust
shielding determines the $A_V$ at which $\CO$ forms. 
Although $\CO$ abundances are similar at
$\xi_\mr{H}=10^{-17}\mr{s^{-1}H^{-1}}$, there are significant differences in the
abundances of other species such as $\CII$ between the \citetalias{NL1999} network and our
network (see Figure
\ref{fig:species_CR1m17_nH}). 
Thus, the apparent similarity in the $\CO$ transition at the low cosmic-ray rate is
deceptive. Using the low cosmic-ray ionization rate,
\citet{GC2012} compared the \citetalias{NL1999} network with the larger \citet{Glover2010}
network, considering the $\CO$ distribution in turbulent clouds. 
They reached the conclusion that the
\citetalias{NL1999} network is sufficiently accurate for simulating $\CO$ formation. 
However, here we show that
with a more realistic cosmic-ray ionization rate,
the \citetalias{NL1999} network significantly underestimates the $\CO$ abundance, and the grain
assisted recombinations of $\CII$ and $\Heplus$ are very important for
obtaining correct $\CO$ abundances.

Figures \ref{fig:species_nH} and \ref{fig:Ni_nH} plot the
abundances of different species as a function of $A_V$ and $N$ 
for the three networks at 
$\xi_\mr{H}=2\times 10^{-16}\mr{s^{-1}H^{-1}}$.  Figure 2 shows the
abundance comparisons for $\xi_\mr{H}=10^{-17}\mr{s^{-1}H^{-1}}$.
Overall, there is a very good
agreement between our chemical network and the PDR code for $\CII$, $\CI$,
$\CO$, and $\mr{H_3^+}$, which are the main observable species in our network.
It is remarkable that our network also reproduces the column
densities $N_i$ at a given $A_V$
of all other species with errors less than a factor of $\sim 2$,
even for the largely simplified pseudo-species $\OHx$ and $\CHx$. This
indicates that our simplified network successfully captures the main chemical
pathways of these species. There are also some differences to be pointed out:
Compared to the PDR code, the $\CO$ abundance is lower at $n=100
\mr{cm^{-3}}$ and $A_V\gtrsim 2$, and higher at $A_V \sim 1$ in our network.
This is partly due to the difference in the grain-assisted recombination rates,
as discussed in Appendix \ref{section:GR1}, and partly due to the approximation of using
pseudo-species $\OHx$, which does not capture all the details of $\OHx$
formation and destruction. 
For similar reasons that lead to the difference in the $\CO$ abundance, 
there are also discrepancies in the abundances of other species such as  
$\CI$, $\Heplus$, $\OHx$, and $\CHx$ at $n=100 \mr{cm^{-3}}$
and $A_V \gtrsim 2$. However, since the differences mainly occur in a limited range of $A_V$, 
the differences in column density in Figure \ref{fig:Ni_nH} are small.

The \citetalias{NL1999} network, however, fails to reproduce the equilibrium
abundances in the PDR code. The $\CO$ abundance is too low as a result of
the high $\CII$ and $\Heplus$ abundances, as discussed above.
The \citetalias{NL1999} network also leads to a much higher $\CHx$ abundance,
due to the absence of the important destruction channel of $\mr{CH}$ 
reacting with $\Ho$.
Therefore, we conclude that our new chemical network is preferred over the \citetalias{NL1999}
network. With one additional species ($\Oplus$) and nineteen additional
reactions, we find that the increase in computational cost is about 50\%.

To cross-check with other PDR codes, we have also run a comparison with the 
publicly available {\sl PyPDR}\footnote{\url{
http://www.mpe.mpg.de/~simonbr/research_pypdr/index.html}} 
code provided by Simon Bruderer. {\sl PyPDR} has been shown 
to agree well with the PDR
bench mark in \citet{Rollig2007}, and therefore is a good point of comparison
with the existing model literature. However, {\sl PyPDR} does not include grain-assisted
recombination of ions, which is important for $\CO$ formation as discussed above.
Appendix \ref{section:PyPDR} shows that our chemical network (without grain
assisted recombinations) agrees very well with the {\sl PyPDR} code,
which further validated our results. 

\section{Sample applications}\label{section:application}
Observations of a wide range of galaxies show a correlation
between the surface density of the
star formation rate (SFR) and the surface density of molecular gas
\citep[e.g.][]{Bigiel2008, Bigiel2011, Saintonge2011b, Genzel2010, Jameson2015}.
Although in some models it has been proposed that star formation is
associated with gas that is shielded enough for $\Ht$ to form \citep{KMT2009},
any relationship with $\Ht$ would have to be coincidental rather than causal,
because $\Ht$ is not an active coolant in cold gas.
In principle, $\Ht$ might be important because it is a prerequisite for $\CO$
formation, and $\CO$ is the dominant coolant in the environment of star
formation, at least when the metallicity is not extremely low \citep{GC2012b}.
However, the cooling provided by $\CII$ can also bring gas to quite low
temperatures, so that it is able to collapse gravitationally at small scales
\citep{GC2012a}. Thus, the association of star formation with the molecular
phase may simply be because both chemical and dynamical timescales are shorter
at higher density, and the shielding that limits photodissociation also 
limits photoheating \citep[see also][]{KLM2011}. 

Regardless of the reason for
the correlation between molecular gas and star formation, quantifying the
relationship between gas and star formation empirically requires tracers of all
gas phases. Atomic gas is easily identified with the 21 cm transition, but 
$\Ht$ is difficult to observe directly due to the high excitation temperature
of its rotational levels. $\CO$ is often used as a tracer of $\Ht$, but
$\CO$ is known to be difficult to detect in metal-poor galaxies, and even when 
detected, the star formation rate per $\CO$
luminosity is much higher than that in the Milky Way-like galaxies 
\citep[e.g.][]{Taylor1998, Leroy2007, Schruba2011, Schruba2012, Hunt2015}.
Moreover, $\CO$ is not necessarily a linear tracer of $\Ht$ \citep{Bolatto2013}.

Here, we use simple slab models to explore the correlations between chemical
state (especially $x(\CO)$ and $x(\Ht)$) and temperature, while also exploring
the dependence of both properties on $n$, $N$, $\chi$, $\xi_\Ho$ and
$Z$. We also explore the mean abundances of various species on average cloud
density and column, and compare with observations. 

\subsection{$\CO$ as a Tracer of $\Ht$ and Cold Gas \label{section:CO_trace_H2}}
Following \citet{KLM2011}, we investigate how well $\CO$ and $\Ht$ trace
cold gas in the ISM with different metallicities.
\citet{KLM2011} used semi-analytic models
to estimate the $\Ht$ and $\CO$ abundances and separately computed the
equilibrium temperature for gas cooled either by $\CII$ or by $\CO$.
Here, we instead use our chemical network and self-consistent cooling to solve
for the equilibrium chemical composition
and temperature of the gas. We run one-dimensional slab
models described in Section \ref{section:numerical_method} at a range of
densities $n = 10^1 - 10^4 \mathrm{cm^{-3}}$, and compute the chemical
and temperature state of the gas in the $A_V$--$n$ plane.
We also run models with different metallicities, incident
radiation field, and cosmic-ray ionization rate to study the dependence of gas
properties on these parameters.

It should be noted that in the realistic ISM,
the timescales of cooling and chemical reactions are different, and the dynamical
timescales set by turbulence may be shorter than these. Thus, the equilibrium
chemical and thermal state does not necessarily apply in the real ISM. This is
particularly an issue for $\Ht$, because the formation time is $t_\Ht \approx
10^7~\mr{yr}~Z_d^{-1} \left(\frac{n}{100\mr{cm^{-3}}}\right)^{-1}$ 
at $T=100\mr{K}$ (see line 24 in Table \ref{table:chem1}), whereas typical dynamical
timescales in dense gas are (see Equation (\ref{eq:v_L})) 
\begin{equation}\label{eq:t_dyn}
    t_\mr{dyn} = L/v_\mr{turb}(L) \sim 1\mr{Myr}~(L/\mr{pc})^{1/2}.
\end{equation}

For transient clouds produced by turbulent compression,
even if the mean shielding column is sufficient
for the equilibrium $\Ht$ abundance to be high, the $\Ht$ formation timescale
may be longer than the cloud lifetime.
For example, the criterion $t_\Ht < t_\mr{dyn}$ is equivalent to
\begin{equation}\label{eq:n_N_H2}
\begin{split}
    Z_d^2 \left(\frac{n}{100~\mr{cm^{-3}}}\right)
    \left(\frac{N}{10^{21}~\mr{cm^{-2}}}\right) &> 31,~\text{or}\\
    Z_d A_V \left(\frac{n}{100~\mr{cm^{-3}}}\right) &>17
\end{split}
\end{equation}
which is well above the equilibrium $\mr{H}$/$\Ht$ transition in Figure
\ref{fig:contour_nH}.

In high column density self-gravitating 
clouds with supersonic turbulence, the mass-weighed density
$\langle n \rangle_M$ is proportional to $N^2$, $\langle n \rangle_M
\propto N^2$ (See Equation (\ref{eq:n_M0}), where the second term dominates
over the first term).  Combining this with Equation (\ref{eq:n_N_H2}), gives
\begin{equation} \label{eq:n_N_H2_cloud}
\begin{split}
    N &\gtrsim 4.3\times 10^{21} Z_d^{-2/3}~\mr{cm^{-2}}, ~\text{or}\\
    A_V &\gtrsim 2.3~Z_d^{-2/3}
\end{split}
\end{equation}
for $t_\Ht < t_\mr{dyn}$. Equation (\ref{eq:n_N_H2_cloud}) suggests that for a
molecular cloud with $A_V\gtrsim 1$, although molecular hydrogen is difficult
to form at an average density $n \sim 100~\mr{cm^{-3}}$ (Equation
(\ref{eq:n_N_H2})), turbulence can compress most of the gas to reach to a
higher density regime and enable the cloud to become molecular in a shorter
timescale.

For gravitationally bound clouds that live longer than several
flow-crossing times, the dynamical cycling of gas to
the unshielded surface may limit the molecular abundance.
The timescale for a fluid element in the cloud to travel through the unshielded
region is $t_\mr{sh} = L_\mr{shield} / v_\mr{turb}(L_\mr{cloud})$, where
$L_\mr{shield}$ is the length scale for $\Ht$ to be self-shielded against the
FUV radiation. Using Equation (52) in \citet{Sternberg2014}, 
and the $\Ht$ formation rate on dust (see line 1 in Table \ref{table:chem2}),
the shielding
length\footnote{This uses the approximation of 
$\ln(\alpha G / 2 + 1)\approx \alpha G /2$ (see
Equation (40) in \citet{Sternberg2014}). For the CNM in
pressure equilibrium (Equation (\ref{eq:nH2G0_G})), $\alpha G/2 \approx 1.3$,
which makes the approximation good up to a factor of $\sim 3$. 
At $\alpha G/2 \approx 1$, most of the $\mr{H}$ column is built up in atomic
regions (see Figure 7 in \citet{Sternberg2014}), and therefore we calculate the
shielding length by $L_\mr{shield} = N/n$. Note, however, in the
limit of $\alpha G/2 \gg 1$, i.e., for low density gas or highly
illuminated PDR such as observed in \citet{Bialy2015}, the dependence changes to 
$N\sim 2/\sigma_g \sim 10^{21} \mr{cm^{-2}}$, independent of $\alpha
G$, giving  $L_\mr{shield} \sim 3\mr{pc}~[n/(100 \mr{cm^{-3}})]^{-1}$. }
\begin{equation}
    L_\mr{shield} = \frac{N}{n} \approx 1~\mr{pc}~\chi 
    \left( \frac{n}{100~\mr{cm^{-3}}} \right)^{-2},
\end{equation}
where $n$ is the average (volume-weighted) density of the cloud.
Adopting $v_\mr{turb}(L) \sim 1~\mr{km/s}~(L/\mr{pc})^{1/2}$, this yields:
\begin{equation}
    t_\mr{sh} = 0.2~\mr{Myr}~\chi \left( \frac{n}{100~\mr{cm^{-3}}} \right)^{-2}
    \left( \frac{L_\mr{cloud}}{10\mr{pc}} \right)^{-1/2}.
\end{equation}
The timescale for $\Ht$ photodissociation is 
$t_\mr{diss, H_2}=1/k_\mr{diss, H_2} = 6\times 10^2~\mr{yr}$
(the photodissociation rate $k_\mr{diss, H_2}=5.6\times 10^{-11}~\mr{s^{-1}}$,
see line 19 in Table \ref{table:chem2}). Setting $t_\mr{sh} < t_\mr{diss, H_2}$ gives
\begin{equation}\label{eq:ts_H2}
    \frac{1}{\chi} \left( \frac{n}{100~\mr{cm^{-3}} } \right)^2
    \left( \frac{L_\mr{cloud}}{10\mr{pc}} \right)^{1/2} > 3\times 10^2;
\end{equation}
when this condition is satisfied, molecules can avoid being destroyed by 
photodissociation over
the timescale to cross the unshielded surface of the cloud. For example, a
cloud of size $10~\mr{pc}$ will require its density
$n > 1.7\times 10^3~\mr{cm^{-3}}$ to be fully molecular.
Equation (\ref{eq:ts_H2}) shows that even in gravitationally bound clouds, the
cycling of gas from the cloud interior to the surface can also limit the $\Ht$
abundance in low density regimes.
If the $\Ht$ abundance is reduced by exposure to UV, other
chemical pathways will also be strongly affected.
Therefore, to fully investigate how well $\CO$ traces molecular gas,
numerical simulations with time-dependent chemistry are needed.
Still, we show that our simple models can
provide insight into this important but complicated problem.

\begin{figure*}[htbp]
    \begin{center}
        \includegraphics[width=0.8\textwidth]{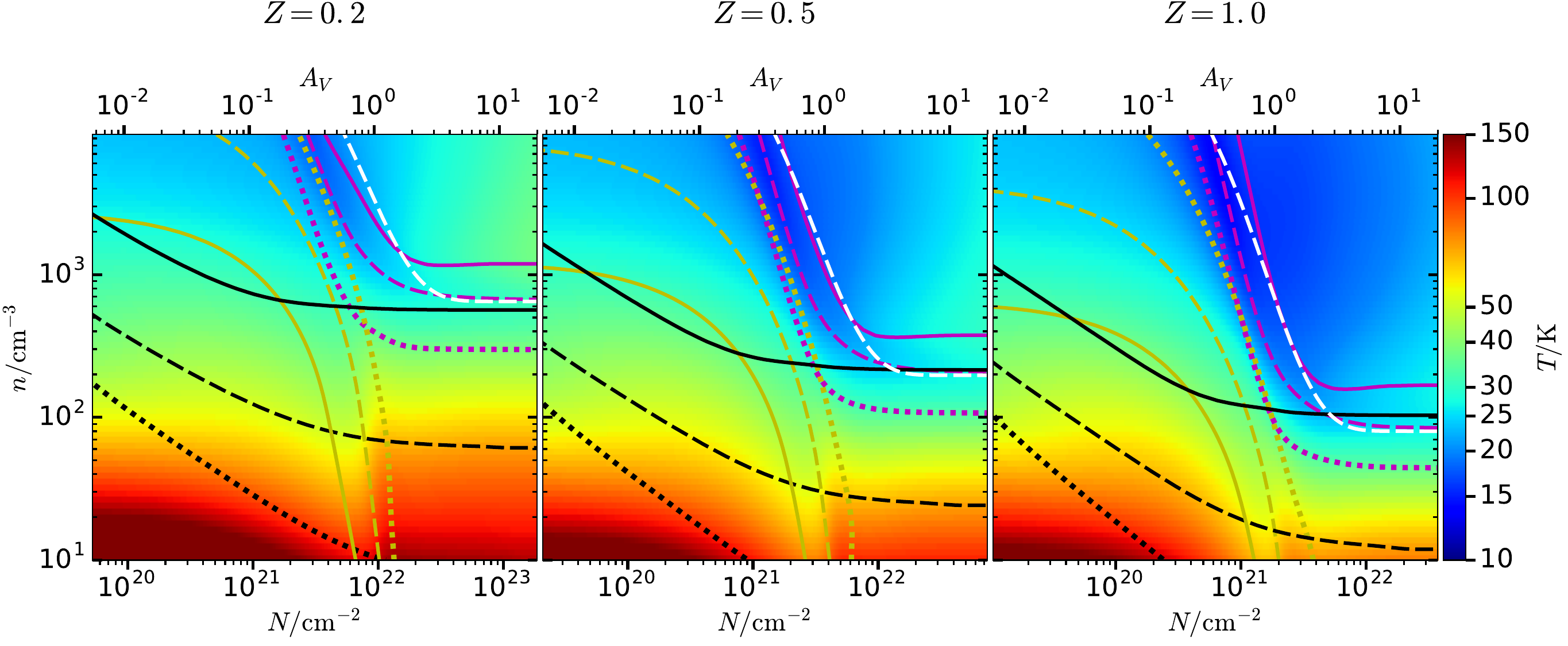}
    \end{center}
    \caption{Equilibrium temperature of the 
        gas as a function of $A_V$ (or $N$) and density $n$.
        The black, yellow, and magenta lines are
        contours of the equilibrium values for
        $2x(\Ht)$, $x(\CII)/x_\mr{C, tot}$ and $x(\CO)/x_\mr{C, tot}$ equal
        to 0.1 (dotted), 0.5 (dashed), and 0.9 (solid). The white dashed line shows
        the fit for the $x(\CO)/x_\mr{C, tot}=0.5$ contour (magenta dashed line) in Equation
        (\ref{eq:CO_fit}). From left to right, the gas metallicities and
        dust abundances are set to be $Z=0.2$, $0.5$, and $1.0$. The incident
        normalized radiation field $\chi=1$, and cosmic-ray ionization rate 
        $\xi_\Ho = 2 \times 10^{-16}\mr{s^{-1} H^{-1}}$.
        \label{fig:KLM_G1_T}
    }
\end{figure*}

Figure \ref{fig:KLM_G1_T} shows the equilibrium gas temperature in color scale, 
in comparison to the contours of equilibrium $\Ht$, $\CO$, and $\CII$ abundances. 
Across a wide range of metallicities $Z=0.2-1$, Figure \ref{fig:KLM_G1_T}
shows that gas that is mostly molecular
is also at $T\lesssim 40 \mr{K}$. However, it is important to note that the coincidence
of high $\Ht$ abundance and low temperature is not causal. For gas in the range
of Figure \ref{fig:KLM_G1_T}, most of the cooling is provided 
by $\CII$, $\CI$, and $\CO$. Even if we artificially turn off the formation of
$\Ht$ and $\CO$, the gas can still be cooled to $T\sim20\mr{K}$ with $\CI$ and
$\CII$ cooling. We discuss this further in Section
\ref{section:molecule_cooling}.  At low metallicity, the minimum
density required for gas to become molecular increases.
This is because $\Ht$ forms less efficiently when the dust
abundance drops. 
At low enough $n$, the equilibrium $x(\Ht)$ is low even in very shielded regions. 
This is due to the destruction of $\Ht$ by cosmic rays. 
The gas temperature is similar at $A_V \lesssim 1$ across $Z=0.2-1$, because
both the photoelectric heating and gas cooling are proportional to $Z$.
At $A_V \gtrsim 1$, and also in low density and low metallicity
regions, the cosmic-ray heating dominates, which is independent of metallicity.
This leads to an increase of the gas temperature in these regions at low
metallicity due to decreased gas cooling.

At solar metallicity ($Z=1$), Figure \ref{fig:KLM_G1_T} shows that 
the region of high $\CO$ abundance (where $\CO$ line cooling dominates)
traces the lowest temperature $T \lesssim 20\mr{K}$
gas very well\footnote{However, there is a subtlety in this: 
even if we artificially force
$\CO$ not to form, $\CI$ would have cooled the gas to similarly low temperatures
(see Section \ref{section:molecule_cooling}).}. At very low metallicities, 
there is less dust shielding and collisional reactions between metal
species occur at a reduced rate, leading to less efficient $\CO$ formation, and 
$\CO$ traces the temperature less well. 

\subsection{Dependence on $\chi$, $\xi_\Ho$, and $Z$}
\begin{figure*}[htbp]
     \begin{center}
        \subfigure[$\chi$]{%
            \includegraphics[width=0.8\textwidth]{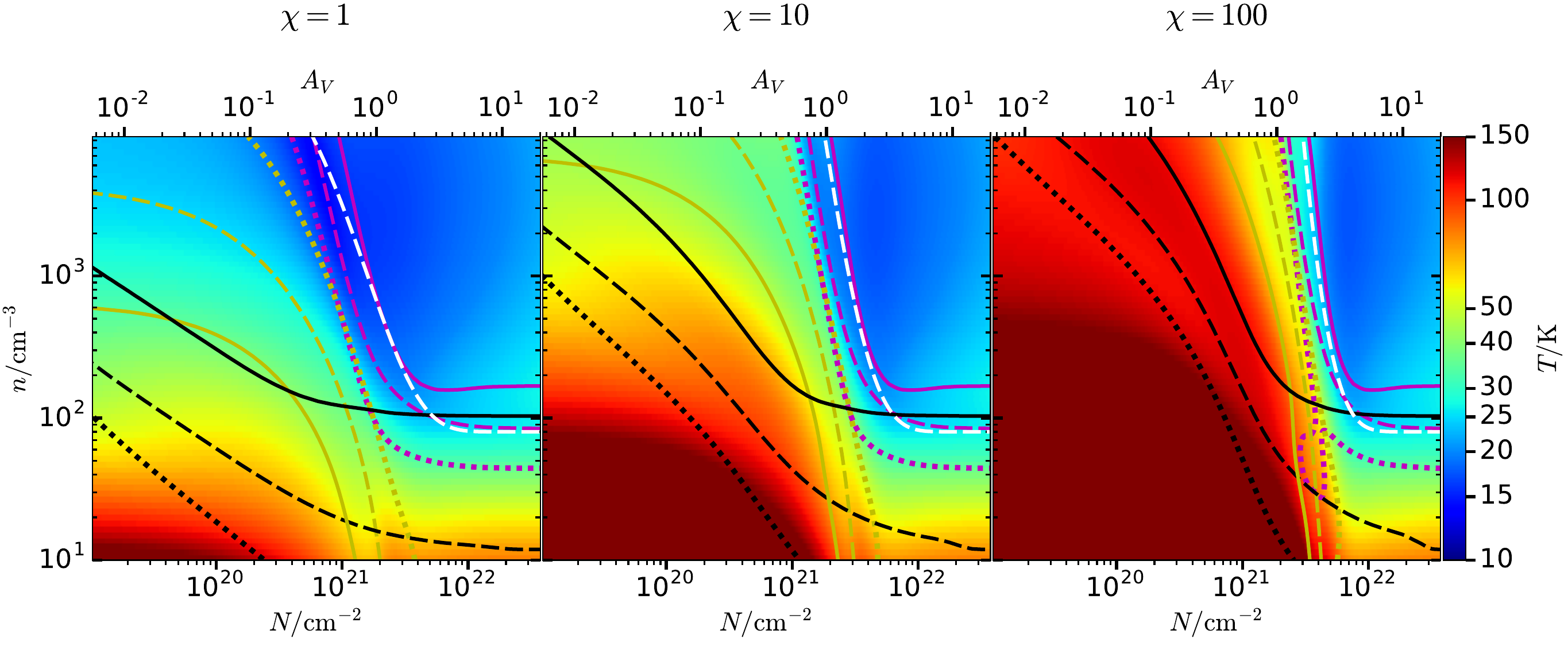}
        }\\
        \subfigure[$\xi_\Ho$]{%
            \includegraphics[width=0.8\textwidth]{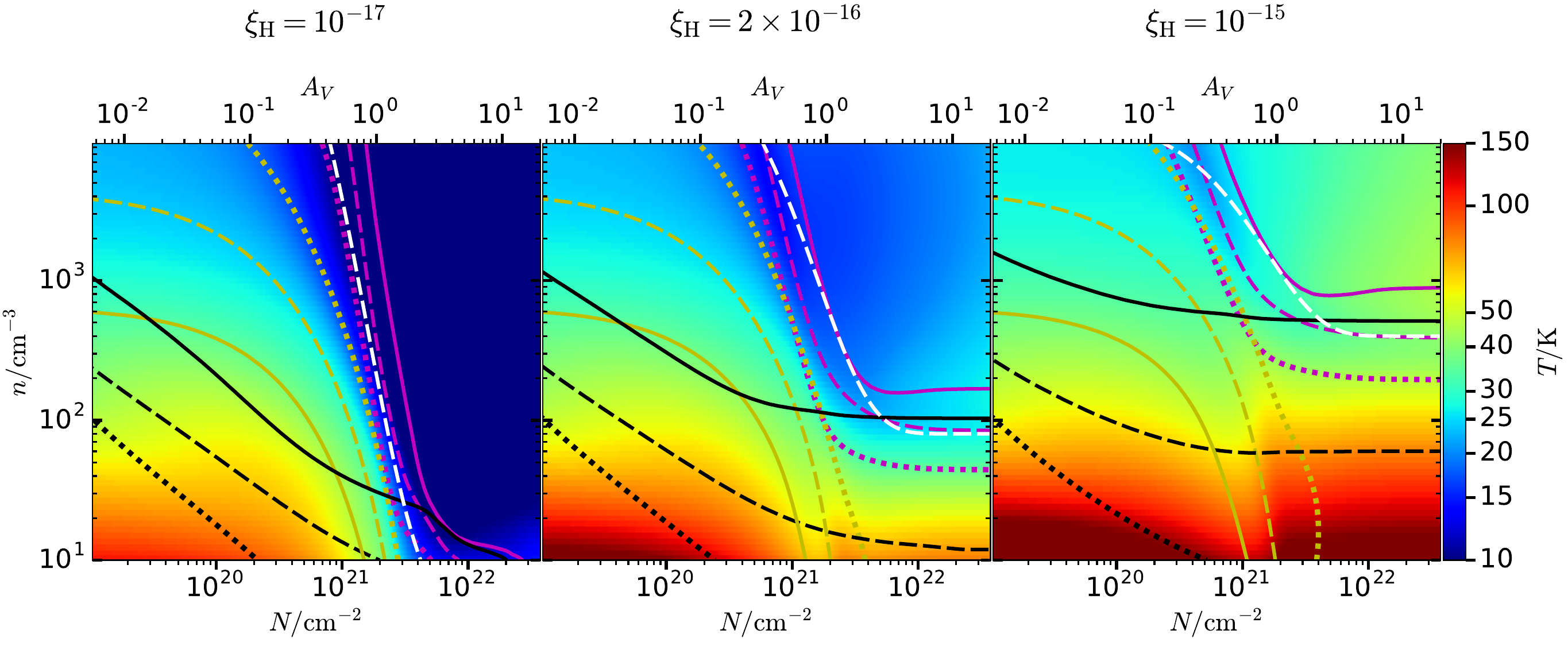}
        }
    \end{center}
    \caption{Gas temperature as a function of $A_V$ (or $N$) and $n$
        with solar metallicity $Z=1$. Panel (a) varies the incident FUV radiation
field strength $\chi=1$, $10$, and $100$ while keeping the cosmic-ray ionization
rate $\xi_\Ho=2\times 10^{-16}~\mr{s^{-1} H^{-1}}$. 
Panel (b) varies the cosmic-ray ionization rate 
$\xi_\Ho=10^{-17}$, $2\times 10^{-16}$, and $10^{-15}~\mr{s^{-1} H^{-1}}$,
while keeping $\chi=1$.
The contours of the $\Ht$, $\CO$, and $\CII$ abundances, and the fit of the
$x(\CO)/x_\mr{C, tot}$=0.5 contour, are also plotted, 
similar to Figure \ref{fig:KLM_G1_T}.
\label{fig:KLM_TCO}
}
\end{figure*}

Figure \ref{fig:KLM_TCO} (a) shows that an increase of $\chi$ pushes the 
$\Ht$-dominated region to higher $A_V$, and that even $\Ht$ dominated gas can be at
$T>100\mr{K}$ if $\chi$ is large enough. Increasing the cosmic-ray rate requires
higher density for the gas to become $\Ht$-dominated, because cosmic rays can
dissociate $\Ht$ (Figure \ref{fig:KLM_TCO} (b)).

An increase (decrease) of $\chi$ moves the boundary of $\CO$-dominated gas to
higher (lower) $A_V$, because $\CO$ only forms in dust-shielded regions (Figure
\ref{fig:KLM_TCO} (a)). Similarly, an increase (decrease) of $\xi_\Ho$ pushes
the minimum density for $\CO$ up (down), because $\Heplus$ and $\CII$ formed in
cosmic-ray reactions are harmful for $\CO$ formation (Figure \ref{fig:KLM_TCO} (b)).
However, in both Figure \ref{fig:KLM_TCO} (a) and Figure \ref{fig:KLM_TCO} (b),
the contours defining the $\CO$-dominated region also clearly delimit the low
temperature ($T\lesssim 20\mr{K}$) gas, except at the highest levels of
$\xi_\Ho$. The temperature of the bulk of the gas in the $\CO$-dominated
regime depends on $\xi_\Ho$ but not on $\chi$. 
This is because when $x(\CO)/x_\mr{C, tot}\approx 1$, the heating and cooling are
respectively dominated by cosmic-ray ionization and $\CO$ rotational
transitions. The temperature is obtained by balancing Equation
(\ref{eq:Gamma_cr}) (with Equation (\ref{eq:q_cr})) with Equation
(\ref{eq:Gamma_CO}). 
We suggest that the temperature of $\CO$, especially the optically thin isotopes,
can be an indicator to probe the cosmic-ray ionization rate in dense molecular
gas. Observations of star forming regions near the galactic center indicate
that $\CO$ gas there is indeed warmer than that in the galactic disk
\citep{GP2004, Ao2013, Bally2014}.
The higher cosmic ray production rate due to higher
SFR in the galactic center may explain the elevated temperature of the $\CO$ gas
there. However, if only the ${}^{12}\CO~(J=1-0)$ line is observed, because it is usually
optically thick, the luminosity of the line is determined mostly by the
temperature at the cloud surface, where photoelectric heating can dominate
over cosmic-ray heating. Therefore, as pointed out by \citet{WHT1993},
the luminosity of the $\CO~(J=1-0)$ line is mostly determined by the strength of UV
radiation, instead of the cosmic-ray ionization rate.

In galactic disks where the warm and cold atomic gas are in pressure
equilibrium, the typical density of the cold neutral medium (CNM) is related to
the strength of the ambient radiation field approximately by $n_\mr{CNM} \propto \chi$ 
\citep{Wolfire2003, OML2010}:
\begin{equation}\label{eq:nH2G0_G}
    \frac{n_\mr{CNM}}{\chi} \approx 23.
\end{equation}
\citet{KMT2009} adopted a similar relation for the density of cold
atomic/molecular complexes.
Assuming the ambient radiation field strength is proportional to the SFR, 
then $\chi \propto \Sigma_\mr{SFR}$, where $\Sigma_\mr{SFR}$ is the average rate
of star formation per unit area in the galactic disk.
In equilibrium, $\Sigma_\mr{SFR}$ is
expected to be proportional to the weight of the ISM, which in general depends
on both the gas surface density $\Sigma_\mr{gas}$
and the stellar density \citep{OML2010, KKO2011, KOK2013}.
In starburst regions where the gas dominates gravity,
$\Sigma_\mr{SFR} \propto \Sigma_\mr{gas}^2$ \citep{OS2011}. Assuming that
$\xi_\Ho \propto \Sigma_\mr{SFR}/\Sigma_\mr{gas}$ \citep{OML2010}, and 
$\Sigma_\mr{SFR}/\Sigma_\mr{gas} \propto \Sigma_\mr{gas} \propto \sqrt{\chi}$,
we consider the case in which
\begin{equation}\label{eq:xi_G0}
\xi_\Ho = 10^{-16}\sqrt{\chi}~\mr{s^{-1} H^{-1}}.
\end{equation}

\begin{figure*}[htbp]
    \begin{center}
         \includegraphics[width=0.8\textwidth]{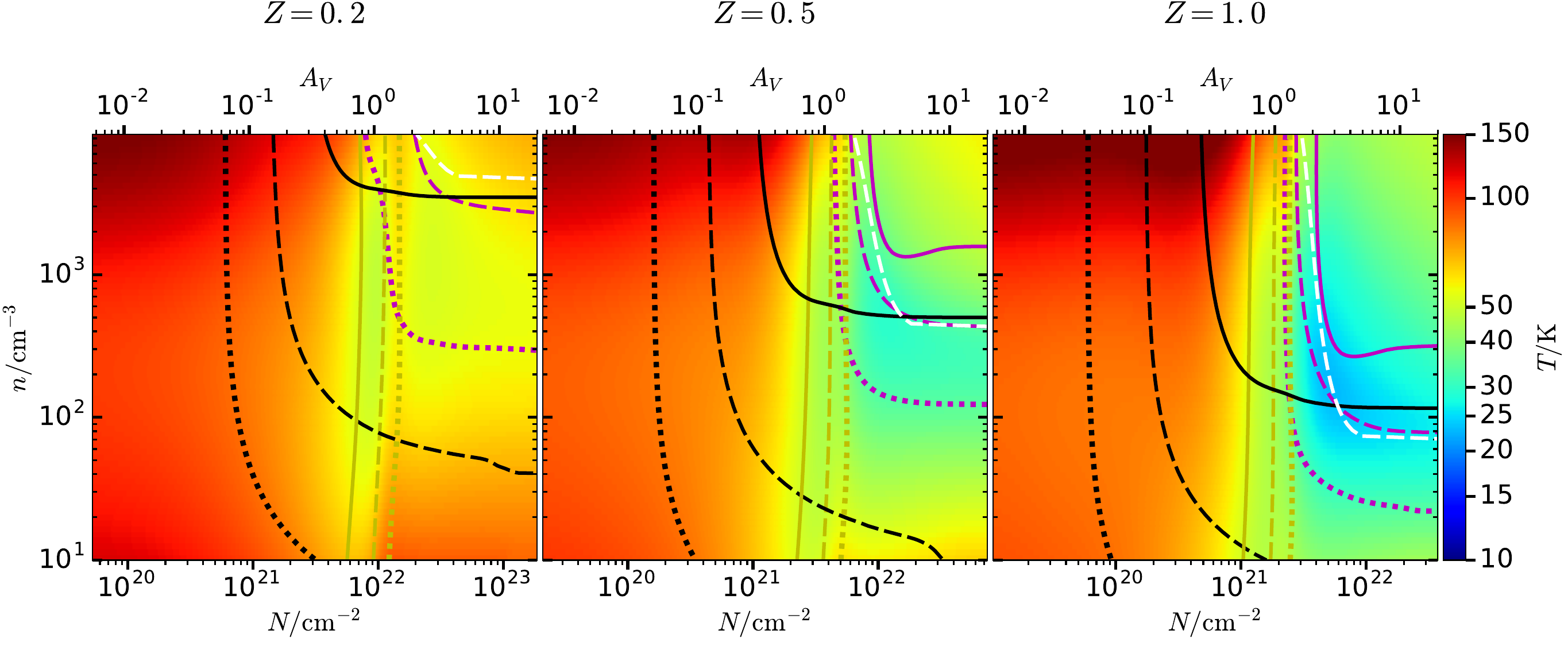}
    \end{center}
    \caption{Temperature of the 
        gas as a function of $A_V$ (or $N$) and density $n$, similar to
    Figure \ref{fig:KLM_G1_T}. The incident radiation field and cosmic-ray
ionization rate are set by Equations (\ref{eq:nH2G0_G}) and (\ref{eq:xi_G0}).
The contours represent the abundances of $\Ht$ (black), $\CO$ (magenta), and
$\CII$ (yellow), and the fit of the
$x(\CO)/x_\mr{C, tot}$=0.5 contour (white dashed), 
similar to  Figures \ref{fig:KLM_G1_T} and \ref{fig:KLM_TCO}.
\label{fig:KLM_Gn} }
\end{figure*}

Using $\chi\propto n$ and $\xi_\Ho \propto \sqrt{\chi}$
based on Equations (\ref{eq:nH2G0_G}) and
(\ref{eq:xi_G0}), Figure \ref{fig:KLM_Gn} plots
the temperature of the gas in the $n$-$N$ plane.
Comparing to Figure
\ref{fig:KLM_G1_T}, Figure \ref{fig:KLM_Gn} shows clearly that it is very
difficult to form $\CO$ in metal-poor gas under typical diffuse-ISM conditions,
implying gas densities must be enhanced by turbulence
or gravity to be well above typical values in the CNM for $\CO$ to be present.
However, at $Z=1$, most of the carbon is in $\CO$ at $A_V \gtrsim 2$ and 
$n \gtrsim 100 ~\mr{cm^{-3}}$. Thus, in regions of galaxies where $A_V$ and
$n$ are high enough, gas need not to be in gravitationally bound clouds to be
molecular or to emit strongly in $\CO$ \citep[see also][]{Elmegreen1993}.
This implies that $\CO$ emission in the galactic center and
starburst galaxies may arise largely from diffuse gas. Indeed, observations of $\CO$
lines in luminous infrared galaxies by \citet{Papadopoulos2012} 
found that most of the $\CO$ emission can come from warm and diffuse gas
in these starburst environments. Figure \ref{fig:KLM_Gn}
also shows that in metal-poor galaxies,
$\CI$ could be the most important tracer for shielded gas, as also noted by
\citet{GC2016}. However, the gas that is traced by $\CI$ is not necessarily
primarily molecular $\Ht$.

We have also found a simple fit for the contour of $x(\CO)/x_\mr{C, tot} = 0.5$
in the plane of $n$ -- $A_V$, given values of incident FUV radiation field,
cosmic-ray ionization rate, and metallicity:
\begin{equation}\label{eq:CO_fit}
    \frac{n_\mr{crit, CO}}{\mr{cm^{-3}}} 
    = \left(4\times10^3 Z \xi_{\Ho,16}^{-2}\right)^{\chi_\CO^{1/3}}
    \left( \frac{50  \xi_{\Ho,16}}{Z^{1.4}}\right),
\end{equation}
where $n_\mr{crit, CO}$ is the critical value above which 
$x(\CO)/x_\mr{C, tot} > 0.5$;
$\xi_{\Ho,16}=\xi_\Ho/(10^{-16}\mr{s^{-1}H^{-1}})$;
$\chi_\CO = \chi \exp(-\gamma_\CO A_V)$ is the effective radiation field for
$\CO$ photodissociation accounting for dust attenuation, 
and $\gamma_\CO = 3.53$ is the dust-shielding
factor of $\CO$ given in Table \ref{table:chem2}.
Figures \ref{fig:KLM_G1_T}-\ref{fig:KLM_Gn} and \ref{fig:KLM_G1_MBE}
shows the fit in Equation
(\ref{eq:CO_fit}) (white dashed line) against the true contour of 
$x(\CO)/x_\mr{C, tot} = 0.5$ (magenta dashed line), demonstrating the good
agreement between the two. We note that this fit is only tested here to be
applicable in the range of $n \approx 10-10^4~\mr{cm^{-3}}$, 
$\xi_\Ho \approx 10^{-17}-10^{-15}~\mr{s^{-1}H^{-1}}$, and $Z=0.2-1$.

\subsection{Are Molecules Necessary for Low Temperatures?
\label{section:molecule_cooling}}
How sensitive is the gas temperature to the detailed chemistry? \citet{GC2012a}
showed that in simulations of turbulent clouds, as long as the gas is shielded,
atomic gas can reach similarly low temperatures as in molecular gas. 
\citet{GC2012} also found that the gas temperature in dense clouds can be very
similar even if the $\CO$ and $\CI$ abundances produced by different chemical
networks vary by more than an order of magnitude. 

\begin{figure*}[htbp]
    \begin{center}
    \includegraphics[width=\textwidth]{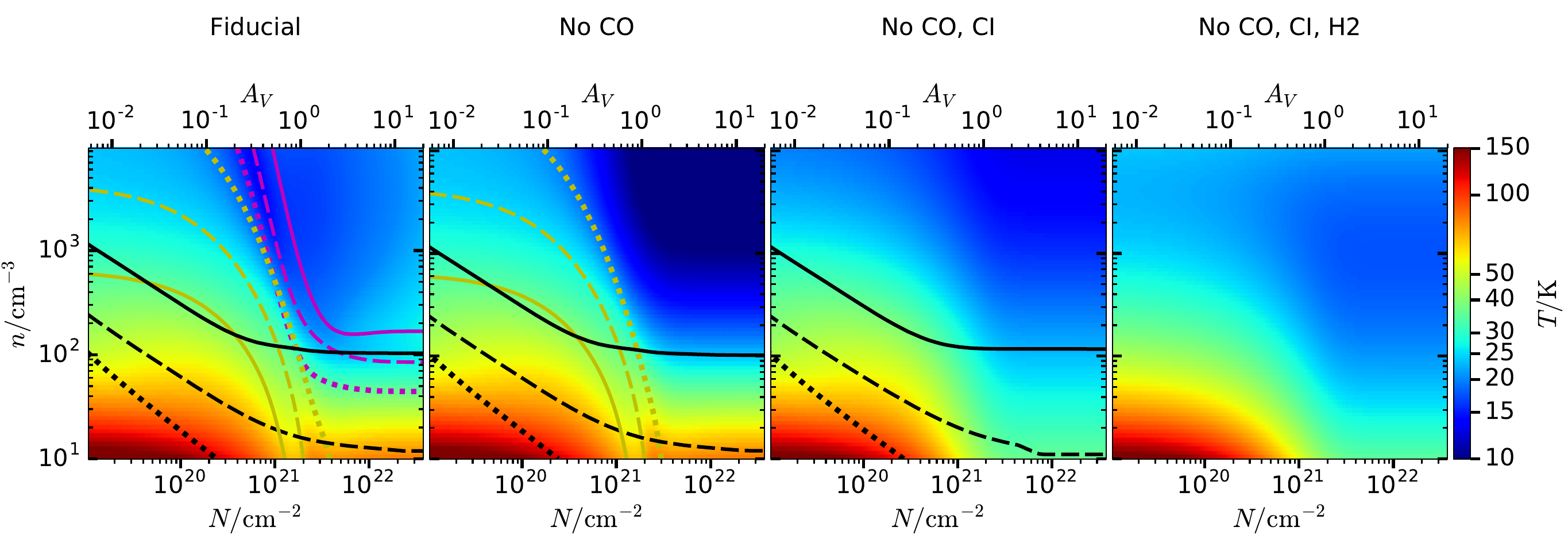}
    \end{center}
    \caption{Comparison of equilibrium gas temperature with different chemical
        compositions. The leftmost panel is the same as the $Z=1$ case in
        Figure \ref{fig:KLM_G1_T}, showing the fiducial model with our 
        chemistry network. The other panels show cases where the formation
        rates of some species ($\CO$, $\CI$, and/or $\Ht$) are artificially
        set to zero, so the gas is forced to be in a state free of these
        molecules or atoms. 
        The contours show the transition of $\Ho$ to $\Ht$ (black),
        $\CII$ to $\CI$ (yellow), and $\CI$ to $\CO$ (magenta), similar to Figure
    \ref{fig:KLM_G1_T}. \label{fig:chemistry_cold_gas}}
\end{figure*}

To directly look into this question, we artificially force the gas into
different chemical states by setting the formation rates of $\CO$, $\CI$, and $\Ht$
to zero (successively). Figure \ref{fig:chemistry_cold_gas} shows that the
equilibrium gas temperatures are very similar with completely different chemical
states. $\CI$ and $\CII$ can cool the gas just as well as
$\CO$.\footnote{Note that we assume the $\CI$ and $\CII$ cooling is optically
thin, whereas the optical depth effect is taken into account in $\CO$ cooling.
This is why $\CI$ and $\CII$ cool the dense gas to lower temperatures than $\CO$ in 
Figure \ref{fig:chemistry_cold_gas}. In reality, $\CI$ and $\CII$ will also be
optically thick at high columns.} $\Ht$ only affects the temperature
slightly by changing the specific heat of the gas. This confirms the
conclusions in \citet{GC2012a}: any correlation between molecular gas and 
low temperatures is likely to be coincidental instead of causal.
Molecular gas traces low temperatures simply because molecules form
in dense and shielded regions that are more efficient at cooling and less exposed
to heating. 

\subsection{Molecular Gas and Star Formation?}
\begin{figure*}[htbp]
    \begin{center}
    \includegraphics[width=0.8\textwidth]{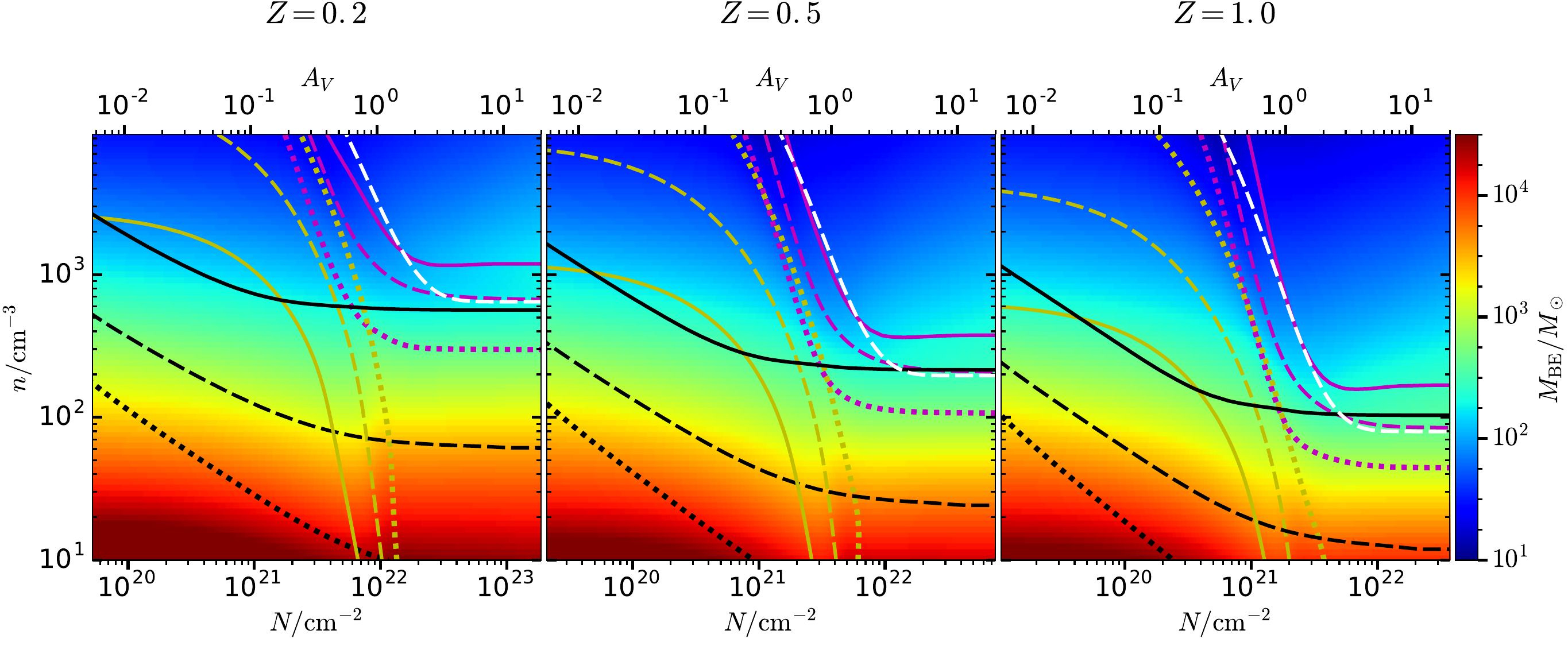}
    \end{center}
    \caption{Similar to Figure \ref{fig:KLM_G1_T}, but for Bonnor--Ebert mass
        of the gas as a function of $A_V$ (or $N$) and density $n$.
        \label{fig:KLM_G1_MBE}
    }
\end{figure*}

\citet{KLM2011} argued that $\Ht$ traces star formation because $\Ht$ forms in
the same regions where the gas is more gravitationally unstable, or,
quantitatively speaking, where the critical 
Bonnor--Ebert mass $M_\mr{BE} = 1.2c_s^3 \left( \frac{\pi^3}{G^3\rho} \right)^{1/2}
= 1.2\left( \frac{\pi^3 k^3 T^3}{G^3 n \mu^4} \right)^{1/2}$
\citep{Bonnor1956, Ebert1955} becomes low.

In Figure \ref{fig:KLM_G1_MBE}, we show the critical 
Bonnor--Ebert mass as a function of gas density and column similar to
\citet{KLM2011}, with the improvement that chemistry and
temperature are calculated self-consistently. 
As noted by \citet{KLM2011},
the contours of the 50\% and 90\% $\Ht$ abundances are somewhat similar to those of 
the Bonnor--Ebert mass of the gas. 
However, we note that the Bonnor--Ebert mass near these contours is
$M_\mathrm{BE}=10^2-10^3~M_\sun$, much larger than the mass of
individual stars. In some regions, the Bonnor--Ebert mass is even larger than
the available mass in the clouds: for example, at $n=1000\mr{cm^{-3}}$ and
$A_V=1$, the total mass of the cloud 
$M\sim R^3 n m_\Ho \sim (N/n)^3 n m_\Ho \sim 5M_\odot$, and is
much smaller than the local Bonnor--Ebert mass $M_\mr{BE}\sim 100 M_\odot$.
This suggests that the correlation of $\Ht$ with star
formation is not a sufficient pre-condition,
but a fairly non-specific coincidence. To reach
Bonnor--Ebert masses comparable to those of individual stars, much higher
densities are needed than the densities required for high equilibrium $\Ht$
abundance.\footnote{Also, as noted above, a small $t_\mr{dyn}/t_\Ht$ 
or $t_\mr{diss}/t_\mr{sh}$ may limit the $\Ht$ abundance at low $A_V$ and $n$.
Non-equilibrium $\Ht$ abundances may be high only in the upper-right corner of each panel,
where $M_\mathrm{BE} <100 M_\odot$.}

\subsection{Comparison with Observations of Diffuse and Translucent
Clouds}

\begin{figure*}[htbp]
     \begin{center}
         \includegraphics[width=0.97\textwidth]{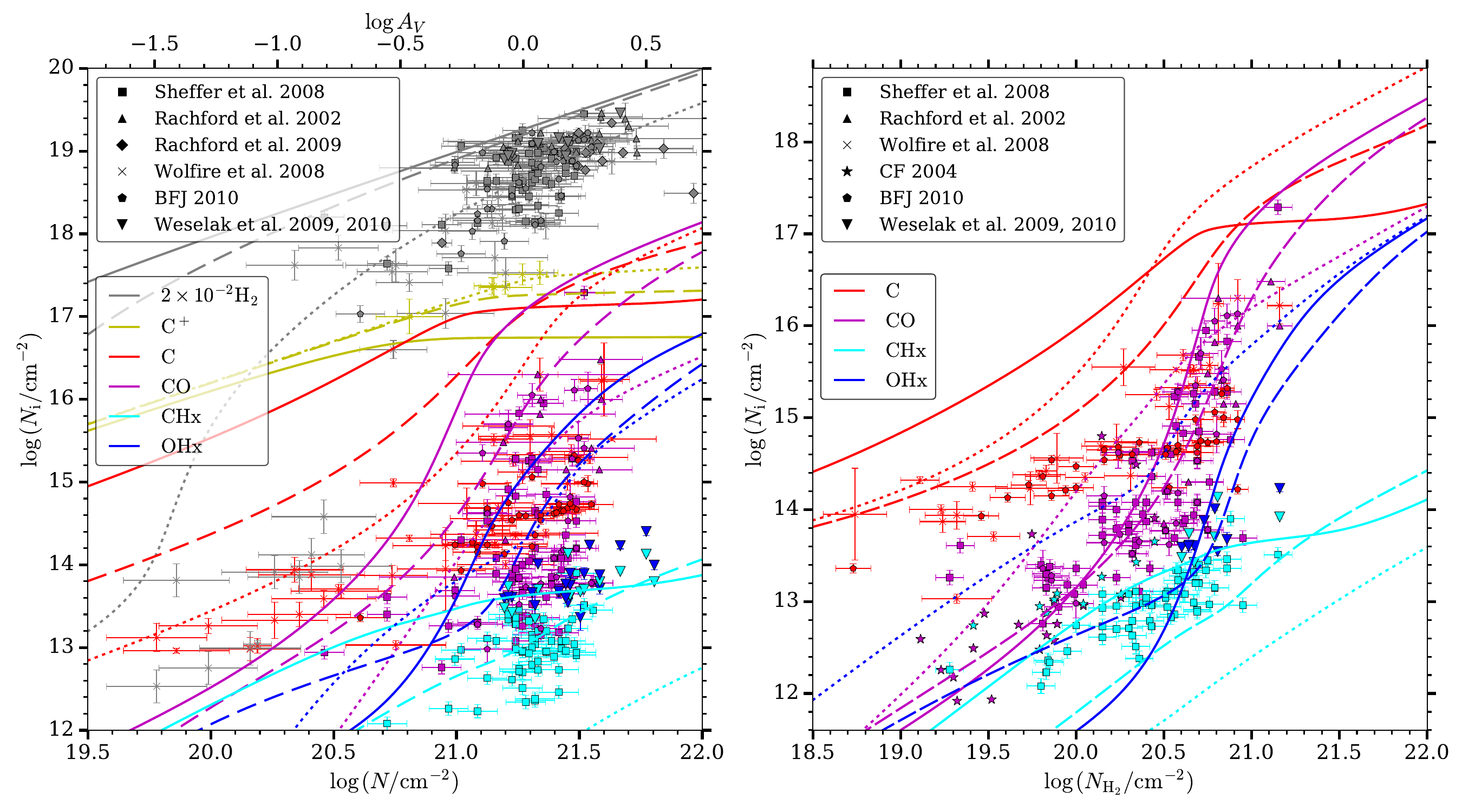}
    \end{center}
    \caption{Comparison between observations and our one-sided slab
        model. The {\sl left} panel plots the column density $N_i$ of 
        chemical species at a given cloud  $A_V$ (or $N$). 
        The lines show the results from our slab model with
        $n=10~\mr{cm^{-3}}$ (dotted), $n=100~\mr{cm^{-3}}$ (dashed), 
        and $n=1000~\mr{cm^{-3}}$ (solid). 
        Different markers plot the observed abundances from the literature.
        Different colors represent different species, as shown in the legends.
        The {\sl right} panel is similar to the left panel, but with the
        $\Ht$ column density on the x-axis instead.
    \label{fig:cloud_obs_NH_NH2}}
\end{figure*}

\begin{figure*}[htbp]
     \begin{center}
         \includegraphics[width=0.97\textwidth]{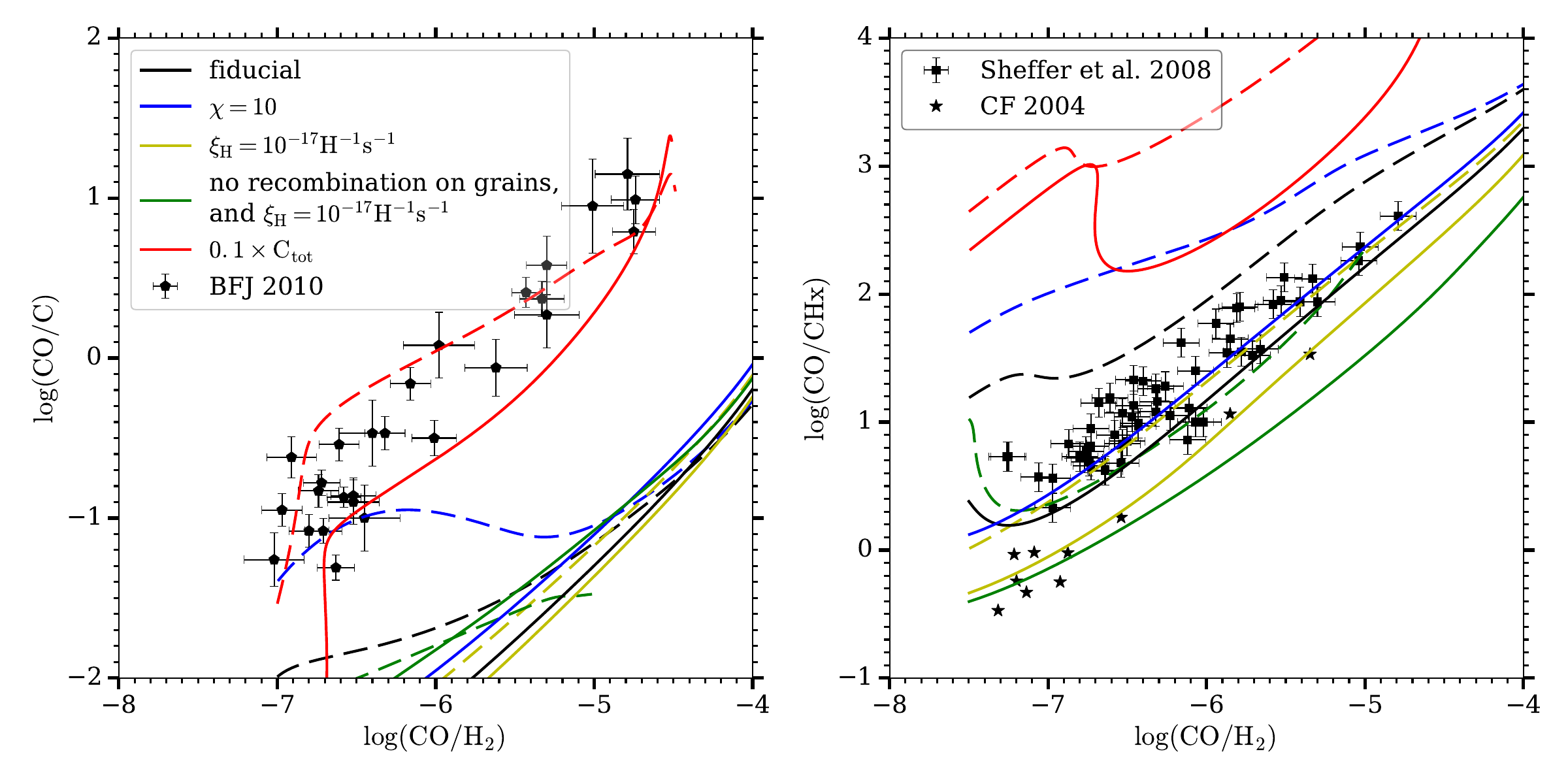}
    \end{center}
    \caption{{\sl Left panel:} abundances of $\CO/\CI$ vs. $\CO/\Ht$. 
        The observational data are taken from \citetalias{BFJ2010}, and lines
        of different colors show the results from different slab models (see
        legend), with $n=100~\mr{cm^{-3}}$ (dashed), 
        and $n=1000~\mr{cm^{-3}}$ (solid). {\sl Right panel}: similar to
        the left panel, but for $\CO/\CHx$ on the y-axis. The observed
        abundance is taken from \citet{Sheffer2008} and \citetalias{CF2004}.
        See also Figure \ref{fig:cloud_obs_nogr_depCI} for a more detailed
        comparison between observations and our model with 
        no grain-assisted recombination, and the depletion of carbon abundance.
    \label{fig:CI_CHx}}
\end{figure*}

The ultimate test for any chemistry model is the comparison with observations.
The one-sided equilibrium slab models that we have used to test our network
are extremely simple and cannot be expected to agree in detail with 
the chemical state in the real ISM, which is characterized by complex 
time-dependent dynamics and morphological structure, both the result of turbulence.
Nevertheless, for illustrative purposes it is useful to compare the abundances obtained
with our network (for idealized slab geometry) with the observed ISM abundances.  
To do so, we compiled observations of $\Ht$, $\CO$, $\CI$,
$\CII$, $\CHx$, and $\OHx$ abundances in the literature 
derived from UV and optical absorption spectra along
different sight lines: $\Ht$, $\CI$, and $\CII$ observations compiled  by
\citet{Wolfire2008}\footnote{In \citet{Wolfire2008}, most of the $\CII$ values
are upper limits. Here we only include the true measurements.};
$\Ht$ and $\CO$ observations in \citet{Rachford2002} and
\citet{Sheffer2008}; $\Ht$ observations in \citet{Rachford2009};
$\Ht$, $\CI$, and $\CO$ observations in 
\citet{BFJ2010} (hereafter \citetalias{BFJ2010});
$\CHx$\footnote{$\CHx = \CH+\mr{CH^+}$ for \citet{Sheffer2008} and \citet{CF2004}.}
abundance in \citet{Sheffer2008} and 
\citet{CF2004} (hereafter \citetalias{CF2004});
and $\OHx$\footnote{$\OHx = \mr{OH}+\mr{OH^+}$ for \citet{Weselak2009}, and 
$\OHx=\mr{OH}$ for \citet{Weselak2010}.} 
abundance in \citet{Weselak2009, Weselak2010}. 
The UV spectroscopy data in these observations mainly come from the {\sl Far
Ultraviolet Spectroscopic Explorer (FUSE) }, the Space Telescope Imaging Spectrograph
(STIS) on board the {\sl Hubble Space Telescope}, 
the {\sl UVES} spectrograph at {\sl European Southern Observatory (ESO)} 
and the {\sl Copernicus}
survey. The optical spectra data used to derive $\CH$ and $\mr{CH^+}$
abundances in
\citet{Sheffer2008} are obtained at the McDonald and European Southern
Observatories. For most sightlines, the total column $N$ is
derived from reddening data using
$N/E(B-V) = 5.8\times 10^{21}~\mr{H~cm^{-2}mag^{-1}}$ \citep{BSD1978,
Rachford2009}, except for sight lines that have $\mr{H}$ abundances directly
observed by Ly$\alpha$ absorption in \citet{BSD1978} (included in the
compilation of observations by \citet{Wolfire2008}).

We construct a simple one-sided slab model with thermal and chemical
equilibrium, as described in Section
\ref{section:CO_trace_H2}, to compare with the
observational data. The fiducial model has $\chi=1$, 
$\xi_\Ho = 2\times 10^{-16}~\mr{s^{-1} H^{-1}}$, and constant densities of 
$n=10$, $100$, and $1000~\mr{cm^{-3}}$. We also explore the effects of
varying
different parameters such as $\chi$, $\xi_\Ho$, the efficiency of 
grain-assisted recombination, and gas-phase carbon abundance.

Figure \ref{fig:cloud_obs_NH_NH2} shows the comparison between the observations and our
slab cloud model in column densities of $\Ht$, $\CO$, $\CI$,
$\CII$ and $\CHx$ and $\OHx$. For the $\mr{H}$ to $\Ht$ transition in the left
panel, low density $n\sim 10~\mr{cm^{-2}}$ is required to match equilibrium
$\Ht$ abundances with observations.
This is consistent with the analysis by \citet{Bialy2015,
Bialy2017} on the $\mr{H}$ to $\Ht$ transition layers in the Perseus molecular
cloud and star forming region W43.\footnote{$n\sim 10~\mr{cm^{-2}}$ is in
fact lower than densities believed to be in the CNM ($n\sim 40~\mr{cm^{-2}}$ 
based on pressures from \citet{JT2011} and temperatures from \citet{HT2003}).
If densities are indeed higher than $n\sim 10~\mr{cm^{-2}}$,
this means that the observed abundances are below equilibrium,
potentially due to dynamical effects 
(as discussed previously in Section \ref{section:CO_trace_H2}).} 

Comparing the right panel with the left, there
is much less dispersion in the observed $\CI$, $\CO$, and $\CHx$ abundances when
plotted against the $\Ht$ column density $N_\Ht$ instead of the total column
$N$ or $A_V$. This is likely due to foreground or background contaminations: low density
atomic ISM is much more ``diffuse'' in spatial distribution than the relatively
``clumpy'' molecular ISM. Therefore, the low density atomic clouds may contribute
a significant fraction of the total column/visual extinction in a given sight line, 
without contributing much to the total column of chemical species 
such as $\Ht$, $\CI$, $\CO$, $\CHx$ and $\OHx$, which
mainly form in higher density regions. 

In spite of the extreme idealizations adopted, there is generally a good agreement
between our model and the observations 
in Figure \ref{fig:cloud_obs_NH_NH2}, especially in the right panel (when
plotted against $N_\Ht$): we can successfully reproduce the range of observed abundances
of $\CO$, $\CHx$ and $\OHx$ with densities between $100~\mr{cm^{-3}}$ and 
$1000~\mr{cm^{-3}}$. It is especially remarkable that 
$\CO$ and $\CHx$ abundances agree over one or two orders of magnitude of $N_\Ht$.

However, there is one significant discrepancy between the simple
model and observations: the predicted
$\CI$ abundance is too high at a given $A_V$ or $N_\Ht$ by about an order of
magnitude (see Figure \ref{fig:cloud_obs_NH_NH2}). 
This is more clearly shown in the left panel of 
Figure \ref{fig:CI_CHx} when plotted with the ratio of $\CO/\CI$
(see also Figure 4 in \citetalias{BFJ2010}). To
investigate the dependence of $\CO/\CI$ on model parameters, we vary the
incident radiation field strength $\chi$, cosmic-ray ionization rate $\xi_\Ho$, 
the efficiency of grain-assisted recombination of ions, and the gas-phase
carbon abundance. The summary of our results is shown in the left panel of
Figure \ref{fig:CI_CHx}. Variations in $\chi$ and $\xi_\Ho$ tend to change the
$\CI$ and $\CO$ abundances in the same direction, giving very a similar
$\CO/\CI$ ratio to that of the fiducial model. Excluding the grain-assisted
recombination of ions does not solve the problem either: it does 
lower the $\CI$ abundance at a given $A_V$, but at the same time, the $\CO$
abundance is much lower as well, and the 
$\CO/\CI$ ratio is still too low (see also the left panel of Figure
\ref{fig:cloud_obs_nogr_depCI}).

Note that our conclusion is different from the conclusion of
in \citet{Liszt2011}, which stated that the absence of grain-assisted
recombination can reproduce the high $\CO/\CI$ observed in
\citetalias{BFJ2010}. We believe this is because \citet{Liszt2011} 
assumed a constant $x_\mr{HCO^{+}}\sim 3\times 10^{-9}$ in their chemistry
model. Without grain-assisted recombination, 
the $\mr{HCO^+}$ abundance is much lower
at $x_\mr{HCO^+} \lesssim 10^{-10}$, leading to much less efficient
$\CO$ formation than calculated in \citet{Liszt2011}\footnote{This is because
    without grain-assisted recombination, the electron abundance is higher. 
Electrons destroy $\mr{H_3^+}$, which forms $\OHx$ and subsequently
$\mr{HCO^+}$. $\CO$ is mainly formed by $\mr{HCO^+ + e \rightarrow CO + H}$. 
To be consistent with \citet{Liszt2011},
We also use a low cosmic-ray ionization rate 
$\xi_\Ho = 10^{-17}~\mr{s^{-1} H^{-1}}$,
for the case without grain-assisted recombination.}.

The only variation of our models that can reproduce 
    the observed $\CO/\CI$ ratio is the one
with gas-phase carbon abundance depleted by an order of magnitude relative to the
fiducial model, i.e. 
$\CI_\mr{tot}/\Ho=0.1(\CI_\mr{tot}/\Ho)_\mr{fiducial}=1.6 \times 10^{-5}$.
This is also found in \citetalias{BFJ2010} (their Figure 4).
However, although carbon depletion can be caused by formation of
grains such as polycyclic aromatic hydrocarbons (PAHs), a
depletion factor as extreme as $0.1$ is unlikely. Even the most carbon-depleted
sight lines observed in \citet{Sofia2011} and \citet{Parvathi2012} have 
$\CI_\mr{tot}/\Ho > 6\times 10^{-5}$ (derived from $\CII$ absorption
measurements), and many of these sight lines are the same as in \citet{BFJ2010}. 
Moreover, as shown in the right panel of Figure \ref{fig:CI_CHx}, the ratio of 
$\CO/\CHx$ in the model with carbon depletion is an order of magnitude higher
than the observed values, whereas other models reproduce $\CO/\CHx$
consistent with observations. Therefore, we conclude that the depletion
of gas-phase carbon is unlikely to be the solution for this problem.

Overpredicting $\CI$ abundance is also an issue among other PDR models.
For example,
\citetalias{BFJ2010} compared their observations with the chemistry model by
\citet{vDB1988}, and found very similar results. \citet{Glover2010} also
predicted $\CO/\CI \ll 1$ for $A_V < 2$ clouds with their extensive 
chemical network (see their Figures 1 and 2). \citet{Bensch2003} observed $\CI$
and $\CO$ emission in the translucent cloud MCLD 123.5+24.9, and also found
$\CO/\CI \gtrsim 1$. They used the PDR code by \citet{Stoerzer1996} to model
their observations, and found that their models also tend to produce 
a $\CO/\CI$ ratio that is too low, unless they assume 
a cloud structure of high density ($n \sim 10^4~\mr{cm^{-3}}$) 
clumps embedded in a low density medium. 

It is worth emphasizing that observations compiled here are all for $A_V\lesssim 1$
diffuse and translucent clouds, for which UV or optical absorption observations
may be used. For these clouds, most carbon ($>90\%$) is still in the
form of $\CII$. The physical properties of these clouds are very different from
GMCs and smaller dark molecular clouds, which typically has $A_V \gg 1$ and 
most carbon is in the form of $\CO$. 

There are some potential solutions to the mismatch of $\CI$ abundance between
equilibrium PDR models and observations. (1) Non-equilibrium chemistry. 
As previously noted, the photoionization and photodissociation timescales
are generally short compared to collisional
reactions or grain-assisted reactions (see Tables \ref{table:chem1} and
\ref{table:chem2}). Dynamical effects such as
limited cloud lifetime and turbulent cycling of gas from a cloud's 
interior to its surface can lower the abundances of species
formed in low-shielding regions such as $\CI$. 
Furthermore, $\CI$ tends to form in
less dense and shielded regions than $\CO$, and can be more subject to these
dynamical effects, which may bring its abundance far from equilibrium.
(2) Improvement of chemical networks and reaction rates. It is possible that the
chemical pathways to form $\CI$ and $\CO$ in translucent clouds are not well
understood, causing discrepancy between chemical models and observations. 
For example, the composition of dust grains and the rates for grain surface
reactions are still uncertain.
In addition, turbulent
dissipation \citep{Godard2014} might convert the dominant reservoir of
carbon (in $\CII$) to CO without much production of $\CI$.
In the future, 3D realistic ISM simulations 
with time-dependent chemistry will give us more insight into the role of
non-equilibrium chemistry.

\section{Summary}
In this paper, we propose a new lightweight and accurate
chemical network for hydrogen and
carbon chemistry in the atomic and molecular ISM. Our network is based on the
\citetalias{NL1999} network in \citet{NL1999} and \citet{GC2012}, with significant
modifications and extensions. We use 1D uniform slab models to compare our 
chemical network in detail with results from a full PDR code 
and also with the original \citetalias{NL1999} network.

Our chemical network shows very good agreement with the much more sophisticated
PDR code in the equilibrium abundances of all species. This agreement holds in
a wide range of densities, visual extinctions, temperature, and metallicities,
implying that our simplified network indeed captures the main chemical pathways. 
The \citetalias{NL1999} network, however, performs poorly when comparing to the
PDR code in detail. In particular, it significantly underproduces
$\CO$ at $n \sim 100-500~\mr{cm^{-3}}$ when using the realistic cosmic-ray
ionization rate $\xi_\Ho = 2\times 10^{-16}~\mr{s^{-1} H^{-1}}$ 
\citep{Indriolo2007, Hollenbach2012} in our galaxy. The \citetalias{NL1999} network 
also fails to reproduce the abundances of other species such as $\CII$. With
only one additional species and 19 additional
reactions, our network has a comparable computational cost to the original
\citetalias{NL1999} network and proves to be much more accurate. Therefore,
we conclude that our new network is preferred over the \citetalias{NL1999}
network for time-dependent numerical simulations of
hydrogen and carbon chemistry, such as $\CO$ formation.

We apply our network to 1D models and obtain the equilibrium temperature and
chemistry in a range of physical conditions, varying the density, incident
radiation field strength, cosmic-ray ionization rate, and metallicity. We find
that at metallicity $Z=1$, the $\CO$-dominated regime
delimits the coldest gas and that the corresponding temperature 
tracks the cosmic-ray ionization rate in molecular clouds. We note, however,
that it is primarily high density and high shielding that leads to low
temperatures, rather than the $\CO$ abundance itself. In metal-poor gas
with $Z\lesssim0.2$, $\CO$ is difficult to form under typical diffuse-ISM
conditions and may only be found in regions where gas density is significantly
enhanced by turbulence or gravity. We provide a simple fit for the locus of 
$\CO$-dominated regions as a function of gas density, column, metallicity, and
cosmic-ray ionization rate.

We also compiled observations of chemical species in diffuse and translucent
clouds, and compared these with our chemistry model predictions,
under the assumption of equilibrium. 
We are able to reproduce the observed $\CO$, $\CHx$ and $\OHx$ abundances with
density $n$ between $100~\mr{cm^{-3}}$ and $1000~\mr{cm^{-3}}$. However,
the predicted $\CI$ abundances are higher than the observed values by an
order of magnitude. Previous equilibrium models have identified a similar
difficulty matching C abundances, suggesting that time dependence or
other more complex effects may be important.
To fully understand the distribution of observed species in the ISM,
detailed 3D simulations
with realistic gas dynamics and non-equilibrium chemistry are required. We plan
to pursue this in the future.

\section{Acknowledgments}
The work of Munan Gong and Eve C. Ostriker was supported in part
by NSF grant AST-1312006 and NASA grant NNX14AB49G.
Mark G. Wolfire was supported in part by NSF grant AST-1411827.  
We thank the referee for helping us to improve
the overall quality of this paper. We also
thank Simon Glover and Edward B. Jenkins for many helpful suggestions and
discussions. We thank Shmuel Bialy, Amiel Sternberg, Thomas Bisbas, 
and Daniel W. Savin for providing suggestions and updated rate information.

\clearpage
\newpage
\appendix

\section{Description of the Chemical Network}\label{section:chem_desp}
A summary of the chemical network is listed in Tables \ref{table:chem1} and
\ref{table:chem2}.
\subsection{Extension of the \citetalias{NL1999} Network\label{section:NL99_extension}}
We have added or modified the following reactions to the \citetalias{NL1999} network:
\begin{enumerate}
    \item $\mr{H_3^+}$ destruction channel 
        $$\mathrm{H_3^+ + e \rightarrow 3 H}$$
        in addition to $$\mathrm{H_3^+ + e \rightarrow H_2 + H}$$ 
        in the original \citetalias{NL1999} network,
        to be more consistent with the hydrogen network adopted in
        \citet{Glover2010} and \citet{Indriolo2007}.
    \item Grain-assisted recombination of $\mr{C^+}$ and $\mr{He^+}$:
        \begin{align*}
            &\mr{C^+ + e + gr \rightarrow C + gr}\\
            &\mr{He^+ + e + gr \rightarrow He + gr}
        \end{align*}
        These two grain surface reactions are the main channels for $\CII$ and
        $\mr{He^+}$ recombination at solar metallicity.
        \citet{GC2012} considered these two reactions, and concluded that
        they made very little difference in determining where $\CO$
        would form. However, we found that using the updated cosmic-ray
        ionization rate $\xi_\Ho=2\times 10^{-16} \mr{s^{-1} H^{-1}}$,
        instead of their rate $\xi_\Ho=10^{-17} \mr{s^{-1} H^{-1}}$,
        these reactions, especially the recombination of
        $\mr{C^+}$ on dust grains, are essential for $\CO$ formation at
        moderate densities $n \lesssim 1000~\mr{cm^{-3}}$. This is because
        electrons from cosmic-ray ionization of $\mr{H}$ and $\CI$
        inhibit $\CO$ formation,
        while $\mr{He^+}$ formed by cosmic rays destroys $\CO$. 
    \item Cosmic-ray ionization of $\Ho$ and $\Ht$ by secondary electrons. The
        measurement of cosmic-ray ionization rate
        $\xi_\mathrm{H}=2.0\times 10^{-16} \mathrm{s^{-1}H^{-1}}$ in
        \citet{Indriolo2007} ($\xi_p$ in their paper) only accounts for the
        initial (primary) cosmic-ray ionization. The secondary electrons
        created by the first ionization event can again ionize $\Ho$ and $\Ht$.
        The secondary ionization rate depends on factors such as electron
        abundance \citep{Dalgarno1999}. Here we apply a simple approach
        motivated by \citet{GL1974}, which has the total ionization rate of
        $\Ho$ 1.5 times the primary rate in atomic regions and 1.15
        times in molecular regions. In regions mixed with atomic and molecular
        gas, we simply scale the total rate with $\Ho$ and
        $\Ht$ abundance as shown in reactions 6 and 7 in Table
        \ref{table:chem2}. The cosmic-ray ionization rate of $\Ht$ is simply
        twice that of $\Ho$.
    \item Removed photodissociation of $\mr{HCO^+}$. We find that the
        destruction of $\mr{HCO^+}$ is dominated by the reaction $\mr{HCO^+ + e
        \rightarrow CO + H}$, and the photodissociation of $\mr{HCO^+}$ has
        very little effect on the whole chemistry network.
    \item $\mr{He^+}$ destruction in the $\Ht$ region:
        \begin{align*}
            &\mr{He^+ + H_2 \rightarrow H_2^+ + He}\\
            &\mr{He^+ + H_2 \rightarrow H^+ + He + H}
        \end{align*}
        The rates of these two reactions are similar to the rate of direct
        recombination of $\mr{He^+}$ with electrons, and can be higher than the
        grain surface recombination of $\mr{He^+}$ abundance at low metallicities. 
    \item $\mr{CH}$ destruction:
        $$\mr{CH + H \rightarrow H_2 + C}$$
        This is one of the main destruction channels of $\mr{CH}$ in shielded
        regions.
    \item Cosmic-ray ionization, and cosmic-ray induced
        photodissociation of $\mr{C}$ and $\CO$: 
        \begin{align*}
            &\mr{cr + C \rightarrow C^+ + e}\\ 
            &\mr{cr + CO \rightarrow CO^+ + e}\\ 
            &\mr{\gamma_{cr} + C \rightarrow C^+ + e}\\ 
            &\mr{\gamma_{cr} + CO \rightarrow C + O}
        \end{align*}
        For $\mr{cr + CO \rightarrow CO^+ + e}$, we assume that
        $\mr{CO^+}$ reacts quickly with $\Ho$/$\Ht$ to form $\mr{HCO^+}$, and
        implement the reaction as 
        $\mr{cr + CO + H \rightarrow HCO^+ + e}$ with the same rate.
        Similar to \citet{CG2015}, 
        the rates for these reactions are assumed to be proportional to
        the cosmic-ray ionization rate of atomic hydrogen $\xi_\Ho$.
        The scaling factors for cosmic-ray ionization
        $\xi_\mr{C, CO}/\xi_\Ho$ are calculated with the rates 
        in \citet{McElroy2013}, and the scaling
        factors for cosmic-ray induced photo-ionization are adopted from
        \citet{Gredel1987}. We also scale the cosmic-ray induced
        photo-ionization to the $\Ht$ fraction $2n(\Ht)/(2n(\Ht)+2n(\Ho))$, 
        because the ionizing photons come
        from the UV light induced by the primary ionization
        and excitation of $\Ht$, and would
        not be present in regions with only atomic hydrogen.
        We have also tested the inclusion of cosmic-ray
        induced photo-reactions of $\mr{CH}$ and $\mr{OH}$ using the
        rates in \citet{Gredel1989}, but found they made very little
        difference for $\CO$ formation.

    \item $\mr{OH_x + O \rightarrow 2O + H}$. This reaction can be important
        for $\OHx$ destruction in dense and shielded regions, where
        photodissociation of $\OHx$ is less efficient. See $\mr{OH_x}$
        pseudo-reactions in Appendix \ref{section:OHx}.
    \item $\mr{He^+ + OH_x \rightarrow O^+ + He + H}$: This can be the major
        channel for $\mr{He^+}$ destruction in shielded regions. 
        See Appendix \ref{section:OHx}. 
    \item $\mr{C^+ + H_2 + e \rightarrow C + 2H}$. See the $\mr{C^+ + H_2}$
       reaction in Appendix \ref{section:CHx}. Note that this is not a 
       three-body reaction but a pseudo-reaction representing a
       two-body reaction followed by a recombination.
    \item $\mr{H_2O^+}$ destruction by reacting with electrons:
        \begin{equation*}
            \mr{H_2O^+ + e \rightarrow 2H + O}
        \end{equation*}
        In regions
        where electron abundance is relatively high, this can be limiting for
        the abundance of $\OHx$ species. This is applied implicitly as the
        branching ratio for $\OHx$ formation in reactions 2-5 in Table
        \ref{table:chem1}. See Appendix \ref{section:OHx} for details.
    \item $\mr{O^+}$ species and reactions: 
        \begin{align*}
            &\mr{H^+ + O \rightarrow O^+ + H}\\ 
            &\mr{O^+ + H \rightarrow H^+ + O}\\ 
        \end{align*}
        In high temperature regions, the first reaction has a higher rate than
        the second one (its reverse reaction). $\mr{O^+}$ formed by change
        exchange with $\mr{H^+}$ can react with $\Ht$, and subsequently from
        $\mr{OH^+}$. See Appendix \ref{section:OHx}.
    \item $\mr{Si}$, $\mr{Si^+}$ species and reactions: 
        \begin{align*}
            &\mr{Si^+ + e \rightarrow Si}\\ 
            &\mr{Si^+ + e + gr \rightarrow Si + gr}\\ 
            &\mr{\gamma_{cr} + Si \rightarrow Si^+ + e}\\ 
            &\mr{\gamma + Si \rightarrow Si^+ + e}\\ 
        \end{align*}
        The \citetalias{NL1999} network include species $\mr{M}$ to represent
        the metals besides $\mr{C}$ and $\mr{O}$. However, the metal abundance
        they are using, $x_\mr{M, tot} = 2\times 10^{-7}$, is much lower than the
        observed abundances in the solar neighborhood: the most abundant gas-phase
        metals are $\mr{S}$ and $\mr{Si}$,
        with $x_\mr{S, tot} = 3.5\times 10^{-6}$ \citep{Jenkins2009}, 
        and $x_\mr{Si, tot}=1.7\times 10^{-6}$ \citep{Cardelli1994}. 
        Because the dust-shielding factor $\gamma_\Si < \gamma_\CI$ (see
        Equation (\ref{eq:R_thick}) and Table \ref{table:chem2}),
        at $A_V \sim 1$, $\Siplus$ can be the main agent for providing
        electrons where $\CI$ is already shielded from photo-ionization
        (see e.g., Figure \ref{fig:species_all_nH500-1000} ).
        Although the total abundance of $\mr{S}$ is higher than $\Si$, we find
        that including $\mr{S}$ and $\mr{S^+}$ has very little effect on the
        electron or $\CO$ abundances, since $\gamma_\mr{S} \approx \gamma_\CI$.
        Therefore, we only include $\mr{Si}$ and $\mr{Si^+}$ for the metals.
\end{enumerate}

\subsection{$\CHx$ Pseudo-reactions\label{section:CHx}}
The pseudo-species $\mr{CH_x}$ includes $\mr{CH}$,  $\mr{CH_2}$,
$\mr{CH^+}$,  $\mr{CH_2^+}$, and $\mr{CH_3^+}$. The creation of
$\mr{CH_x}$ is from two reactions. The first one is
\begin{equation*}
    \mr{H_3^+ + C \rightarrow CH_x + H_2}.
\end{equation*}
This represents the sum of two reactions:
$\mr{H_3^+ + C \rightarrow CH^+ + H_2}$, 
and $\mr{H_3^+ + C \rightarrow CH_2^+ + H}$. We use the sum of both rates for
this pseudo-reaction. $\mr{CH^+}$ reacts with
$\Ht$ to form $\mr{CH_2^+}$ and $\Ho$, and $\mr{CH_2^+}$ again reacts with $\Ht$ to form
$\mr{CH_3^+}$ and $\Ho$. 
The second one,
\begin{equation*}
    \mr{C^+ + H_2 \rightarrow CH_x + H},
\end{equation*}
represents the reaction of $\mr{C^+ + H_2 \rightarrow CH_2^+}$.
$\mr{CH_2^+}$ quickly reacts with $\Ht$ and turns into $\mr{CH_3^+}$ and $\Ho$,
which then reacts with an electron. 70\% of the reaction
$\mr{CH_3^+ + e}$ gives $\mr{CH}$ or $\mr{CH_2}$, and 30\% turns
back into $\mr{C}$. Therefore, we use the rate in \citet{KIDA2010}
multiplied by 0.7 for $\mr{C^+ + H_2 \rightarrow CH_x + H}$, and by 0.3
for $\mr{C^+ + H_2 + e \rightarrow C + 2H}$.

The recombinations of $\mr{CH^+}$, $\mr{CH_2^+}$, and $\mr{CH_3^+}$ with electrons
form $\mr{CH}$ and  $\mr{CH_2}$, which can be destroyed in three
pathways:
\begin{align*}
    &\mr{CH_x + O \rightarrow CO + H}\\
    &\mr{CH_x + H \rightarrow C + H_2}\\
    &\mr{\gamma + CH_x \rightarrow C + H}
\end{align*}
For the reaction $\mr{CH_x + O \rightarrow CO + H}$, $\mr{CH}$ and
$\mr{CH_2}$ have similar rates. Here we use the rate of $\mr{CH}$. 
For the reactions $\mr{CH_x + H \rightarrow C + H_2}$ and 
$\mr{\gamma + CH_x \rightarrow C + H}$, $\mr{CH_2}$ will react with $\Ho$
to form $\mr{CH}$ and with a photon to form $\mr{CH}$ and $\mr{CH_2^+}$.
Therefore, we again use the rate assuming $\mr{CH_x}$ are all in $\mr{CH}$.

\subsection{$\OHx$ Pseudo-reactions\label{section:OHx}}
The pseudo-species $\OHx$ represents $\mr{OH}$,  
$\mr{H_2O}$, $\mr{OH^+}$,  $\mr{H_2O^+}$,  $\mr{H_3O^+}$.
The formation of $\mr{OH_x}$ is mainly through
\begin{equation}\label{eq:H3+_O}
    \mr{H_3^+ + O \rightarrow OH_x + H_2}
\end{equation}
This represents two reactions, the formation of $\mr{OH^+}$ and
$\mr{H_2O^+}$. The sum of both rates gives the rate of this 
pseudo-reaction. Another channel for $\OHx$ formation is 
\begin{equation}\label{eq:O+_H2}
    \mr{O^+ + H_2 \rightarrow OH_x + H},
\end{equation}
and this represents the reaction $\mr{O^+ + H_2 \rightarrow OH^+ + H}$.
$\Oplus$ mainly comes from the charge exchange $\mr{O + H^+ \rightarrow O^+ + H}$.

$\mr{OH^+}$ formed by the two channels above reacts with $\Ht$ to 
form $\mr{H_2O^+}$. The $\mr{H_2O^+}$ molecule has two fates: it can
again react with $\Ht$ to form $\mr{H_3O^+}$:
\begin{equation*}
    \mr{H_2O^+ + H2 \rightarrow H_3O^+ + H},
    ~k_\mathrm{H_2O^+, H_2} = 6.0\times 10^{-10} \mr{cm^{3}s^{-1}},
\end{equation*}
and then $\mr{H_3O^+}$ combines with electrons, eventually 
forming more stable $\mr{OH}$ and $\mr{H_2O}$.
Or, the $\mr{H_2O^+}$ molecule can be destroyed by electrons:
\begin{equation}\label{eq:H2O+_e}
    \mr{H_2O^+ + e \rightarrow 2H + O},
    ~k_\mathrm{H_2O^+, e} = 5.3\times 10^{-6} T^{-0.5} \mr{cm^{3}s^{-1}}.
\end{equation}
At low density and low $A_V$ regions where $\Ht$ abundance is low and electron
abundance is high, $\mr{H_2O^+}$ destruction by 
Equation (\ref{eq:H2O+_e}) can limit the formation of $\OHx$ by 
Equations (\ref{eq:H3+_O}) and (\ref{eq:O+_H2}). Therefore, we multiply the
reaction rates of Equations (\ref{eq:H3+_O}) and (\ref{eq:O+_H2}) by a branching factor
$r = k_\mathrm{H_2O^+, H_2} x_\mathrm{H_2} / (
k_\mathrm{H_2O^+, H_2} x_\mathrm{H_2} + k_\mathrm{H_2O^+, e}x_\mr{e})$, and
also add the reactions 3 and 5 in Table \ref{table:chem1}, which is
equivalent to the combined reactions of Equations (\ref{eq:H3+_O}) and (\ref{eq:O+_H2})
with Equation (\ref{eq:H2O+_e}).

We assume most $\mr{OH_x}$ are in $\mr{OH}$, and use the rate of
$\mr{OH}$ reactions
for the following pseudo-reactions of $\mr{OH_x}$ destruction:
\begin{align*}
    &\mr{C^+ + OH_x \rightarrow HCO^+}\\
    &\mr{OH_x + C \rightarrow CO + H}\\
    &\mr{\gamma + OH_x \rightarrow O + H}\\
    &\mr{OH_x + O \rightarrow 2O + H}\\
    &\mr{He^+ + OH_x \rightarrow O^+ + He + H}
\end{align*}
For the reaction $\mr{C^+ + OH_x \rightarrow HCO^+}$, 
we use the rate of $\mr{C^+ + OH \rightarrow CO^+ + H}$, because $\mr{CO^+}$
will quickly react with $\Ht$ to form $\mr{HCO^+}$. We use the rate of
reaction $\mr{OH + O \rightarrow O_2 + H}$ for $\mr{OH_x + O
\rightarrow 2O + H}$, assuming most oxygen will be in the form of atomic
$\mr{O}$.

\newpage
\begin{table*}[htbp]
    \caption{List of Collisional Chemical Reactions}
    \label{table:chem1}
    \begin{tabular}{l l l l l}
        \tableline
        \tableline
        No. &Reaction &Rate Coefficient\tablenotemark{a} &Notes
        &Reference\tablenotemark{b}\\ 
        \tableline
        1 &$\mathrm{H_3^+ + C \rightarrow CH_x + H_2}$ 
        &$1.04\times 10^{-9}(300/T)^{0.00231} + $ &
        $\mr{H_3^+ + C \rightarrow CH^+ + H_2}$ &1\\
        & &$T^{-3/2}\Sigma_{i=1}^{4}c_i\exp(-T_i/T)$ 
        & and $\mr{H_3^+ + C \rightarrow CH_2^+ + H}$ &\\
        & &$c_i = [3.40\times 10^{-8}, 6.97\times 10^{-9}, 
          1.31\times 10^{-7}, 1.51\times 10^{-4}]$ &  &\\
         & &$T_i = [7.62, 1.38, 2.66\times 10^{1}, 8.11\times 10^{3}]$ &  &\\
        2 &$\mathrm{H_3^+ + O \rightarrow OH_x + H_2}$ 
        &$1.99\times10^{-9} T^{-0.190} \times r$ & &20\\ 
        3 &$\mr{H_3^+ + O + e \rightarrow H2 + O + H}$ 
        &$1.99\times10^{-9} T^{-0.190} \times (1-r)$ 
        &$\mr{H_2O^+ + e \rightarrow 2H + O}$ &20\\
        4 &$\mr{O^+ + H_2 \rightarrow OH_x + H}$ &$1.6\times 10^{-9} \times r$ & &22\\
        5 &$\mr{O^+ + H_2 + e\rightarrow O + H + H}$ 
        &$1.6\times 10^{-9} \times (1-r)$ & &22\\
        & & $r = k_\mathrm{H_2O^+, H_2} x_\mathrm{H_2} / (
          k_\mathrm{H_2O^+, H_2} x_\mathrm{H_2} + k_\mathrm{H_2O^+, e}
          x_\mathrm{e})$
          &see Appendix \ref{section:OHx} &\\ 
          & &$ k_\mathrm{H_2O^+, H_2} = 6.0\times 10^{-10} $ & &22\\    
          & &$ k_\mathrm{H_2O^+, e} = 5.3\times 10^{-6} T^{-0.5}$ & &22\\    
        6 &$\mathrm{H_3^+ + CO \rightarrow HCO^+ + H_2}$ &$1.7\times 10^{-9}$
          & &2\\
        7 &$\mathrm{He^+ + H_2 \rightarrow H^+ + He + H}$ 
          &$1.26\times 10^{-13}\exp\left(-\frac{22.5}{T}\right)$ & &26\\
        8 &$\mathrm{He^+ + CO \rightarrow C^+ + O + He}$
          &$1.6\times10^{-9}$ & &4, 5\\
        9 &$\mathrm{C^+ + H_2 \rightarrow CH_x + H}$ 
        &$2.31\times 10^{-13}T^{-1.3}\exp(-\frac{23}{T})$ 
        &$\mathrm{C^+ + H_2 \rightarrow CH_2^+}$ &22\\
        10 &$\mathrm{C^+ + H_2 + e \rightarrow C + H + H}$ 
        &$0.99\times 10^{-13}T^{-1.3}\exp(-\frac{23}{T})$ 
        &$\mathrm{C^+ + H_2 \rightarrow CH_2^+}$ &22\\
        11 &$\mathrm{C^+ + OH_x \rightarrow HCO^+}$ 
        &$9.15\times 10^{-10}( 0.62 + 45.41 T^{-1/2} )$ 
        &$\mathrm{C^+ + OH \rightarrow CO^+ + H}$ &22\\
        12 &$\mathrm{CH_x + O \rightarrow CO + H}$ &$7.7\times 10^{-11}$ 
        &$\mathrm{CH + O \rightarrow CO + H}$ &22\\
        13 &$\mathrm{OH_x + C \rightarrow CO + H}$ 
        &$7.95\times 10^{-10} T^{-0.339} \exp(\frac{0.108}{T})$ 
        &$\mathrm{OH + C \rightarrow CO + H}$ &21, 22\\
        14 &$\mathrm{He^+ + e \rightarrow He}$ 
           &$10^{-11}T^{-0.5}\times[11.19$ &Case B &6, 7\\
        & &$~-1.676 \log T-0.2852(\log T)^2 + 0.04433(\log T)^3]$ & & \\
        15 &$\mathrm{H_3^+ + e \rightarrow H_2 + H}$ 
           &$4.54\times 10^{-7}T^{-0.52}$ & &8\\
        16 &$\mathrm{H_3^+ + e \rightarrow 3 H}$ 
           &$8.46\times 10^{-7}T^{-0.52}$ & &8\\
        17 &$\mathrm{C^+ + e \rightarrow C}$ 
        &$k_\mr{rr} = \frac{2.995\times 10^{-9}}
        {\alpha (1.0+\alpha)^{1.0-\gamma} (1.0+\beta)^{1.0+\gamma}}$ 
        &Radiative and&9, 23, 27\\
           & &$k_\mr{dr} = T^{-3/2}\times [ 6.346\times 10^{-9}
                \exp(\frac{-12.17}{T})$ &dielectronic recombination & \\
           & &$~ + 9.793\times 10^{-9} \exp(\frac{-73.8}{T}) 
                + 1.634\times 10^{-6} \exp(\frac{-15230}{T})]$ & &\\
           & &$k = k_\mr{rr} + k_\mr{dr}, 
        ~\alpha \equiv \sqrt{\frac{T}{6.670\times 10^{-3}}}, 
        ~\beta \equiv \sqrt{\frac{T}{1.943\times 10^{6}}},$ & &\\
           & &$~\gamma \equiv 0.7849 + 0.1597\exp(\frac{-49550}{T})$ & &\\
        18 &$\mathrm{HCO^+ + e \rightarrow CO + H}$ 
        &$1.06\times10^{-5}T^{-0.64}$ & &10\\
        19 &$\mathrm{H_2^+ + H_2 \rightarrow H_3^+ + H}$
        &$1.76\times 10^{-9}T^{0.042}\exp\left(-\frac{T}{46600}\right)$ & &7, 11\\
        20 &$\mathrm{H_2^+ + H \rightarrow H^+ + H_2}$
        &$6.4\times 10^{-10}$ & &22\\
        21 &$\mathrm{H^+ + e \rightarrow H}$ 
        &$2.753\times 10^{-14}\left( \frac{315614}{T}\right)^{1.500} 
        \left[1.0 + \left( \frac{115188}{T} \right)^{0.407}\right]^{-2.242}$
        & Case B &7, 12\\
        22 &$\mathrm{H_2 + H \rightarrow 3 H}$
        &$k_{16,l} = 6.67\times 10^{-12}\sqrt{T}\exp\left[-\left(1.0 +
        \frac{63590}{T}\right)\right]$ & Density dependent, see &13, 14 \\
        & &$k_{16,h} = 3.52\times 10^{-9}\exp\left(-\frac{43900}{T}\right)$ 
        &\cite{GM2007}&13, 15\\
        & &$n_\mathrm{cr, H} = 
        \mathrm{dex}[3.0 - 0.416\log T_4 - 0.327(\log T_4)^2]$ & &13, 15, 16\\
        & &$n_\mathrm{cr, H_2} =  \mathrm{dex}[
        4.845 - 1.3\log T_4 + 1.62(\log T_4)^2]$ & &13, 17\\
        23 &$\mathrm{H_2 + H_2 \rightarrow H_2 + 2 H}$
        &$k_{17, l} = \frac{5.996\times 10^{-30}T^{4.1881}} {
        \left(1.0 + 6.761\times 10^{-6}T\right)^{5.6881}  } 
        \exp\left(\frac{-54657.4}{T}\right)$
        &Density dependent, see &13, 18\\
        & &$k_{17, h} = 1.3\times 10^{-9}\exp\left(-\frac{53300}{T}\right)$
        &\cite{GM2007}&13, 17\\
        24 &$\mathrm{H + e \rightarrow H^+ + 2 e}$
        &$\exp[ -3.271396786\times 10^1 +1.35365560\times10^{1}\ln T_e $ & &19\\  
        & &$~- 5.73932875(\ln T_e)^2 +1.56315498 (\ln T_e)^3 $ & &\\
        & &$~- 2.877056\times 10^{-1}(\ln T_e)^4 +3.48255977\times 10^{-2}(\ln T_e)^5$
        & &\\
        & &$~-2.63197617\times 10^{-3}(\ln T_e)^6 $ & &\\
        & &$~+ 1.11954395\times 10^{-4}(\ln T_e)^7$ & &\\
        & &$~-2.03914985\times 10^{-6}(\ln T_e)^8]$ & &\\
        25 &$\mr{He^+ +H_2 \rightarrow H_2^+ +He}$ 
        &$7.20\times 10^{-15}$ & &3, 4\\
        26 &$\mr{CH_x + H \rightarrow H_2 + C}$ &$2.81\times 10^{-11}T^{0.26}$ & &22\\
        27 &$\mr{OH_x + O \rightarrow 2O + H}$ &$3.5\times 10^{-11}$ 
        &$\mr{OH + O \rightarrow O_2 + H}$ &24\\
        28 &$\mr{Si^+ + e \rightarrow Si}$ &$1.46\times 10^{-10} T^{-0.62}$ & &4\\
        29 &$\mr{He^+ + OH_x \rightarrow O^+ + He + H}$ &$1.35\times 10^{-9} (
        0.62 + 45.41 T^{-1/2})$ & &22\\
        30 &$\mr{H^+ + O \rightarrow O^+ + H}$ &$( 1.1\times 10^{-11} T^{0.517}
        + 4.0\times 10^{-10} T^{0.00669} ) \exp\left(\frac{-227}{T}\right)$ & &25\\
        31 &$\mr{O^+ + H \rightarrow H^+ + O}$ &$4.99\times 10^{-11}T^{0.405} +
        7.5\times 10^{-10} T^{-0.458}$ & &25\\

        \tableline
    \end{tabular}
    \tablenotetext{1}{Rate coefficients are in units of
        $\mathrm{cm^{3}s^{-1}}$. The
    temperature $T$ is in Kelvin, and $T_4 = T/(10^4\mathrm{K})$. $T_e$ is
    temperature in the unit of $\mathrm{eV}/k$ where $k$ is the Boltzmann
    constant, i.e., $T_e = 8.6173\times10^{-5}T$. $x_i=n_i/n$ is the
    abundance of species $i$. $\chi$ is the flux of the
    incident radiation field in $6-13.6\mathrm{eV}$ relative to the standard
    radiation field ($\chi=1$, 
    $J_\mr{FUV} = 2.7 \times 10^{-3} \mr{erg~cm^{-2} s^{-1}}$)
    in \citet{Draine1978} units. This field has a strength of
    $G_0=1.7$ in \citet{Habing1968} units. $A_V$ is the visual extinction.}
    \tablenotetext{2}{References: (1)\citet{Vissapragada2016}, (2)\citet{Kim1975},
    (3)\citet{Barlow1984}, (4)\citet{McElroy2013}, (5)\citet{AH1986},
    (6)\citet{HS1998}, (7)\citet{Glover2010}, (8)Fit by \citet{Woodall2007} to data
    from \citet{McCall2004}, (9)\citet{Badnell2003} (Fitting coefficients are
    taken for $T\lesssim10^4~\mr{K}$ from their
    website http://amdpp.phys.strath.ac.uk/tamoc/DR/), (10)\citet{Geppert2005},
    (11)\citet{Linder1995} (12)\citet{Ferland1992}, (13)\citet{GM2007},
    (14)\citet{MS1986}, (15)\citet{LS1983}, (16)\citet{Martin1996},
    (17)\citet{SK1987}, (18)\citet{Martin1998}, (19)\citet{Janev1987}, 
(20)Fit to \citet{deRuette2015}, accurate to $\sim 10\%$ at $T=10-1000\mathrm{K}$,
(21)\citet{Zanchet2009}, (22)\citet{KIDA2010}, (23)\citet{Badnell2006},
(24)\citet{Carty2006}, (25)\citet{Stancil1999}, (26)Fit to \citet{Schauer1989}, 
(27)\citet{Bryans2009}.}
\end{table*}

\begin{table*}[htbp]
    \caption{List of Grain-assisted Reactions, Cosmic-ray Reactions,
    and Photodissociation Reactions}
    \label{table:chem2}
    \begin{tabular}{l l l l l}
        \tableline
        \tableline
        No. &Reaction &Rate coefficient\tablenotemark{a} &Notes
        &Reference\tablenotemark{b}\\ 
        \tableline
        \multicolumn{5}{l}{Grain-assisted reactions:}\\
        1 &$\mathrm{H + H + gr \rightarrow H_2 + gr}$ 
        &$3.0\times 10^{-17}$ & &1, 2\\
        2 &$\mathrm{H^+ + e + gr \rightarrow H + gr}$
        &$12.25\times 10^{-14} [1 + 8.074\times 10^{-6}\psi^{1.378} \times$ &
        &3, 4\\
        & &$~
        (1 + 508.7T^{0.01586}\psi^{-0.4723 - 1.102\times 10^{-5}\ln T})]^{-1}$
        & &\\
        3 &$\mathrm{C^+ + e + gr \rightarrow C + gr}$
        &$45.58\times 10^{-14} [1 + 6.089\times 10^{-3}\psi^{1.128} \times$ &
        &3\\
        & &$~
        (1 + 433.1T^{0.04845}\psi^{-0.8120 - 1.333\times 10^{-4}\ln T})]^{-1}$
        & &\\
        4 &$\mathrm{He^+ + e + gr \rightarrow He + gr}$
        &$5.572\times 10^{-14} [1 + 3.185\times 10^{-7}\psi^{1.512} \times$ &
        &3\\
        & &$~
    (1 + 5115T^{3.903\times 10^{-7}}\psi^{-0.4956 - 5.494\times 10^{-7}\ln T})]^{-1}$
        & &\\
        5 &$\mr{Si^+ + e + gr \rightarrow Si + gr}$ 
        &$2.166\times 10^{-14} [1 + 5.678\times 10^{-8}\psi^{1.874} \times$ &
        &3\\
        & &$~
    (1 + 43750T^{1.635\times 10^{-6}}\psi^{-0.8964 - 7.538\times 10^{-5}\ln T})]^{-1}$
        & &\\

        & &$\psi \equiv \frac{1.7\chi\exp(-1.87 A_V)\sqrt{T}}{n_e/\mathrm{cm^{-3}}}$
        & &\\

        \multicolumn{5}{l}{Cosmic-ray ionization or cosmic-ray induced
        photodissociation:}\\
        6 &$\mathrm{cr + H  \rightarrow H^+ + e}$ 
         & $2.3x_\mathrm{H_2} + 1.5x_\mathrm{H}$ 
         &primary and  &8\\ 
        7 &$\mathrm{cr + H_2 \rightarrow H_2^+ + e}$ 
          &$2 \times (2.3 x_\mathrm{H_2} + 1.5x_\mathrm{H})$
         &secondary ionization &8\\ 
        8 &$\mathrm{cr + He \rightarrow He^+ + e}$ & 1.1 & &5, 6\\
        9 &$\mr{cr + C \rightarrow C^+ + e}$ &3.85 & &5, 6\\
        10 &$\mr{cr + CO + H \rightarrow HCO^+ + e}$ &6.52 
        &$\mr{cr + CO \rightarrow CO^+ + e}$ &5, 6\\
        11 &$\mr{\gamma_{cr} + C \rightarrow C^+ + e}$ &560 & &7\\
        12 &$\mr{\gamma_{cr} + CO \rightarrow C + O}$ &90 & &7\\
        13 &$\mr{\gamma_{cr} + Si \rightarrow Si^+ + e}$ &8400 & &7\\

        \multicolumn{5}{l}{Photoionization and photodissociation reactions\tablenotemark{c}:}\\
        14 &$\mathrm{\gamma + C \rightarrow C^+ + e}$
        &$3.5\times 10^{-10} \chi \exp(-3.76A_V)
        f_{s,\mathrm{C}}(N_\mathrm{C},N_\mathrm{H_2})$ & &7, 9\\
        15 &$\mathrm{\gamma + CH_x \rightarrow C + H}$
        &$9.1\times 10^{-10} \chi \exp(-2.12A_V)$ 
        &$\mathrm{\gamma + CH \rightarrow C + H}$ &7\\
        16 &$\mathrm{\gamma + CO \rightarrow C + O}$
        &$2.4\times 10^{-10} \chi \exp(-3.88A_V)
        f_{s, \mathrm{CO}}(N_\mathrm{CO}, N_\mathrm{H_2})$ & &7, 10\\
        17 &$\mathrm{\gamma + OH_x \rightarrow O + H}$
        &$3.8\times 10^{-10} \chi \exp(-2.66 A_V)$ 
        &$\mathrm{\gamma + OH \rightarrow O + H}$ &7\\
        18 &$\mathrm{\gamma + Si \rightarrow Si^+ + e}$
        &$4.5\times 10^{-9}\chi \exp(-2.61 A_V)$
        & &7\\
        19 &$\mathrm{\gamma + H_2 \rightarrow H + H}$
        &$5.7\times 10^{-11}\chi \exp(-4.18
        A_V)f_{s,\mathrm{H_2}}(N_\mathrm{H_2})$ & &7, 11\\
        \tableline
    \end{tabular}
    \tablenotetext{1}{Rate coefficients are in units of 
    $\mathrm{cm^{3}s^{-1}}Z_d^{-1}$ for grain-assisted
    reactions, where $Z_d$ is the dust abundance relative to the solar
    neighbourhood (the reaction rates for grain-assisted reactions are 
    $x_i n k_\mr{gr}$, where $k_\mr{gr}$ is the listed rate coefficient,
    and $x_i=x_\Ht$ for reaction 1 and $x_i=x_\mr{ion}$ for reactions 2-5);  
    $\xi_\mathrm{H}$ for cosmic-ray ionization where
    $\xi_\mathrm{H}=2.0\times 10^{-16} \mathrm{s^{-1}H^{-1}}$ is the local
    primary cosmic-ray ionization rate for atomic $\Ho$ per $\Ho$ atom
    \citep{Indriolo2007}; and
    $\mathrm{s^{-1}}$ for photoreactions. $n_\mathrm{e}$ is the number
    density of electrons. $x_i=n_i/n$ is the
    abundance of species $i$. $\chi$ is the flux of the
    incident radiation field in $6-13.6\mathrm{eV}$ relative to the standard
    radiation field ($\chi=1$, 
    $J_\mr{FUV} = 2.7 \times 10^{-3} \mr{erg~cm^{-2} s^{-1}}$
    ) in \citet{Draine1978}. This field has a strength of
    $G_0=1.7$ in \citet{Habing1968} units. $A_V$ is the visual extinction.}
    \tablenotetext{2}{References: (1) \citet{Wolfire2008}, (2)\citet{Hollenbach2012},
        (3)\citet{WD2001b}, (4)\citet{Draine2003}, (5)\citet{Glover2010}, 
        (6)\citet{LeTeuff2000}, (7)\citet{Heays2017}, (8)\citet{GL1974},
        (9)\citet{TH1985}, (10)\citet{Visser2009}, (11)\citet{DB1996}.}
\tablenotetext{3}{The self-shielding of $\mathrm{H_2}$ and self-shielding plus
shielding from $\mathrm{H_2}$ of $\mathrm{CO}$ and $\mathrm{C}$ are included in the factors
$f_{s, \mathrm{H_2}}(N_\mathrm{H_2})$,
$f_{s, \mathrm{CO}}(N_\mathrm{CO}, N_\mathrm{H_2})$, and
$f_{s,\mathrm{C}}(N_\mathrm{C},N_\mathrm{H_2})$.
This is described in detail in Section \ref{section:photo-chemstry}.} 
\end{table*}

\section{Heating and Cooling Processes}\label{section:heating_cooling}
A summary of all the heating and cooling processes is listed in Table
\ref{table:thermo}.
\subsection{Heating}
\subsubsection{Cosmic-Ray Ionization}
When the gas is ionized by cosmic rays, the primary and secondary
electrons thermalize with and heat up the gas. 
The heating rate per $\Ho$,
\begin{equation}\label{eq:Gamma_cr}
    \Gamma_\mr{cr} = \xi q_\mr{cr},
\end{equation}
where $\xi = \xi_\Ho x(\Ho) +  \xi_\Ht x(\Ht) + \xi_\He x(\He)$, and
$q_\mr{cr}$ is the energy added to the gas per primary ionization.
We use the result fitted by \citet{Draine2011} to the data
of \citet{DM1972} in atomic regions:
\begin{equation}
    q_\mr{cr, H} = 6.5 + 26.4 \left( \frac{x(\mr{e})}{ x(e)+0.07} \right)^{0.5}~
    \mr{eV}, 
\end{equation}
and the fit by \citet{DESPOTIC} to the data of \citet{Glassgold2012} in
molecular regions:
\begin{equation}\label{eq:q_cr}
    \frac{q_\mr{cr, H_2}}{\mr{eV}} =
    \begin{cases}
        10, ~&\log n < 2\\
        10 +  \frac{3(\log n - 2)}{2} , ~& 2\leq \log n < 4\\
        13 +  \frac{4(\log n - 4)}{3} , ~& 4\leq \log n < 7\\
        17 + \frac{\log n - 7}{3} , ~& 7\leq \log n < 10\\
        18, ~&\log n \geq 10
    \end{cases}.
\end{equation}
Similar to \citet{DESPOTIC}, we simply assume that the total heating rate can be
calculated by summing the heating rates in the atomic and molecular regions
weighted by their number fraction:
\begin{equation}
    q_\mr{cr} = x(\Ho) q_\mr{cr, H} + 2x(\Ht) q_\mr{cr, H_2}.
\end{equation}

\subsubsection{Photoelectric Effect on Dust Grains}
One main heating source in the ISM comes from the photoelectric effect on dust
grains, where electrons ejected from dust grains by FUV radiation thermalize
with the gas.
We use the results in \citet{WD2001b} Equation (44) to calculate the heating from
the photoelectric effect on dust grains.
The rates are extracted from their Table 2, 
using the values for $R_v=3.1$, $b_c=4.0$, and radiation field with ISRF spectrum. 

\subsubsection{$\Ht$ Formation on Dust Grains, and UV Pumping of $\Ht$}
We follow \citet{HM1979}, using their Equation (6.43) for
heating
from $\Ht$ formation on dust grains and their Equation (6.46) for 
UV pumping of $\Ht$. Both processes are more efficient at higher densities. 
The rate of $\Ht$ formation on grains increases with density. 
For UV pumping, the energy is initially in the
excited rotational/vibrational levels of $\Ht$, and subsequently turns
into heat by collisional de-excitation. 

\subsubsection{$\Ht$ Photodissociation}
The photodissociation of $\Ht$ releases $\sim 0.4 \mr{eV}$ heat per reaction
\citep{BD1977, HM1979}.

\subsection{Cooling}
\subsubsection{Atomic lines: $\mr{C^+}$, $\mr{C}$, and $\mr{O}$ Fine Structure
lines, and the Ly$\alpha$ line}
To calculate atomic line cooling, $\mr{C^+}$ and $\Ho$ (Ly$\alpha$)
are considered as two-level systems, and $\mr{C}$ and $\mr{O}$ as three-level
systems. The calculation of level populations follows the standard procedure,
including collisional excitation and de-excitation, photoabsorption, and
spontaneous and stimulated emission \citep[e.g.][]{Draine2011}. The species
considered for collisional excitation and de-excitation are listed in Table
\ref{table:thermo}.


\subsubsection{$\CO$ Rotational Lines}
The $\CO$ rotational lines are often optically thick in regions where $\CO$
cooling is important. To calculate the cooling rate by $\CO$ rotational lines,
we use the cooling functions in \citet{Omukai2010} 
(their Appendix C and Table B2). They use the large velocity
gradient (LVG) approximation, similar to the approach by \citet{NK1993}. 
The cooling rate per $\Ho$ is given by
\begin{equation}\label{eq:Gamma_CO}
    \Gamma_\CO = L_\CO x(\CO) n_\Ht
\end{equation}
where $L_\CO$ is given in terms of $n_\Ht$, $\tilde{N}(\CO)$, and $T$ by
Equation (B1) in \citet{Omukai2010}.
In \citet{NK1993} and \citet{Omukai2010}, 
they assume $\Ht$ is the only species responsible for the collisional
excitation of the $\CO$ rotational levels, which is appropriate in
fully molecular regions. To take into account of the collisional excitation by
atomic hydrogen and electrons, we replace $n_\Ht$ in
Equation (\ref{eq:Gamma_CO}) with an effective number density, 
following \citet{Yan1997}, \citet{MS2005} and \citet{Glover2010}:
\begin{equation}
    n_\mr{eff, CO} = n_\Ht + \sqrt{2} \left( \frac{\sigma_\Ho}{\sigma_\Ht}
    \right)n
    + \left( \frac{1.3\times 10^{-8}~\mr{cm^3 s^{-1}}}{\sigma_\Ht v_\mr{e} }
    \right)n_\mr{e},
\end{equation}
where $\sigma_\Ho = 2.3\times 10^{-15}~\mr{cm^2}$, 
$\sigma_\Ht = 3.3\times 10^{-16} (T/1000\mr{K})^{-1/4}~\mr{cm^2}$
and $v_\mr{e} = 1.03\times 10^{4}(T/1\mr{K})^{1/2}~\mr{cm~s^{-1}}$.

In LVG approximation , the escape probability of a photon emitted by $\CO$
is related to the effective column density parameter
\begin{equation}
    \tilde{N} (\CO) = \frac{n(\CO)}{|\nabla v|},
\end{equation}
where $n(\CO)$ is the local $\CO$ number density and $\nabla v$ is the local
gas velocity gradient. In \citet{Omukai2010}, they used the approximation 
$\tilde{N}(\CO) = N(\CO)/ \sqrt{2kT/m(\CO)}$, where $T$ is the gas temperature,
$N(\CO)$ the total $\CO$ column density, and $m(\CO)$ the mass of a
single $\CO$ molecule. This assumes thermal motions dominate the
velocity of $\CO$ molecules. However, in realistic molecular clouds, 
the gas velocity will be largely determined by the supersonic turbulence. 
For Milky Way molecular clouds, the observed turbulence
spectrum obeys the line-width size relation \citep{Larson1981, Solomon1987, HB2004}:
\begin{equation}\label{eq:v_L}
    v(L) \sim 1~\mr{km/s}~\left( \frac{L}{\mr{pc}} \right)^{1/2}.
\end{equation}
In our slab models, we use the velocity determined using Equation (\ref{eq:v_L}) to
calculate the effective $\CO$ column density parameter 
$\tilde{N} (\CO) = N(\CO)/ v(L)$, where $N(\CO)$ is the
column density of $\CO$ from the local position to the edge of the one-sided
slab. This assumes that for thicker clouds, the global turbulent velocity will
be larger as implied by Equation (\ref{eq:v_L}). We have also tried using a constant
velocity $v=1\mr{km/s}$ to calculate $\tilde{N}$, and the results for the equilibrium
temperature in $\CO$-dominated regions are very similar.

Care needs to be taken when the temperature $T < 10\mr{K}$. 
\citet{Omukai2010} only gives fitting parameters down to $T=10\mr{K}$, and
naive extrapolation may give a finite cooling rate even when the temperature is
close to zero. Moreover, $\CO$ can freeze-out onto dust grains in regions with
high density and $T \lesssim 10 \mr{K}$, leading to depleted $\CO$ abundance
\citep{Bergin1995, Acharyya2007}.
For simplicity, we artificially turn off $\CO$ cooling at temperatures below
$10\mr{K}$. \citet{Omukai2010} also only gives fitting parameters up to
$T=2000\mr{K}$. For $T>2000\mr{K}$, we simply use the $\CO$ cooling rate at
$T=2000\mr{K}$. In reality, $\CO$ will likely be destroyed by collisional
dissociation and ionization at such high temperatures. 

\subsubsection{$\Ht$ Rotational and Vibrational Lines}
We use the results in \citet{GA2008} to calculate cooling by $\Ht$ rotational and
vibrational lines. The cooling rate per $\Ht$ molecule $\Lambda_\Ht$
is approximated by 
\begin{equation}
    \Lambda_\Ht =
    \frac{\Lambda_\mr{H_2, LTE}}{1 + \Lambda_\mr{H_2, LTE}/\Lambda_\mr{H_2,
    n\rightarrow 0}},
\end{equation}
where $\Lambda_\mr{H_2, LTE}$ is the cooling rate at high densities when $\Ht$
level populations are in local thermal equilibrium, and 
$\Lambda_\mr{H_2, n\rightarrow 0}$ is the cooling rate in the low density limit. We
use Equations (6.37) and (6.38) in \citet{HM1979} to calculate 
$\Lambda_\mr{H_2, LTE} = \Lambda_\mr{H_2, LTE, rot} + \Lambda_\mr{H_2, LTE, vib}$
for $\Ht$ rotational and vibrational lines. $\Lambda_\mr{H_2, n\rightarrow 0}$
is obtained by the fitting functions provided in \citet{GA2008}, including
collisional species $\Ho$, $\Ht$, $\He$, $\mr{H^+}$, and $\mr{e}$. We use 
the fitting parameters in their
Table 8, assuming a fixed ortho-to-para ratio of $\Ht$ of 3:1.
\citet{GA2008} only gives the fit for $\Ht$ cooling in the temperature range
$20\mr{K} < T < 6000\mr{K}$. Similar to the treatment in $\CO$ rotational line
cooling, we artificially cut off $\Ht$ cooling below $10\mr{K}$, and use the
$\Ht$ cooling rate at $T=6000\mr{K}$ for $T > 6000\mr{K}$. In reality, $\Ht$
cooling rate is likely to be negligible at $T<10\mr{K}$.

\subsubsection{Dust Thermal Emission}
By exchanging energy with dust via collision, gas can either heat or cool. The
rate of gas--dust energy exchange is given by \citep{DESPOTIC}:
\begin{equation}
    \Psi_\mr{gd} = \alpha_\mr{gd} 
    n Z_\di T_\mr{g}^{1/2} (T_\mr{d} - T_\mr{g}), 
\end{equation}
where $Z_d$ is the dust abundance relative to the Milky Way, and
$\alpha_\mr{gd}$ the dust--gas coupling coefficient. 
Positive $\Psi_\mr{gd}$  means heating of the gas, and negative means cooling of the
gas. In the realistic ISM, dust almost always acts to cool the ISM with $T_\di <
T_\mr{g}$. We use $ \alpha_\mr{gd}=3.2\times 10^{-34} \mr{erg~s^{-1}cm^3 K^{-3/2}}$
for $\Ht$-dominated regions \citep{Goldsmith2001}. $\alpha_\mr{gd}$ is expected
to be a factor $\sim 3$ higher in $\mr{H}$-dominated regions
\citep{KLM2011}, but dust cooling is likely to be only important in dense
regions dominated by $\Ht$.

The dust temperature $T_\mr{d}$ necessary for evaluating $ \Psi_\mr{gd}$ is
determined by the total rate of change of dust specific energy per $\Ho$ nucleus
\citep{DESPOTIC}:
\begin{equation}\label{eq:edust}
    \frac{\di e_\mr{d, sp}}{\di t} = \Gamma_\mr{ISRF} + \Gamma_\mr{d,line}
    + \Gamma_\mr{d, CMB} + \Gamma_\mr{d, IR} - \Lambda_\mr{d} - \Psi_\mr{gd} 
\end{equation}
Because the dust specific heat is very small compared to the gas, we can always
assume the dust temperature is in equilibrium, $\di e_\mr{d, sp}/\di t = 0$.
Note that for very small dust with size $a\lesssim 10$\AA, 
they are not in thermal equilibrium, and can have temperature spikes 
\citep{Draine2011}. Here we mainly consider bigger grains.
In principle, given all the terms in Equation (\ref{eq:edust}), one can solve for
$T_\di$. However, because Equation (\ref{eq:edust}) is a non-linear function of
$T_\di$, this procedure will involve a costly root-finding algorithm. For the
purposes of this paper, we assume a fixed dust temperature $T_\mr{d}=10~\mr{K}$. 
In the density range $n<10^4~\mr{cm^{-3}}$ that we consider, dust cooling
is never important and contributes to less than $\sim 5\%$ of the total cooling
rate.

\subsubsection{Recombination of $\mr{e}$ on PAHs}
At high temperatures $T\sim 10^3-10^4 \mr{K}$, cooling by electron
recombination on PAHs can be important relative to the photoelectric heating
on dust. Following \citet{WD2001b} Equation (45), we use the parameters in
their Table 3 ($R_v=3.1$, $b_c=4.0$, ISRF) to calculate the cooling rate 
from electron recombination on PAHs. 

\subsubsection{Collisional Dissociation of $\Ht$ and Collisional Ionization of $\Ho$}
The collisional dissociation of $\Ht$ and ionization of $\Ho$ takes
$4.48\mr{eV}$ and $13.6\mr{eV}$ energy per reaction \citep{KROME}.

\begin{table*}[htbp]
    \caption{List of Heating and Cooling Processes}
    \label{table:thermo}
    \begin{tabular}{l  l}
        \tableline
        \tableline
        Process &Reference\\ 
        \tableline
        \multicolumn{2}{l}{Heating:}\\
        Cosmic-ray ionization of $\Ho$, $\Ht$ and $\He$ 
        &Cosmic-ray ionization rate -- See Table \ref{table:chem2}\\
         & Heating rate per primary ionization in atomic region --
        \citet{DM1972}\\
         & Heating rate per primary ionization in molecular region -- 
        \citet{Glassgold2012} \\
        Photoelectric effect on dust grains & \citet{WD2001a}\\
        $\Ht$ formation on dust grains &\citet{HM1979}\\
        UV pumping of $\Ht$ &\citet{HM1979}\\
        $\Ht$ photodissociation &\citet{BD1977, HM1979}\\
        \\

        \multicolumn{2}{l}{Cooling:}\\
        $\mr{C^+}$ fine structure line & Atomic data -- \citet{SV2002}\\
                                       & Collisional rates with $\Ho$ --
          \citet{Barinovs2005}\\
           & Collisional rates with $\Ht$ -- \citet{WG2014} ($T<500\mr{K}$), and\\
           & $\qquad$\citet{GJ2007} ($T > 500\mr{K}$)\\
            & Collisional rates with $\mr{e}$ -- \citet{Keenan1986}\\
        $\mr{C}$ fine structure line & Atomic data -- \citet{SV2002}\\
                                     & Collisional rates with $\Ho$ -- fit to
          \citet{AKD2007} by \citet{Draine2011} \\
          & Collisional rates with $\Ht$ -- fit to \citet{Schroder1991} by
          \citet{Draine2011}\\
           & Collisional rates with $\mr{e}$ -- \citet{JBK1987} \\
        $\mr{O}$ fine structure line & Atomic data -- \citet{SV2002}\\
                                     & Collisional rates with $\Ho$ -- fit to
          \citet{AKD2007} by \citet{Draine2011}  \\
          & Collisional rates with $\Ht$ -- fit to \citet{Jaquet1992} by
          \citet{Draine2011}\\
          & Collisional rates with $\mr{e}$ -- fit to \citet{Bell1998} \\
          Ly$\alpha$ line & Atomic data -- \citet{Draine2011}\\
                            & Collisional rates with $\mr{e}$ -- fit to
          \citet{Vrinceanu2014} \\
        $\CO$ rotational lines &\citet{Omukai2010}\\
        $\Ht$ rotational and vibrational lines & \citet{GA2008, HM1979}\\
        Dust thermal emission &\citet{DESPOTIC}\\
          Recombination of $\mr{e}$ on PAHs & \citet{WD2001a}\\
          Collisional dissociation of $\Ht$ & \citet{KROME}\\
        Collisional ionization of $\Ho$ & \citet{KROME}\\
        \tableline
    \end{tabular}
\end{table*}

\section{Additional Plots and Discussions for Code Test and Comparison}
\label{section:add_plots}
\subsection{Chemistry with Constant Temperature\label{section:T20}}
\begin{figure*}[htbp]
     \begin{center}
        \subfigure[$n=50~\mr{cm^{-3}}$]{%
            \includegraphics[width=0.97\textwidth]{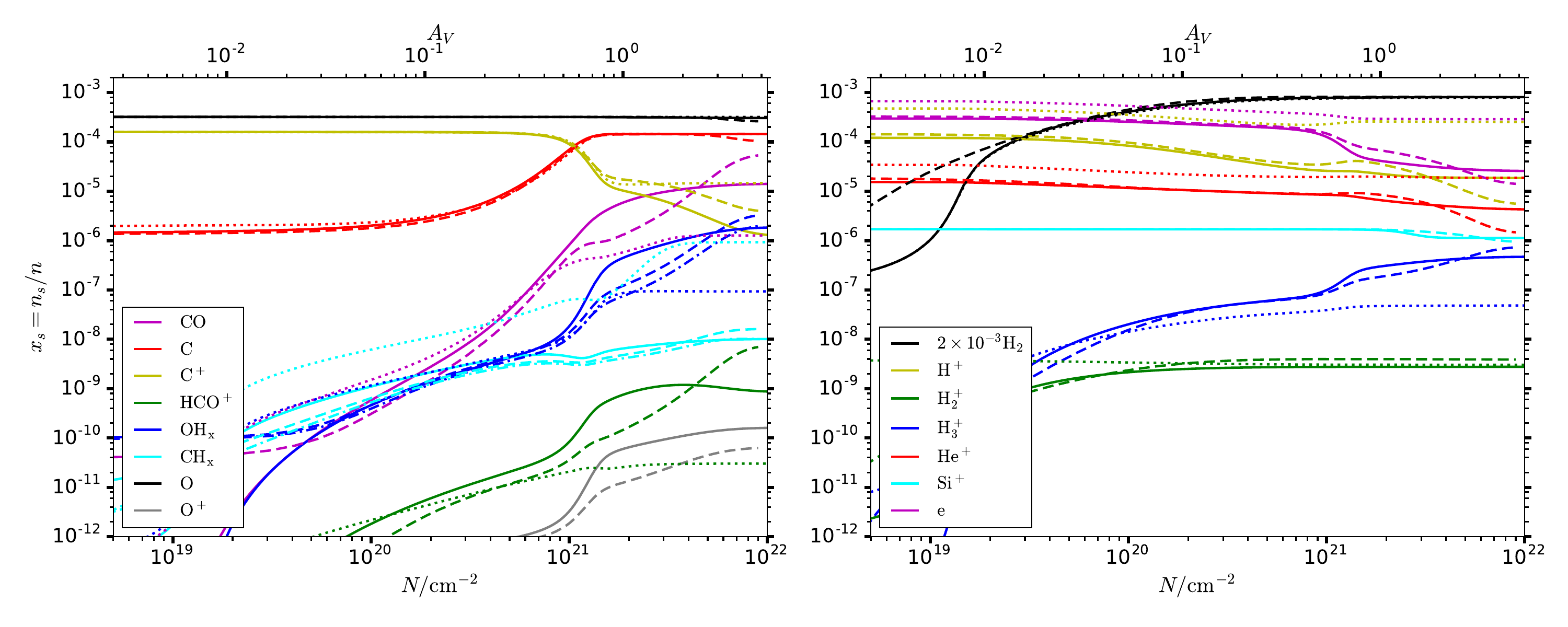}
        }
        \subfigure[$n=100~\mr{cm^{-3}}$]{%
           \includegraphics[width=0.97\textwidth]{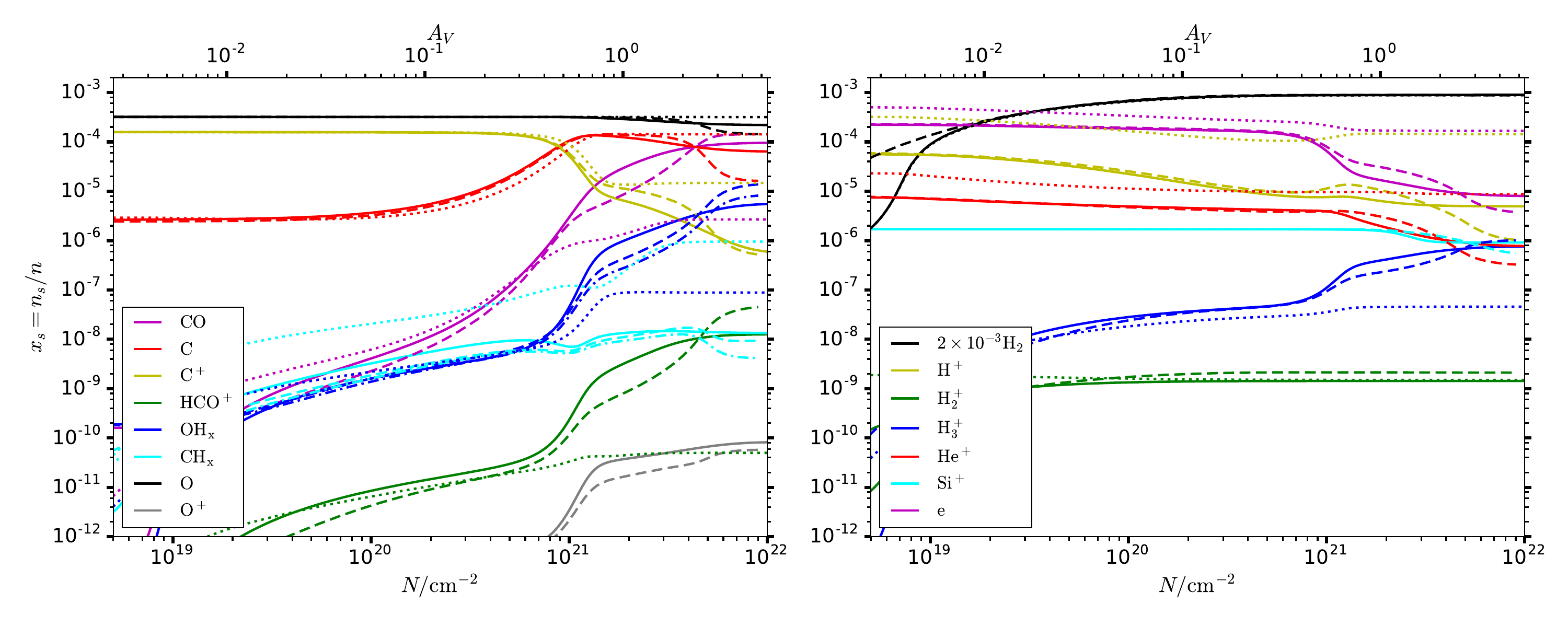}
        }
        \subfigure[$n=200~\mr{cm^{-3}}$]{%
           \includegraphics[width=0.97\textwidth]{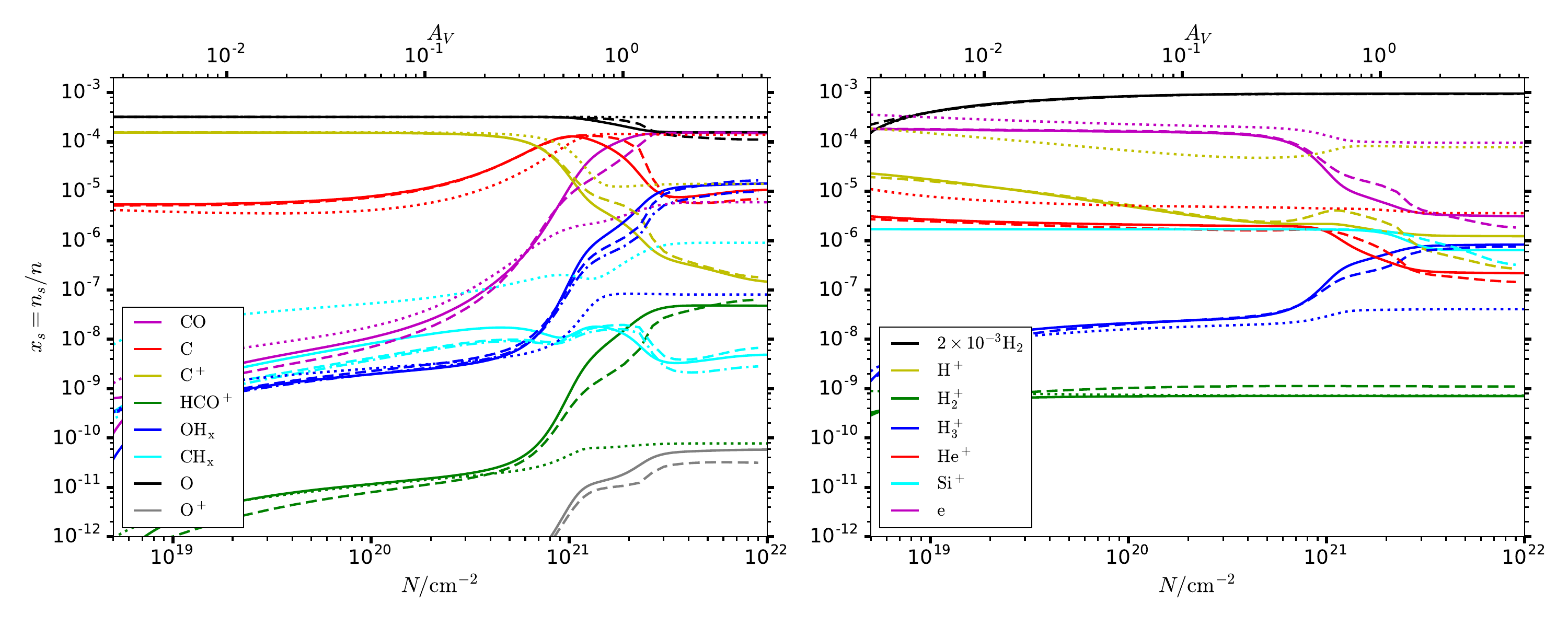}
        }

    \end{center}
    \caption{Abundances of all independent species at
        densities $n=50-200~\mr{cm^{-3}}$ in our network (solid), 
        the PDR code (dashed), and the \citetalias{NL1999} network
        (dotted), as a function of $A_V$/$N$.
        These figures are similar to Figure \ref{fig:species_nH}.
        Different species are represented
    by different colors as indicated in the legends. The abundance of $\Ht$ is
    multiplied by a factor of $2\times 10^{-3}$, i.e., if all hydrogen is in
    $\Ht$, the black line will be at $10^{-3}$.  We have also plotted the
    abundances of $\mr{CH}$ and $\mr{OH}$ species from PDR code (cyan and
    blue dash-dotted lines), in addition to the sum of all $\CHx$ and $\OHx$ species.
    The cosmic-ray ionization rate $\xi_\mr{H}=2\times
    10^{-16}~\mr{s^{-1}H^{-1}}$, and gas temperature is fixed at
    $T=20~\mr{K}$.
\label{fig:species_all_nH50-200}
}
\end{figure*}

\begin{figure*}[htbp]
     \begin{center}
        \subfigure[$n=50~\mr{cm^{-3}}$]{%
            \includegraphics[width=0.97\textwidth]{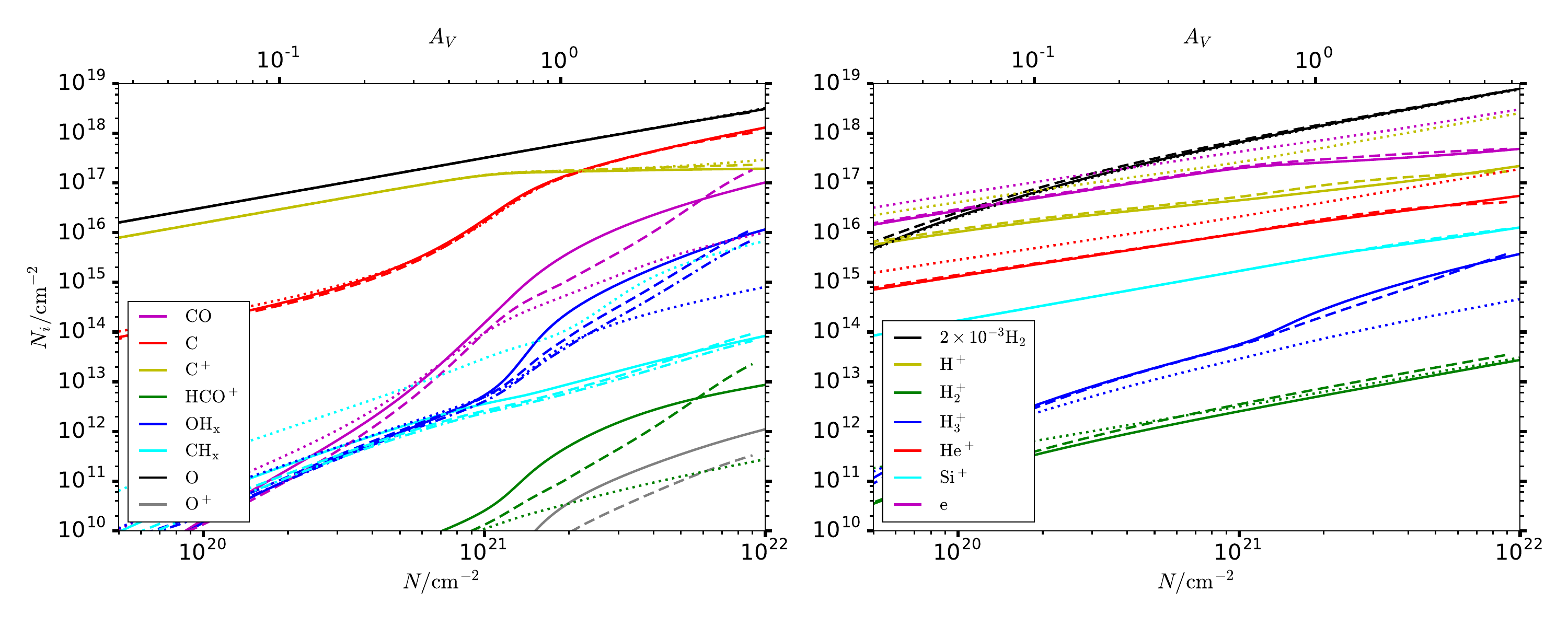}
        }
        \subfigure[$n=100~\mr{cm^{-3}}$]{%
            \includegraphics[width=0.97\textwidth]{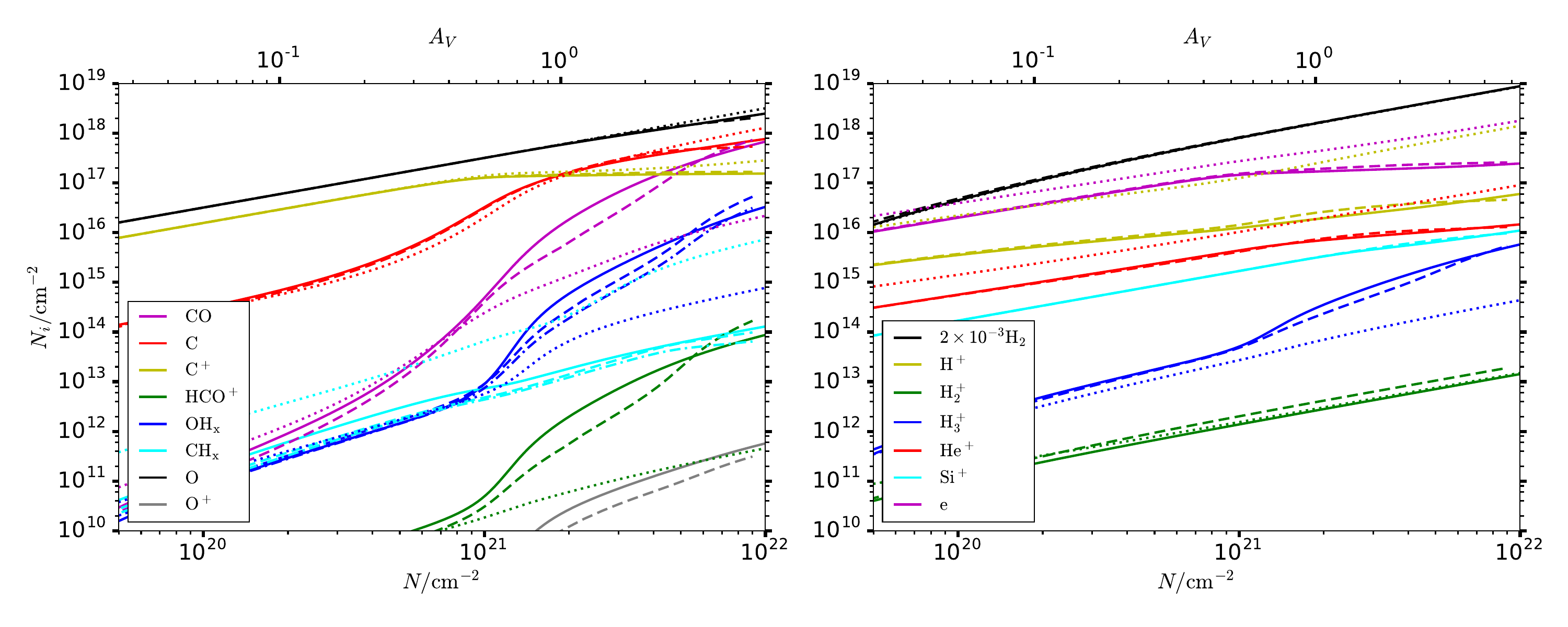}
        }
        \subfigure[$n=200~\mr{cm^{-3}}$]{%
            \includegraphics[width=0.97\textwidth]{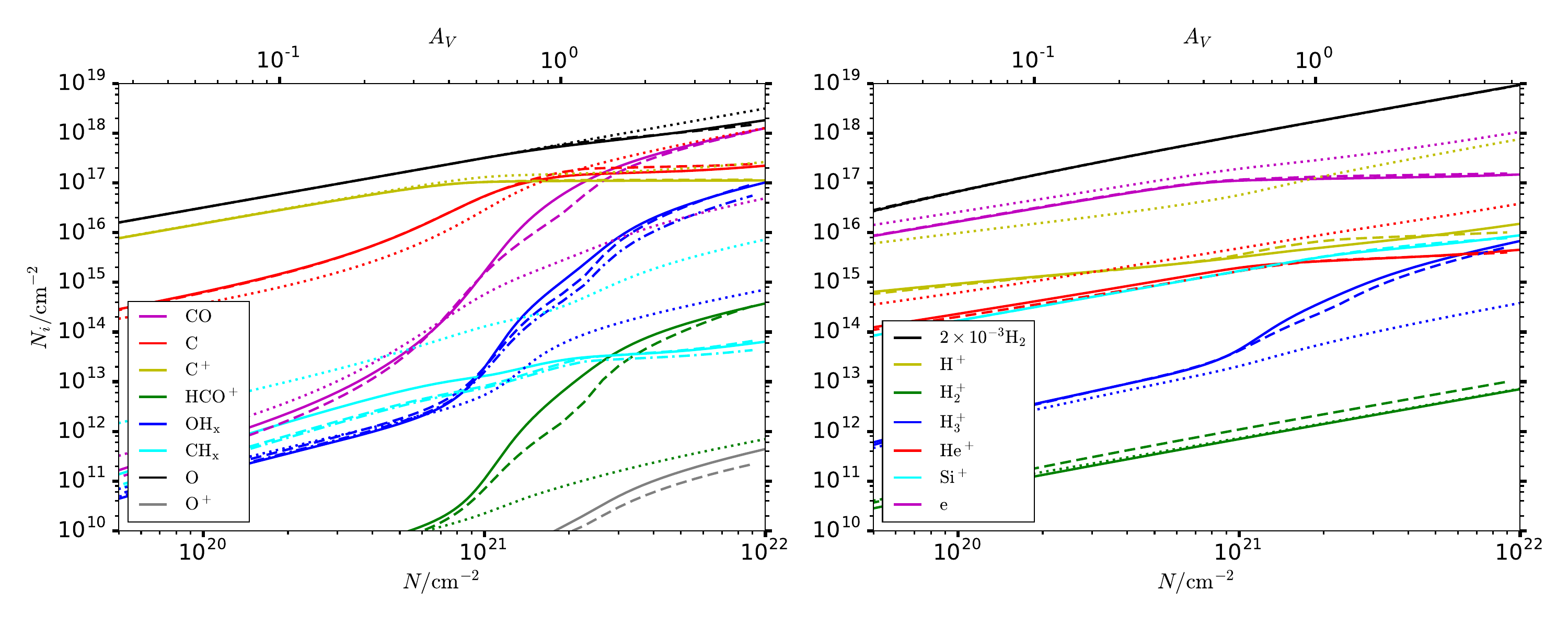}
        }
    \end{center}
    \caption{Integrated column densities of all independent species at
        densities $n=50-200~\mr{cm^{-3}}$ as a function of $A_V$/$N$, 
    similar to Figure \ref{fig:Ni_nH}. Line colors and styles are the
    same as in Figure \ref{fig:species_all_nH50-200}. 
\label{fig:Ni_all_nH50-200}}
\end{figure*}

\begin{figure*}[htbp]
     \begin{center}
        \subfigure[$n=500~\mr{cm^{-3}}$]{%
            \includegraphics[width=0.97\textwidth]{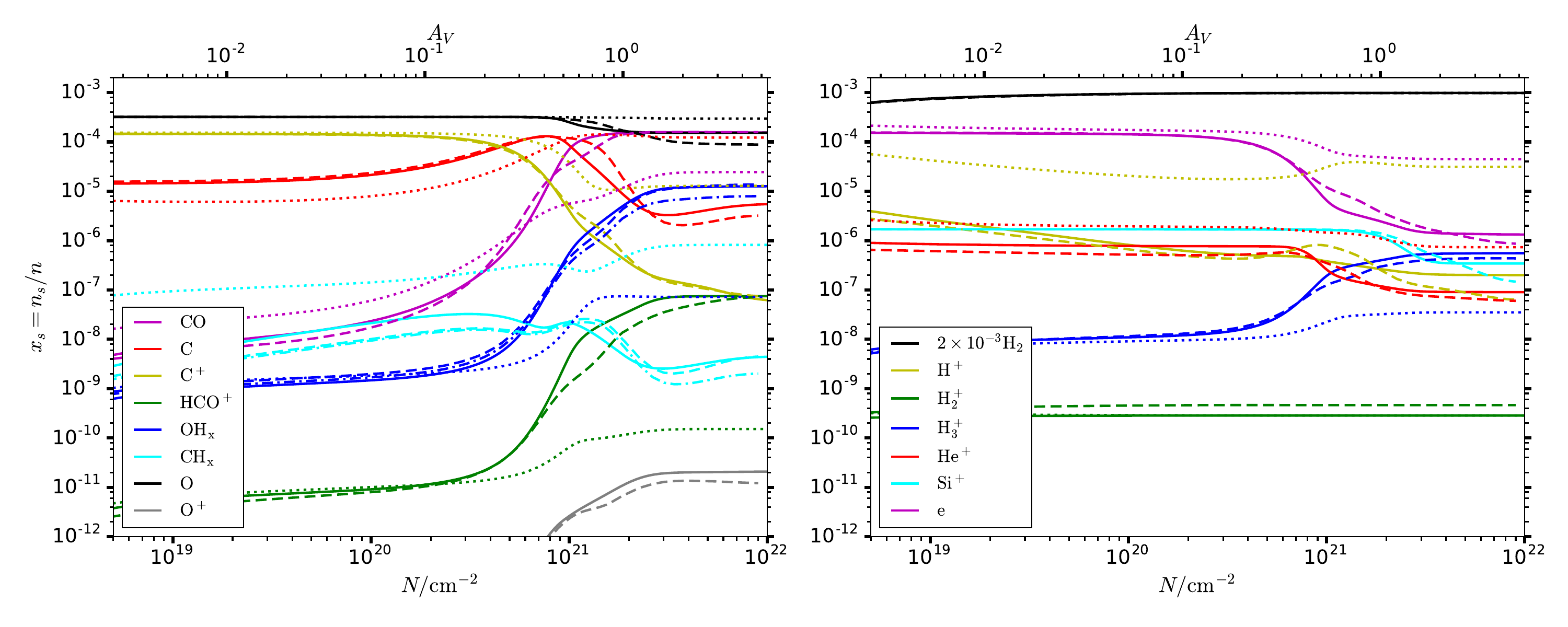}
        }
        \subfigure[$n=700~\mr{cm^{-3}}$]{%
           \includegraphics[width=0.97\textwidth]{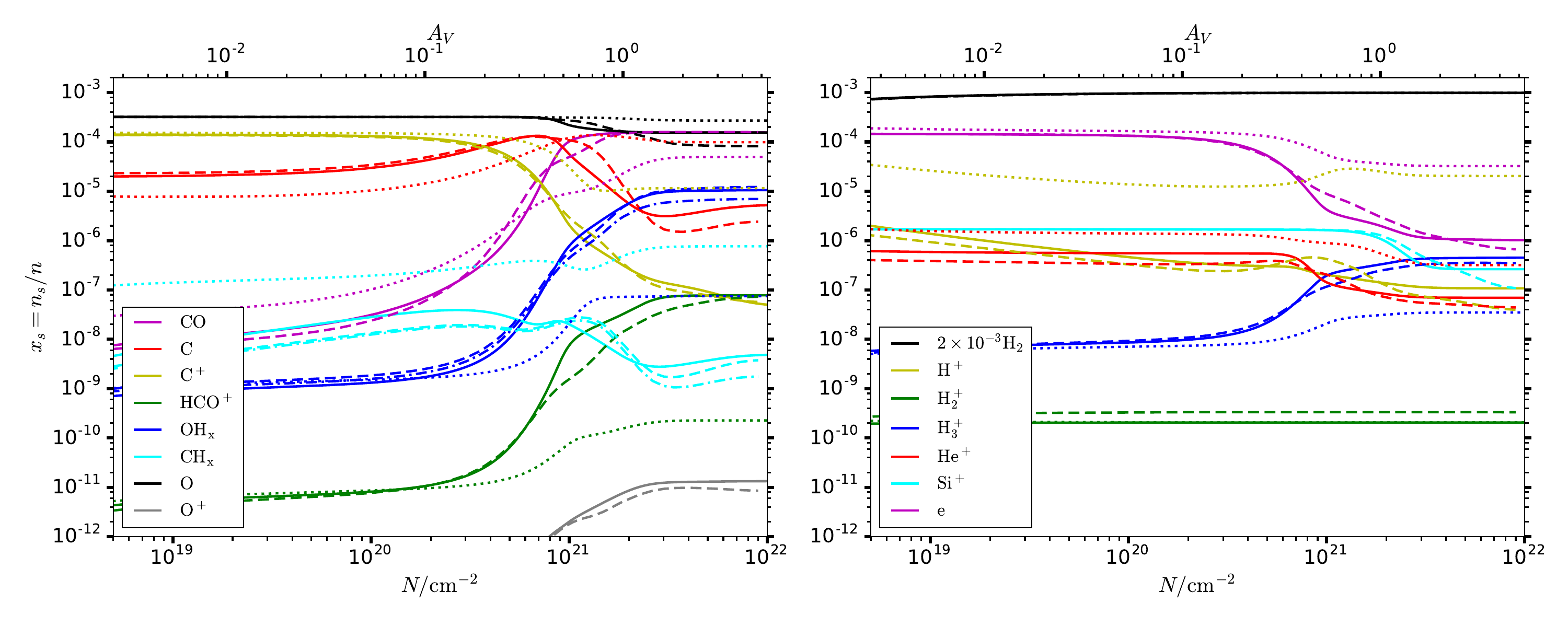}
        }
        \subfigure[$n=1000~\mr{cm^{-3}}$]{%
           \includegraphics[width=0.97\textwidth]{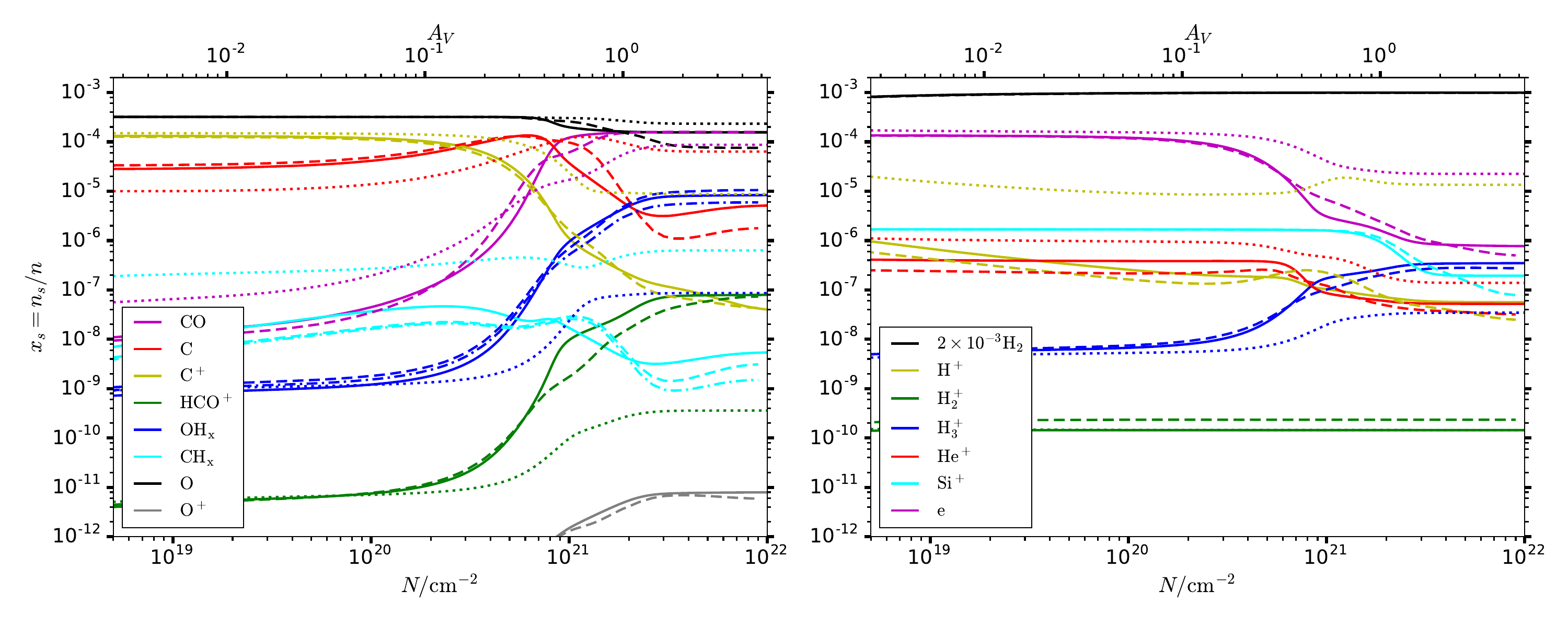}
        }

    \end{center}
    \caption{Abundances of all independent species at
        densities $n=500-1000~\mr{cm^{-3}}$, similar to Figure
        \ref{fig:species_all_nH50-200}.
\label{fig:species_all_nH500-1000}
}
\end{figure*}

\begin{figure*}[htbp]
     \begin{center}
        \subfigure[$n=500~\mr{cm^{-3}}$]{%
            \includegraphics[width=0.97\textwidth]{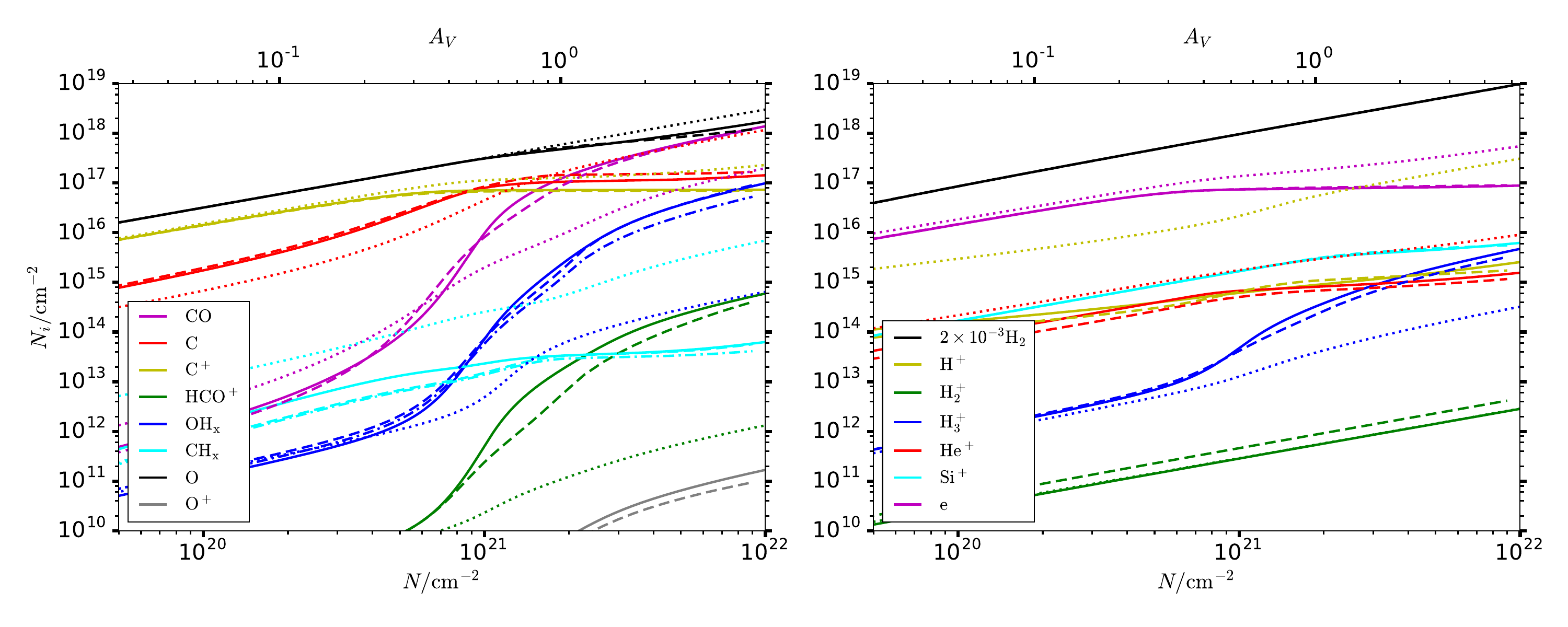}
        }
        \subfigure[$n=700~\mr{cm^{-3}}$]{%
            \includegraphics[width=0.97\textwidth]{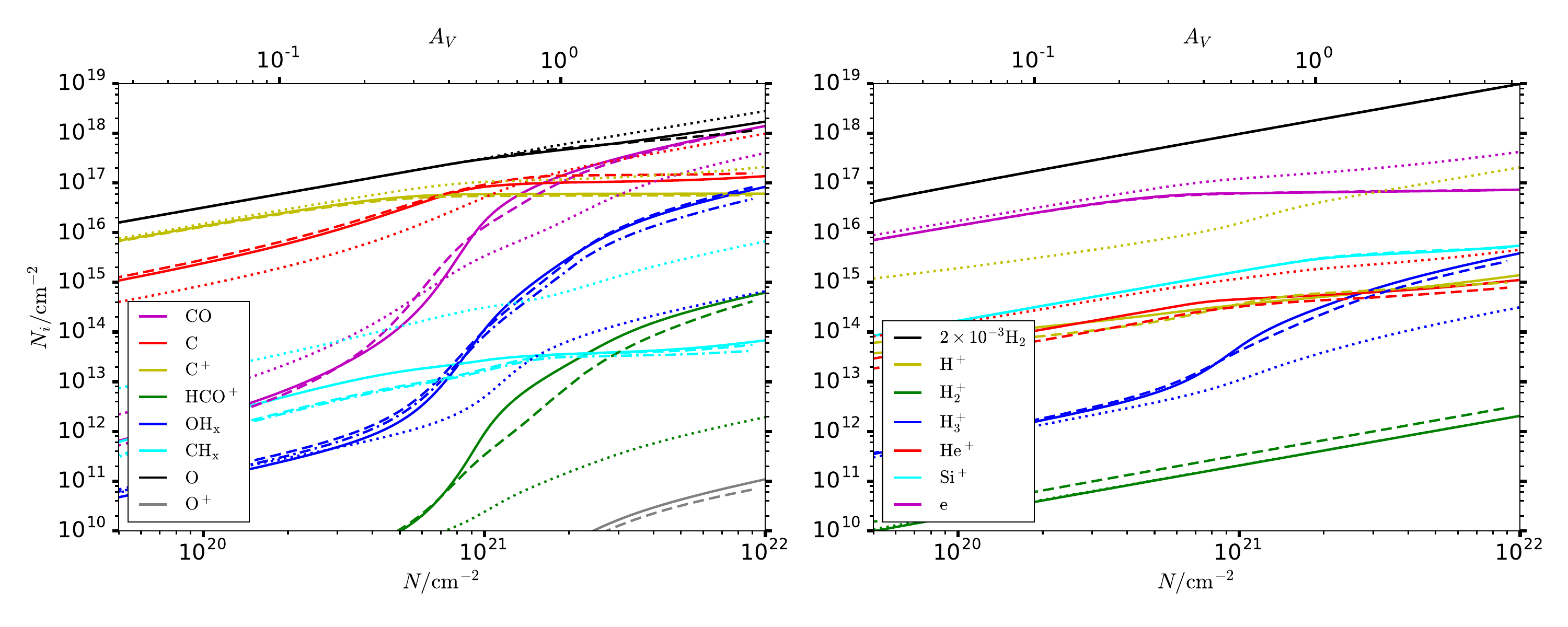}
        }
        \subfigure[$n=1000~\mr{cm^{-3}}$]{%
            \includegraphics[width=0.97\textwidth]{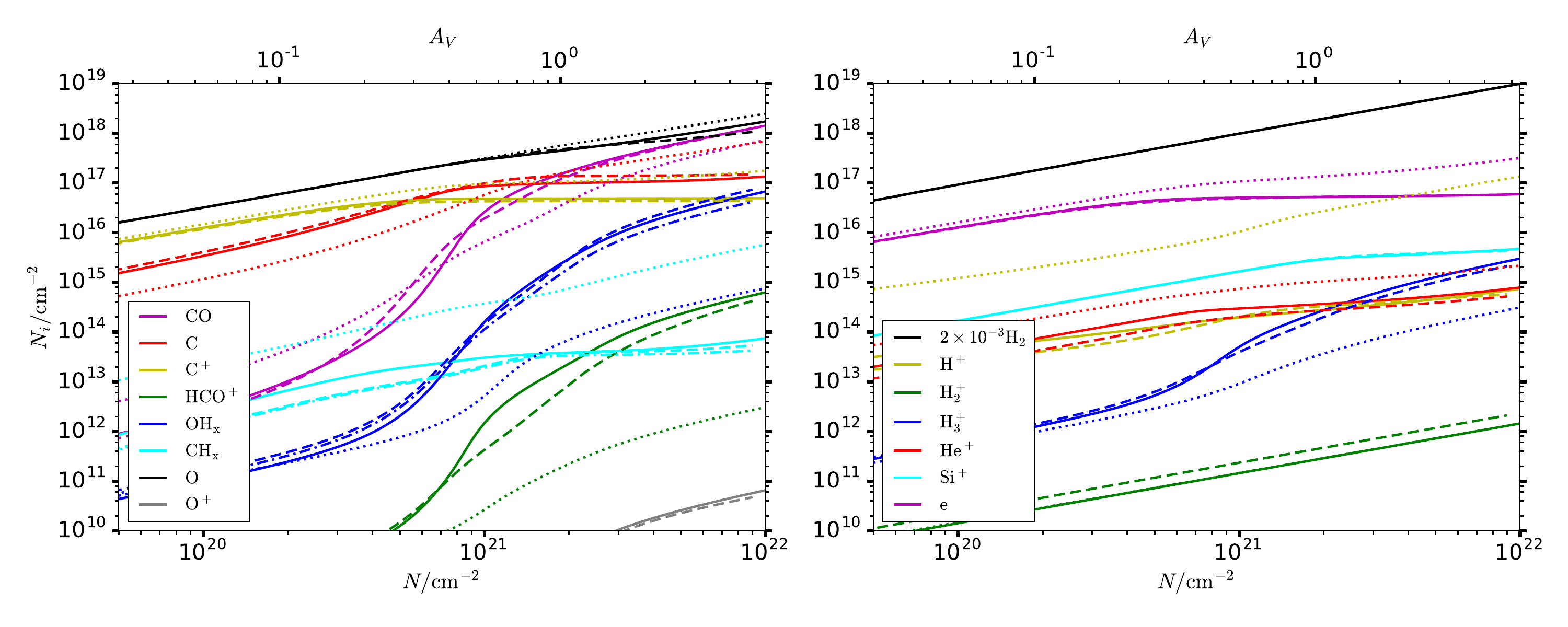}
        }
    \end{center}
    \caption{Integrated column densities of all independent species at
        densities $n=500-1000~\mr{cm^{-3}}$, similar to Figure 
        \ref{fig:Ni_all_nH50-200}.
\label{fig:Ni_all_nH500-1000}}
\end{figure*}

Here we show the comparison with the PDR code, our chemical network, and
the \citetalias{NL1999} network. The chemistry is evolved until it reaches
equilibrium, and we fix the temperature at $T=20\mr{K}$.
All independent species are shown between densities
$n=50~\mr{cm^{-3}}$ and $n=1000~\mr{cm^{-3}}$,
supplementing Figures \ref{fig:species_nH} and \ref{fig:Ni_nH}. Figures 
\ref{fig:species_all_nH50-200} and \ref{fig:species_all_nH500-1000} plot the
abundances,
and Figures \ref{fig:Ni_all_nH50-200} and \ref{fig:Ni_all_nH500-1000}
plot the integrated column densities of different species.

\subsection{Chemistry and Temperature\label{section:TTB}}
\begin{figure*}[htbp]
     \begin{center}
        \subfigure[$n=100~\mr{cm^{-3}}$]{%
            \includegraphics[width=0.97\textwidth]{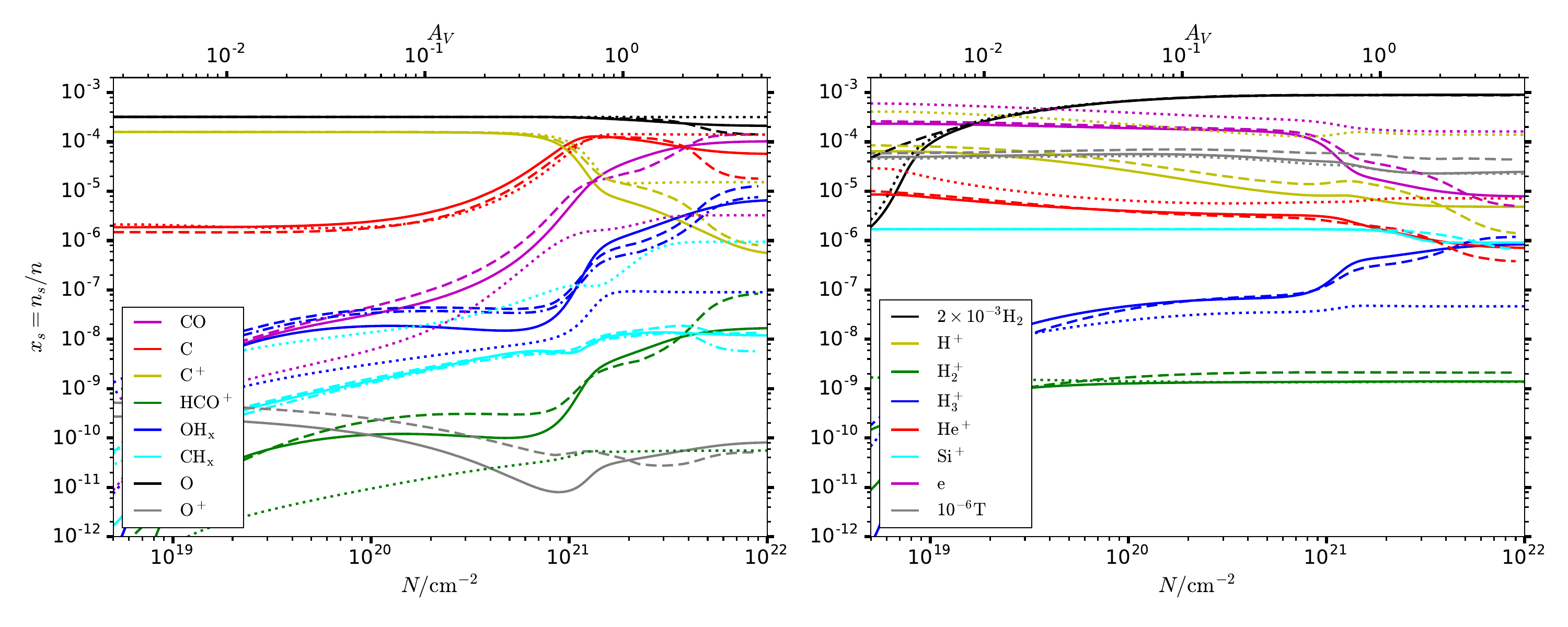}
        }
        \subfigure[$n=1000~\mr{cm^{-3}}$]{%
           \includegraphics[width=0.97\textwidth]{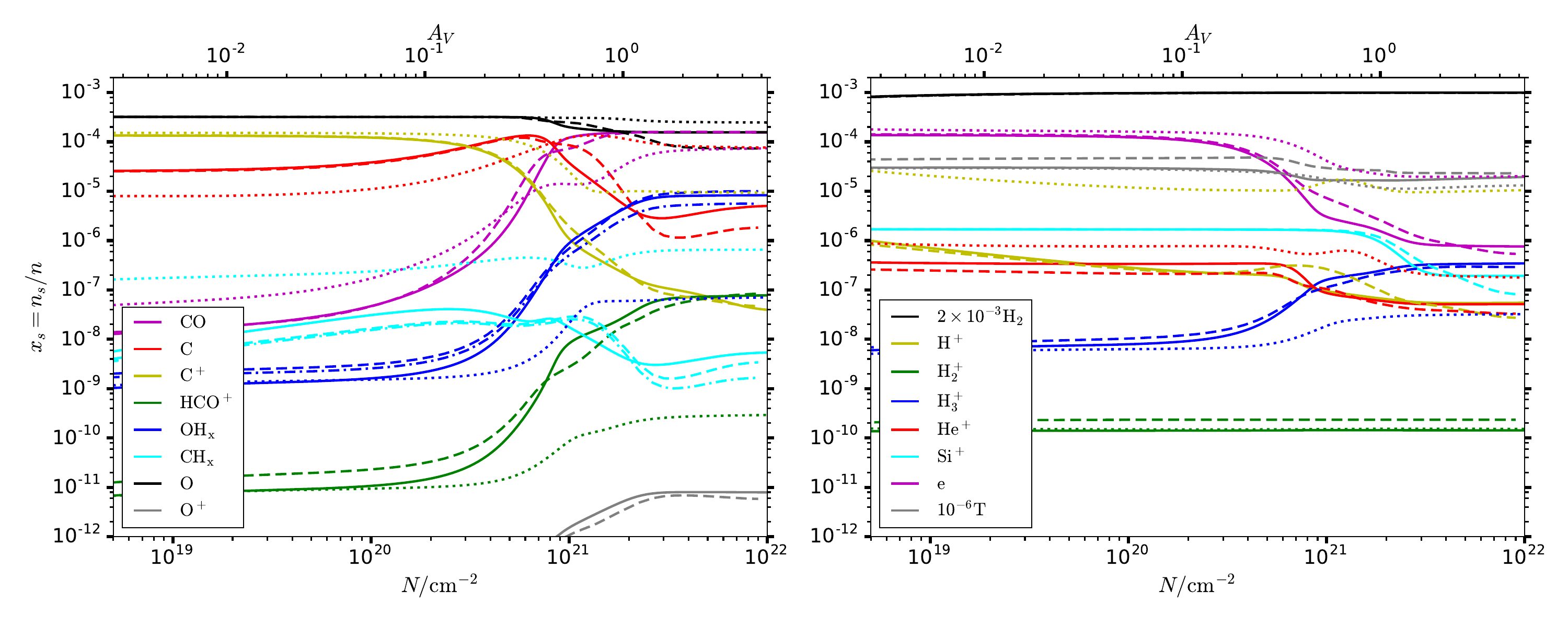}
        }

    \end{center}
    \caption{Abundances of species, similar to Figure
        \ref{fig:species_all_nH50-200} and \ref{fig:species_all_nH500-1000},
        but for temperature in thermal equilibrium.
\label{fig:species_all_TTB}
}
\end{figure*}

\begin{figure*}[htbp]
     \begin{center}
        \subfigure[$n=100~\mr{cm^{-3}}$]{%
            \includegraphics[width=0.97\textwidth]{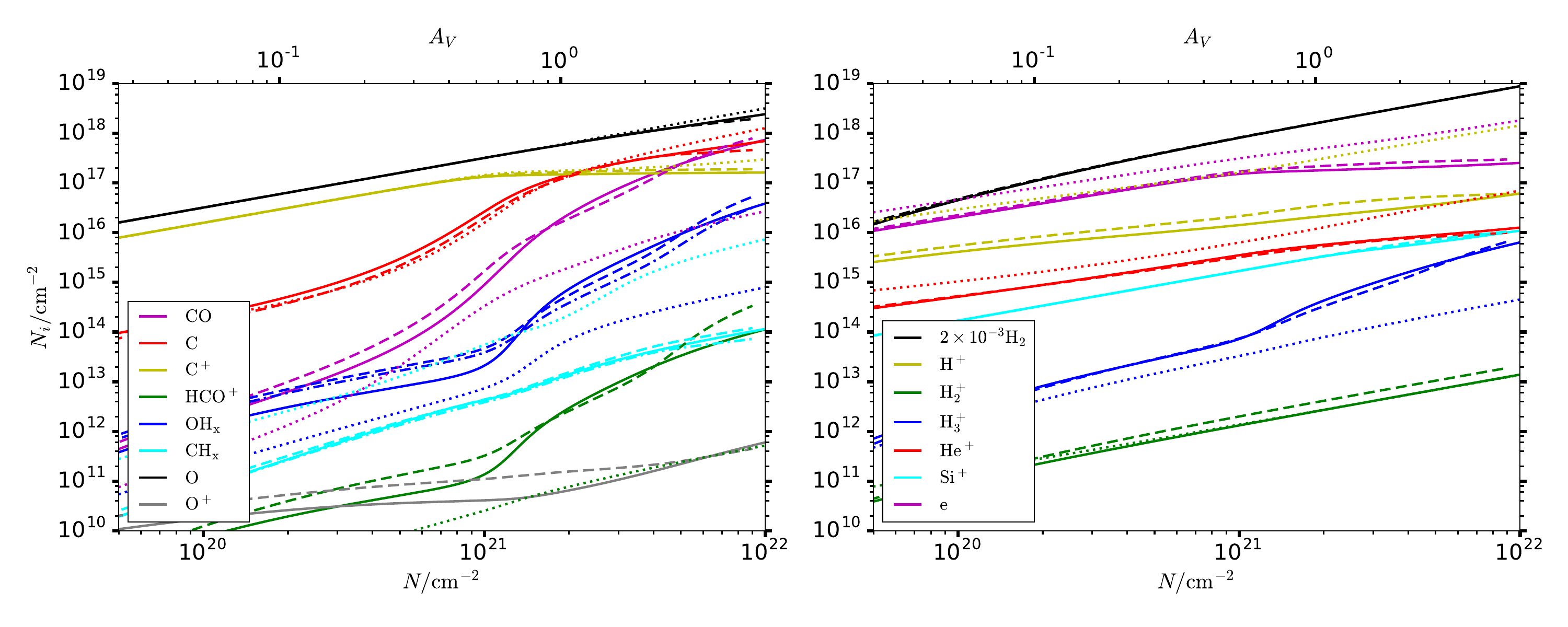}
        }
        \subfigure[$n=1000~\mr{cm^{-3}}$]{%
            \includegraphics[width=0.97\textwidth]{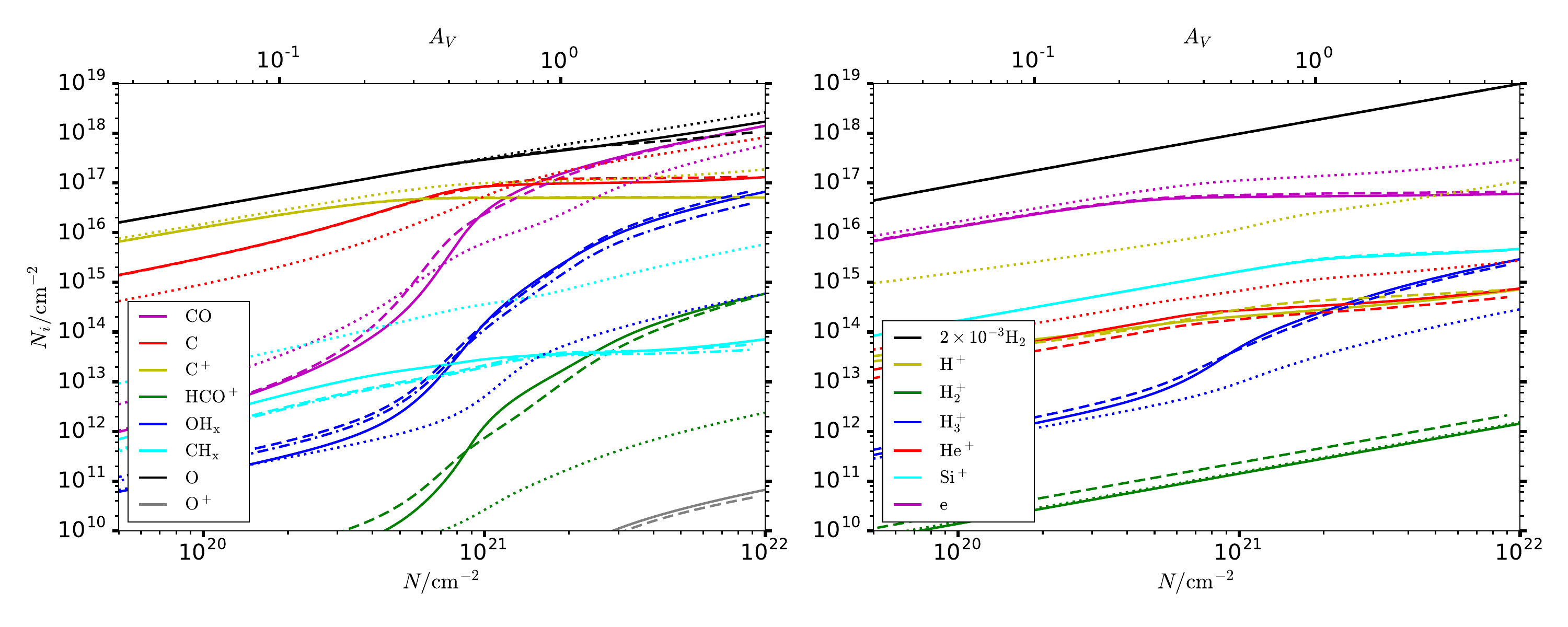}
        }

    \end{center}
    \caption{Integrated column densities of species in Figure 
        \ref{fig:species_all_TTB}.
\label{fig:Ni_all_TTB}}
\end{figure*}

We carry out comparisons similar to Appendix \ref{section:T20}, but here we 
solve the chemistry and temperature simultaneously, until they both reach
equilibrium. The results are shown in Figures \ref{fig:species_all_TTB} and
\ref{fig:Ni_all_TTB}. Again, our network agrees well with the PDR code, whereas
the \citetalias{NL1999} network shows significant disagreement. However, 
the equilibrium temperature in the \citetalias{NL1999} network is very similar to
that in our network. For example, at $n = 100~\mr{cm^{-3}}$ and $A_V>1$,
although there is much less $\CO$ in the \citetalias{NL1999} network, $\CI$
provides most cooling, resulting in similar equilibrium temperatures. As noted
by \citet{GC2012}, temperature is insensitive to detailed chemistry, 
although shielding is important because it sets the photoelectric heating rate.

\subsection{Comparison at Low Metallicity\label{section:Z0p1}}
\begin{figure*}[htbp]
     \begin{center}
        \subfigure[$n=100~\mr{cm^{-3}}$]{%
            \includegraphics[width=0.97\textwidth]{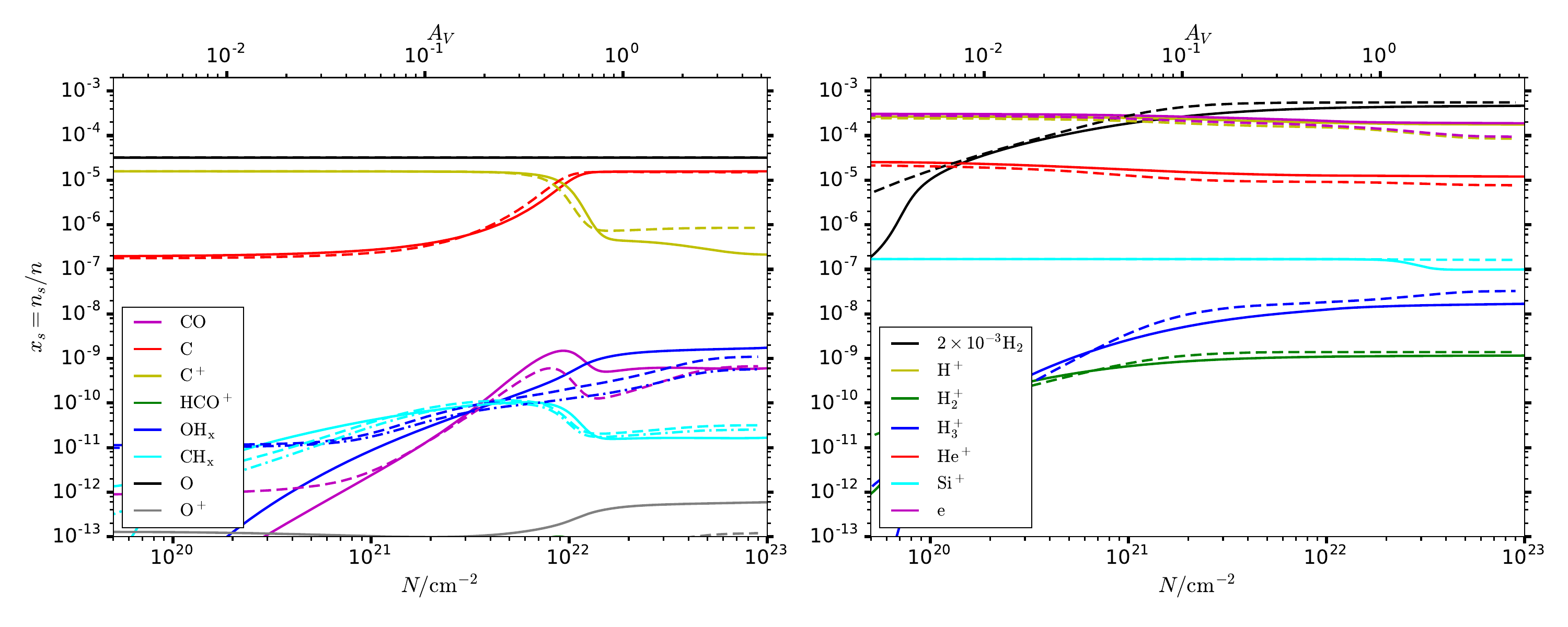}
        }
        \subfigure[$n=1000~\mr{cm^{-3}}$]{%
           \includegraphics[width=0.97\textwidth]{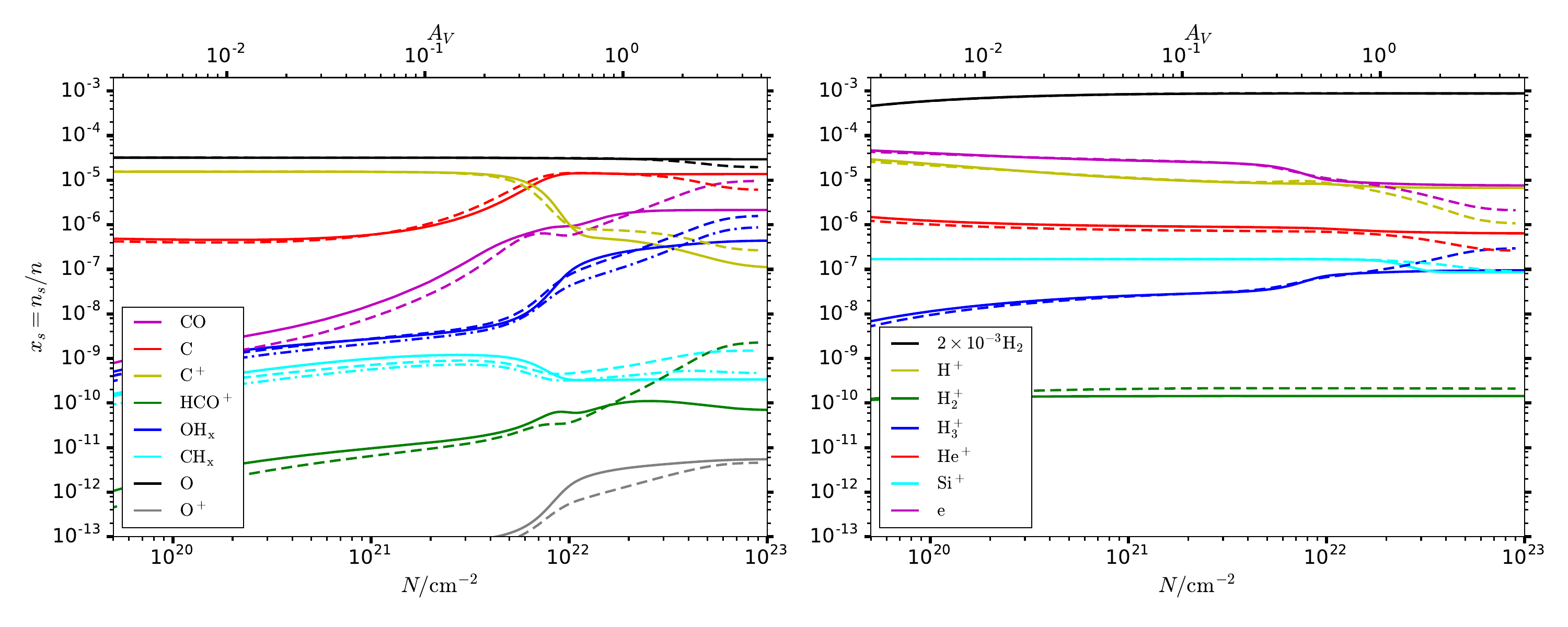}
        }
        \subfigure[$n=5000~\mr{cm^{-3}}$]{%
           \includegraphics[width=0.97\textwidth]{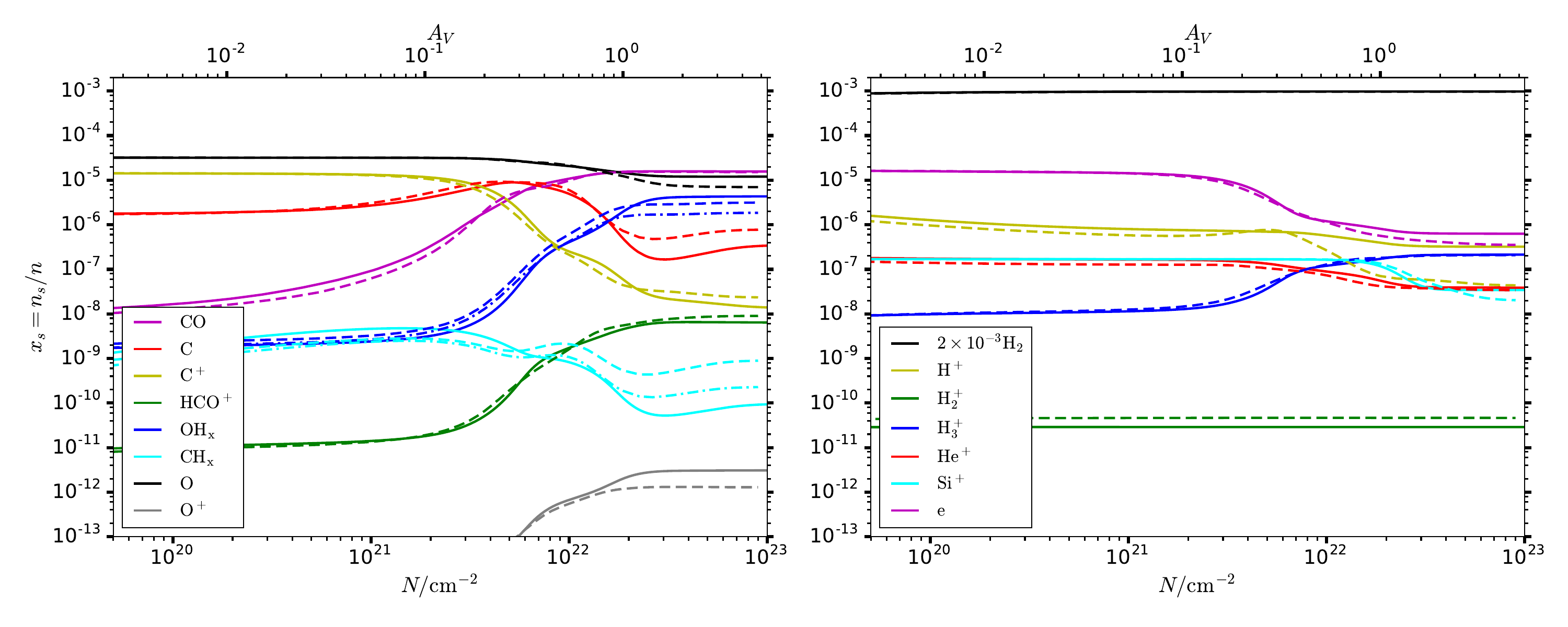}
        }

    \end{center}
    \caption{Abundances of species, similar to Figures 
        \ref{fig:species_all_nH50-200} and \ref{fig:species_all_nH500-1000},
        but for $Z=0.1$. Note that the \citetalias{NL1999} network is not shown
        here.
    \label{fig:species_all_Z0p1}
}
\end{figure*}

\begin{figure*}[htbp]
     \begin{center}
        \subfigure[$n=100~\mr{cm^{-3}}$]{%
            \includegraphics[width=0.97\textwidth]{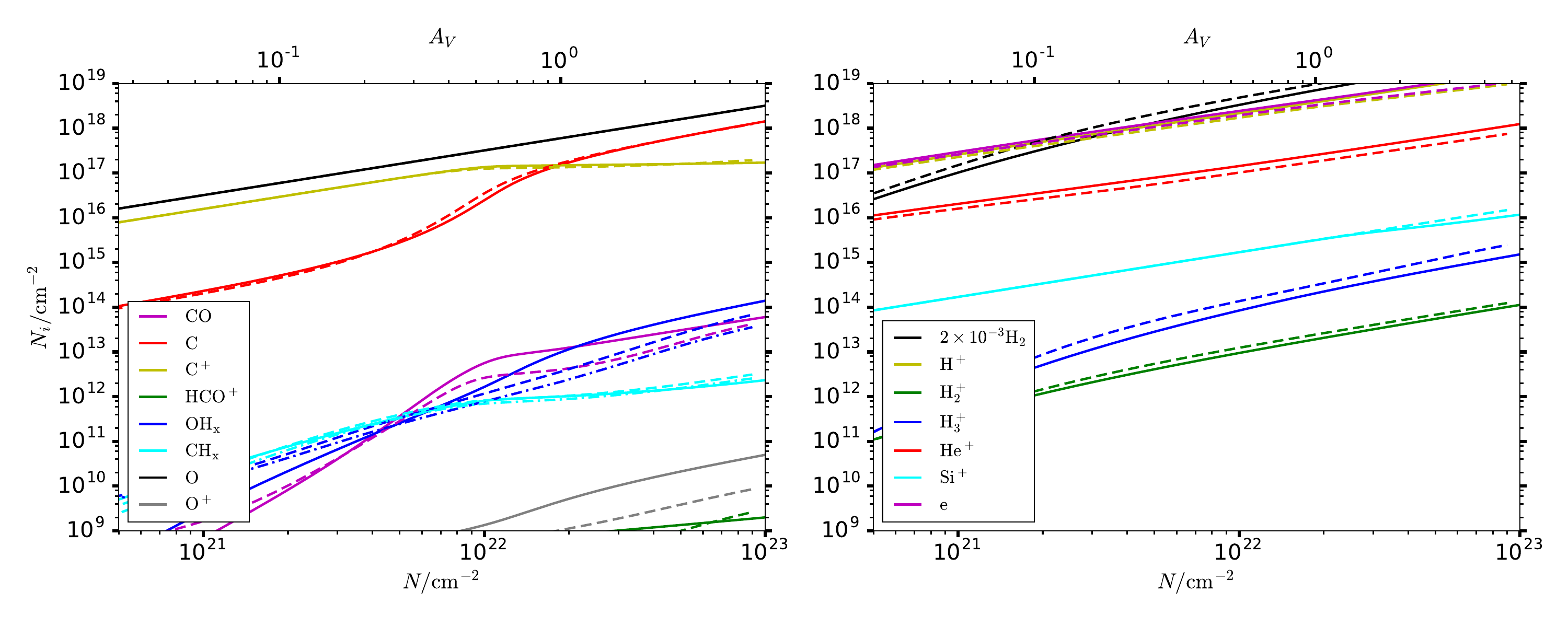}
        }
        \subfigure[$n=1000~\mr{cm^{-3}}$]{%
            \includegraphics[width=0.97\textwidth]{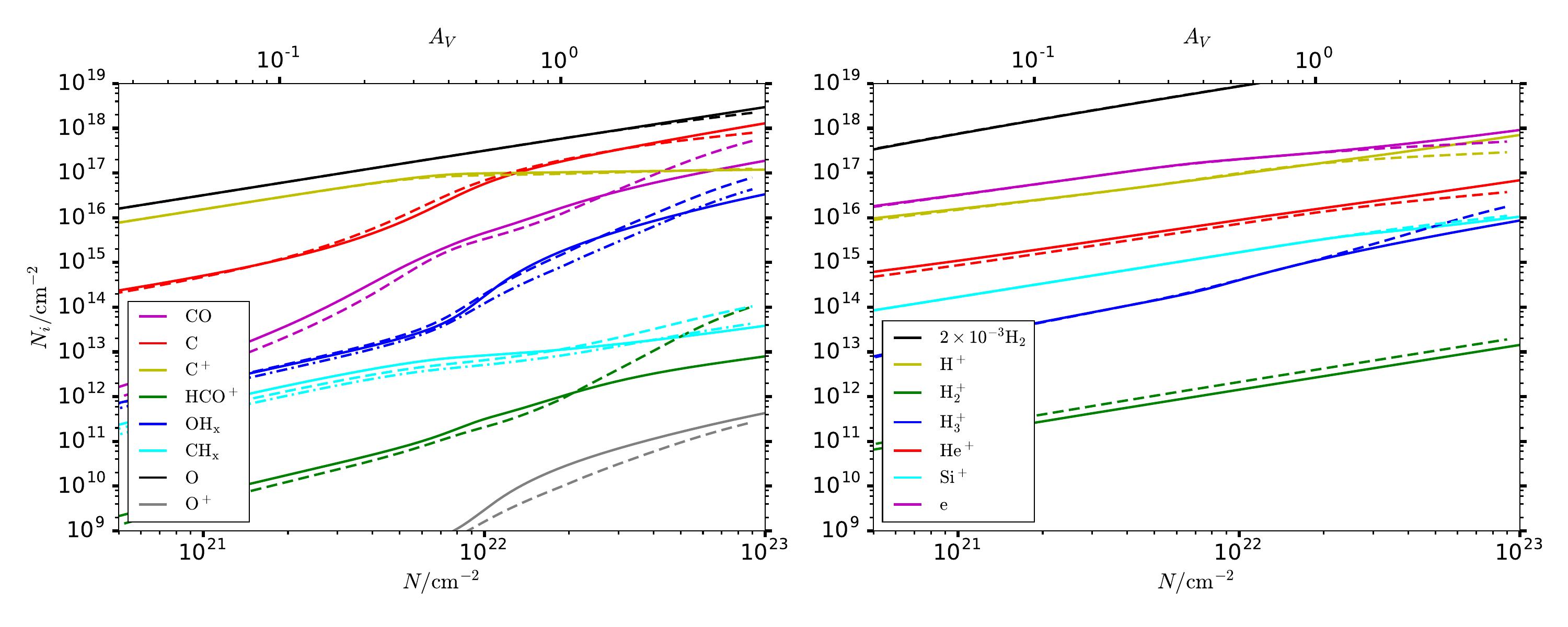}
        }
        \subfigure[$n=5000~\mr{cm^{-3}}$]{%
            \includegraphics[width=0.97\textwidth]{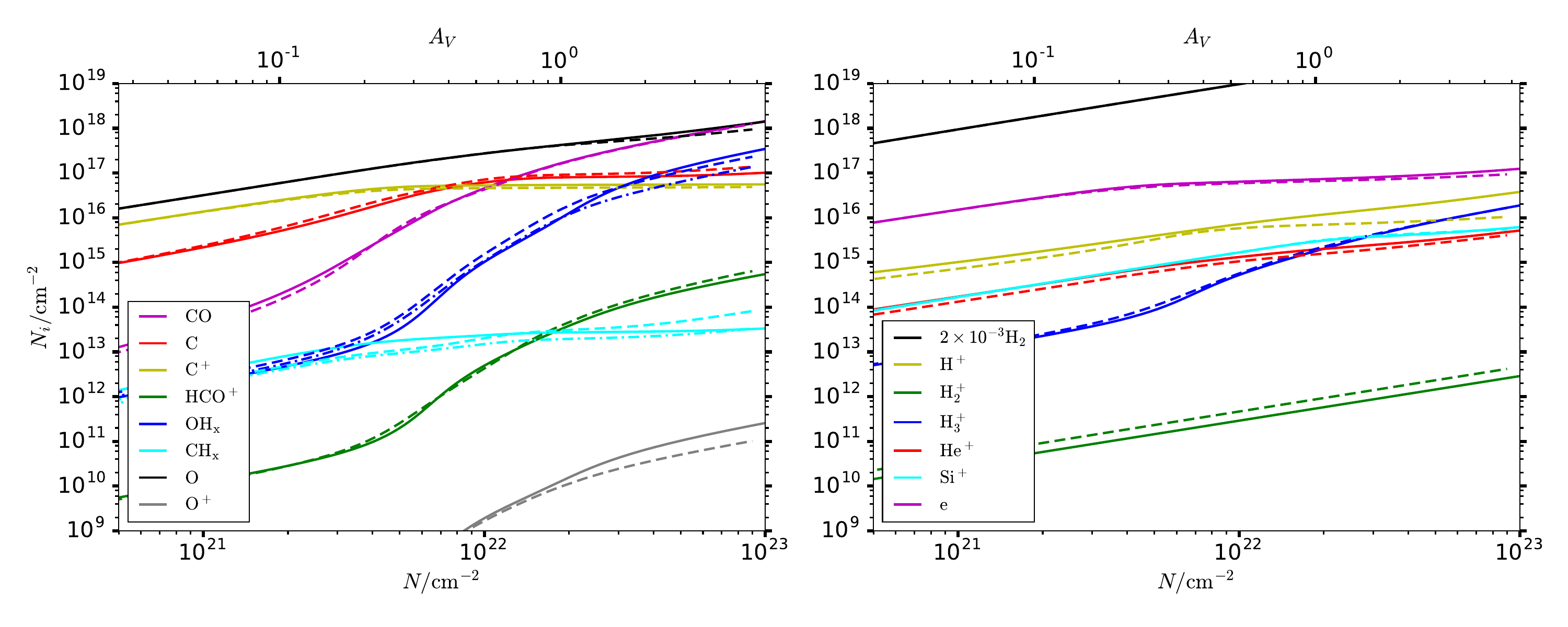}
        }

    \end{center}
    \caption{ Integrated column densities of species in Figure \ref{fig:species_all_Z0p1}.
    \label{fig:Ni_all_Z0p1}}
\end{figure*}

Figures \ref{fig:species_all_Z0p1} and \ref{fig:Ni_all_Z0p1} plot comparisons
at $Z=0.1$ between our network and the PDR code. Overall, there is a good
agreement. Comparing to $Z=1$ cases, $\CO$ forms at a much higher density 
$n\gtrsim 1000~\mr{cm^{-3}}$.
We note that at $n=1000~\mr{cm^{-3}}$, our network 
overproduces $\CO$ in shielded regions. This is partly due to the difference
in grain-assisted recombination rates, which is discussed in Appendix
\ref{section:GR1}.

\subsection{Grain-assisted Recombination Rates\label{section:GR1}}
Ions such $\Hplus$, $\CII$, $\Heplus$, and $\Siplus$ can recombine when they
collide with PAHs.
The recombination rate on dust grains is often higher than direct recombination with
electrons, and therefore can have a major effect on chemistry. We use the
results of \citet{WD2001b} for grain-assisted recombination rates in our
network (see Table \ref{table:chem2}). In \citet{WD2001b}, they assume a
certain size distribution of PAHs and calculate the charge distribution
on PAHs. The final grain-assisted recombination rates are obtain by averaging
over the PAH population. The PDR code uses a slightly different approach in
\citet{Wolfire2008}. The PAHs are assumed to have a single size, and three
charge states, $\mr{PAH^+}$, $\mr{PAH^-}$, and $\mr{PAH^0}$. 
By comparing with observations, \citet{Wolfire2008} calibrate their grain-assisted 
recombination rate of $\CII$ with the parameter $\phi_\mr{PAH}=0.4$.

\begin{figure*}[htbp]
     \begin{center}
         \includegraphics[width=0.5\textwidth]{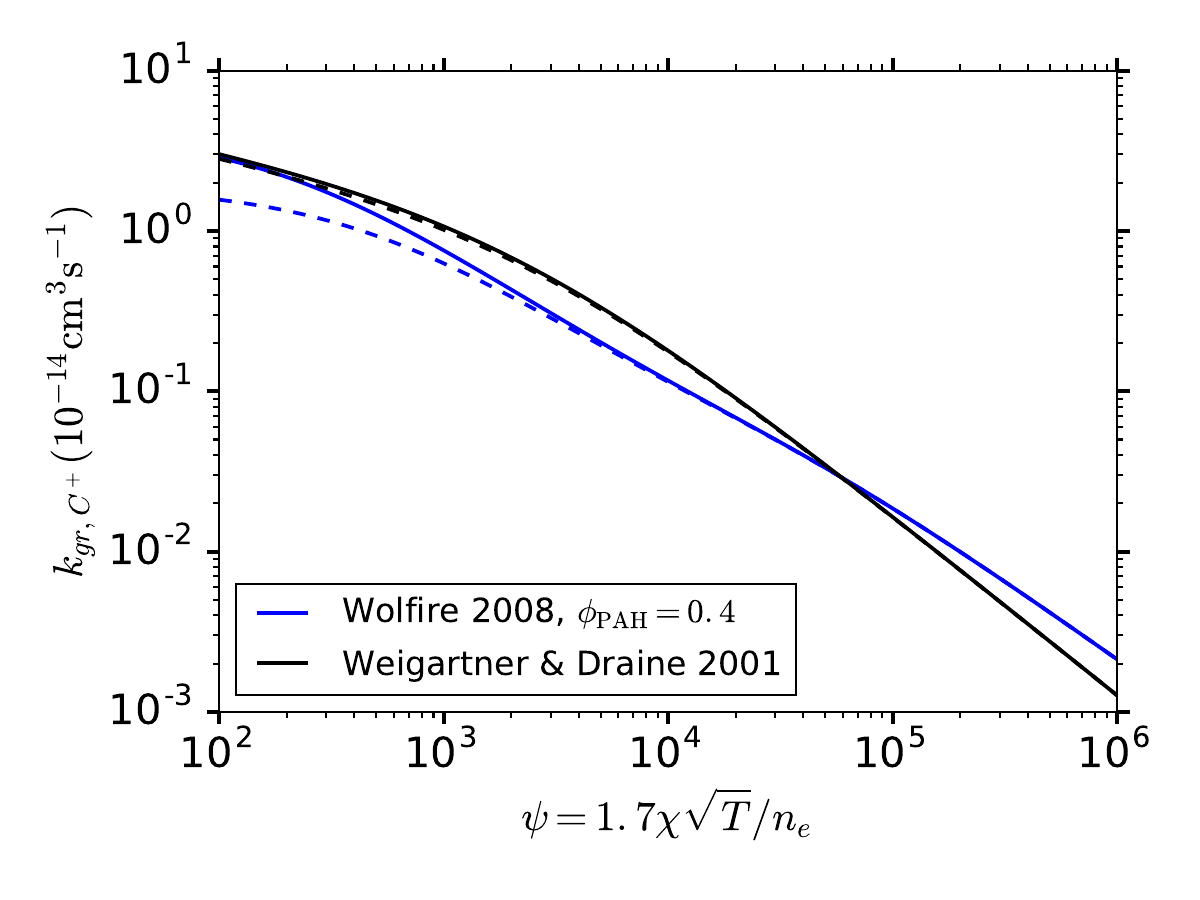}
    \end{center}
    \caption{Comparison of the grain-assisted 
        recombination rate of $\CII$ between \citet{Wolfire2008} (blue) 
        and \citet{WD2001b} (black). The solid and dashed lines are results
        with $T=20~\mr{K}$ and $T=100~\mr{K}$. The temperature $T$ is in Kelvin, and
        the electron density $n_\mr{e}$ is in $\mr{cm^{-3}}$ in $\psi$.
\label{fig:Cplus_gr_rate}
}
\end{figure*}

\begin{figure*}[htbp]
     \begin{center}
        \subfigure[$n=100~\mr{cm^{-3}}$]{%
            \includegraphics[width=0.97\textwidth]{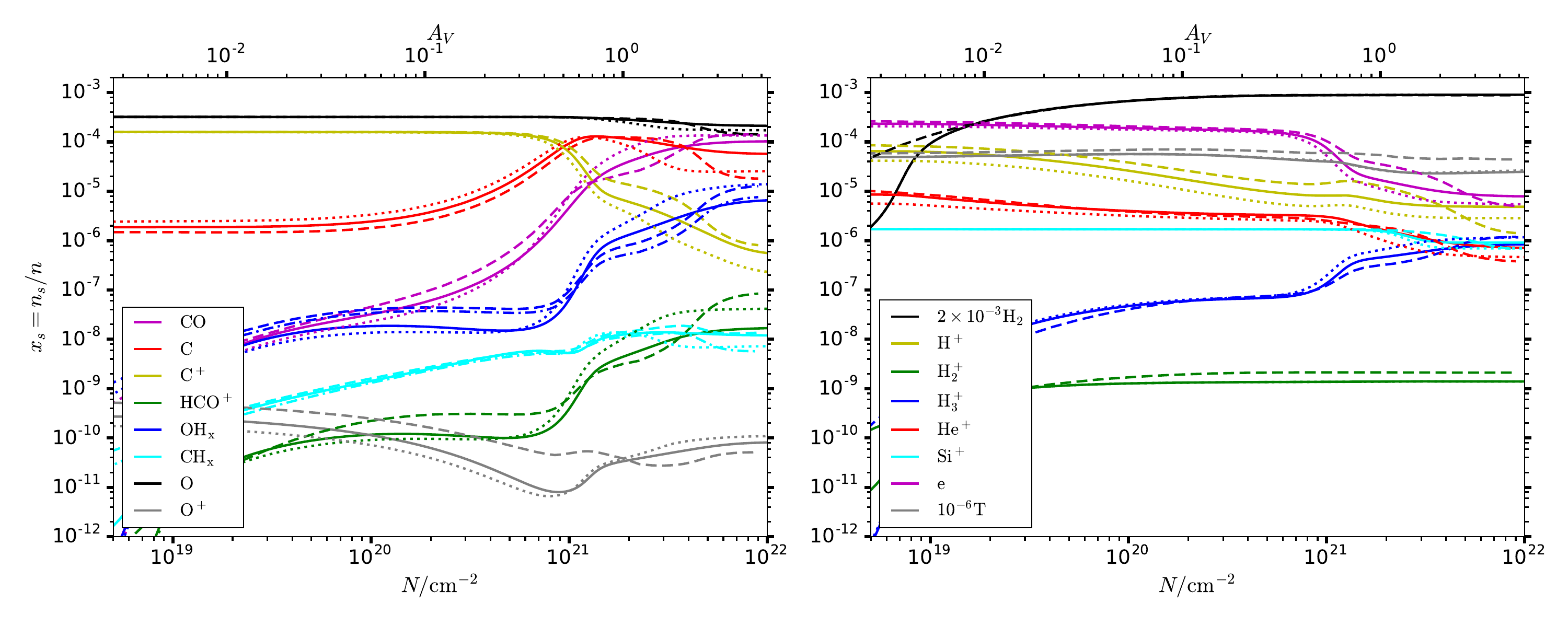}
        }
        \subfigure[$n=1000~\mr{cm^{-3}}$]{%
           \includegraphics[width=0.97\textwidth]{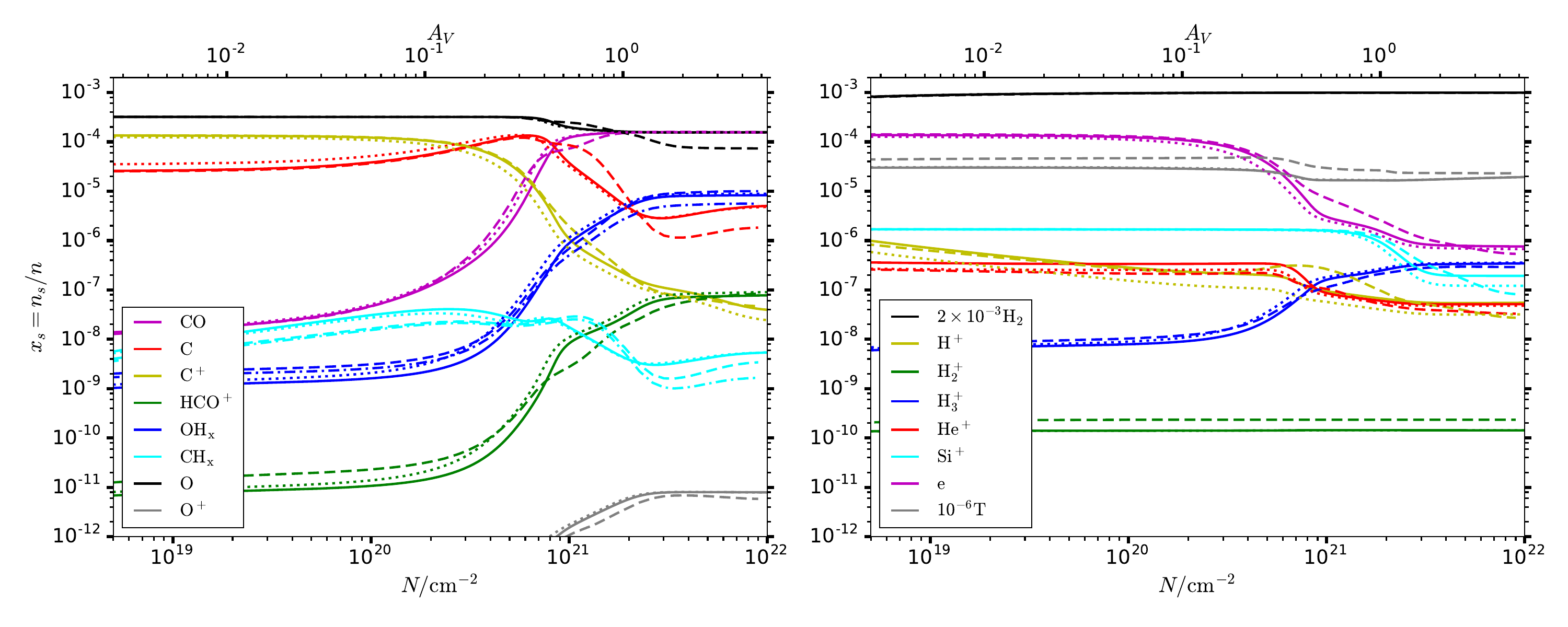}
        }
    \end{center}
    \caption{Comparison of chemical abundances in our network with 0.6 times
        the grain-assisted recombination rates in \citet{WD2001b} (solid), the
        original \citet{WD2001b} rates (dotted), and the PDR code (dashed).
        The temperature is in thermal equilibrium.
        \label{fig:species_all_GR1}
}
\end{figure*}

\begin{figure*}[htbp]
     \begin{center}
        \subfigure[$n=100~\mr{cm^{-3}}$]{%
            \includegraphics[width=0.97\textwidth]{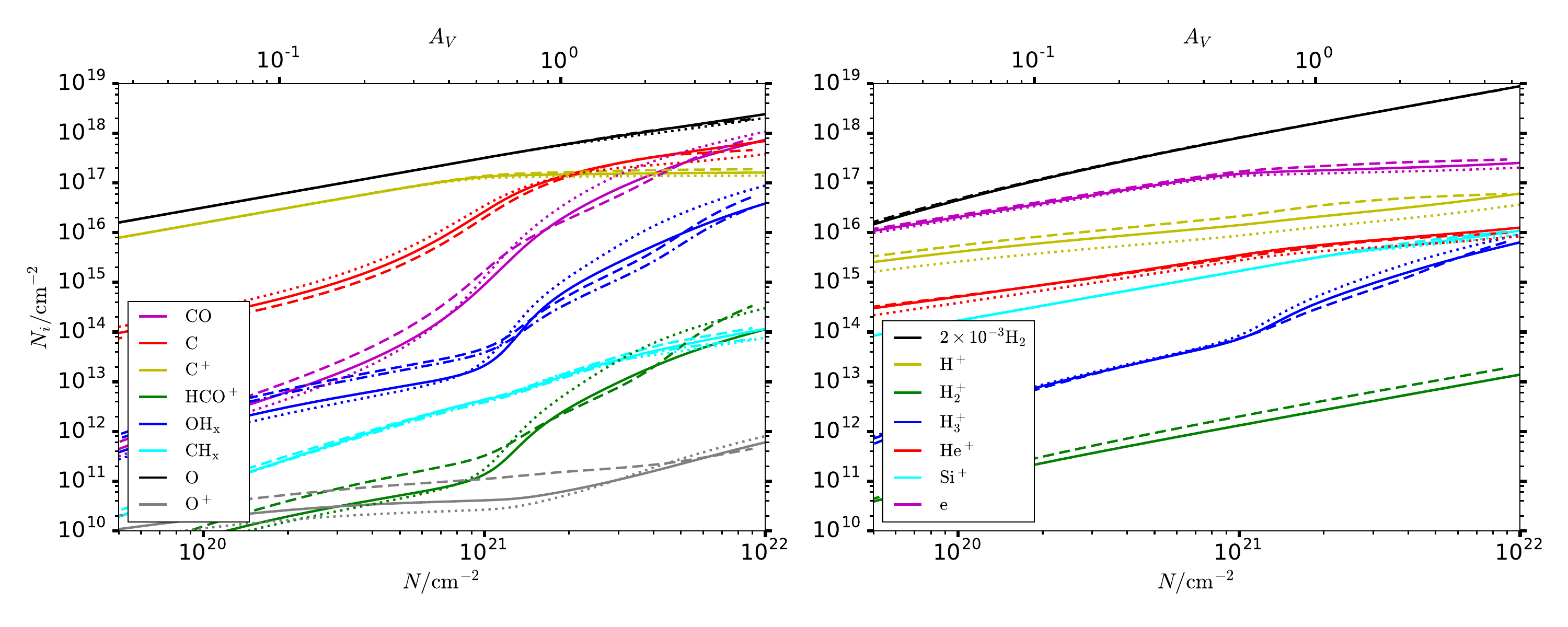}
        }
        \subfigure[$n=1000~\mr{cm^{-3}}$]{%
            \includegraphics[width=0.97\textwidth]{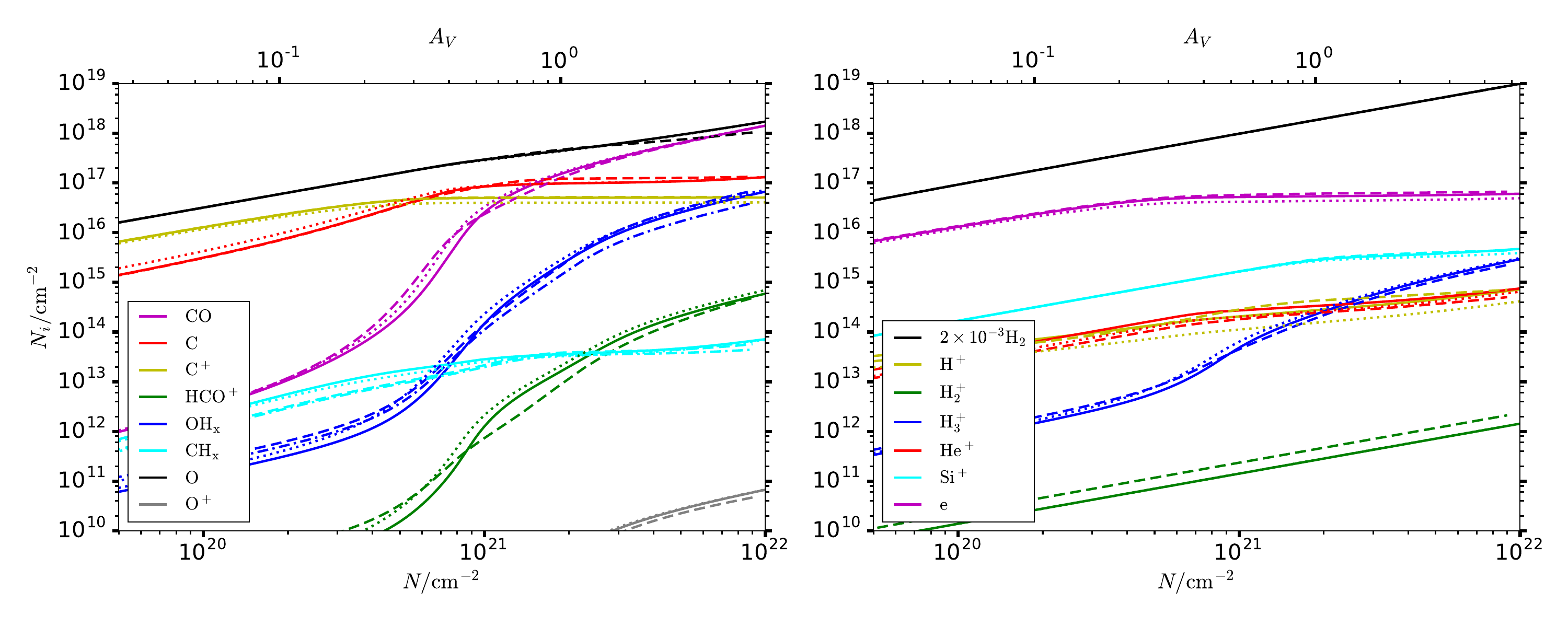}
        }

    \end{center}
    \caption{Integrated column densities of species in Figure 
        \ref{fig:species_all_GR1}.
\label{fig:Ni_all_GR1}}
\end{figure*}

Figure \ref{fig:Cplus_gr_rate} shows the comparison of the grain-assisted 
recombination rate of $\CII$ between \citet{Wolfire2008} and \citet{WD2001b}. 
The PAH
populations in \citet{Wolfire2008} are assumed to be in charge equilibrium
states
with reactions described in Appendix C2 of \citet{Wolfire2003}, and the total
PAH abundance is $n_\mr{PAH}/n=2\times 10^{-7}$. The reaction
rate depends mainly on the parameter $\psi = 1.7\chi \sqrt{T}/n_e$, and only has a
very weak dependence on temperature. As can be seen in Figure
\ref{fig:Cplus_gr_rate}, the \citet{Wolfire2008} and \citet{WD2001b} rates
can differ by a factor of $\sim 2$. In the main part of this paper, we use 0.6
times the \citet{WD2001b} rates to match with \citet{Wolfire2008} at
$\psi \sim 10^4$. Figure \ref{fig:species_all_GR1} and \ref{fig:Ni_all_GR1}
show comparisons with the original \citet{WD2001b} rates. At
$n=100~\mr{cm^{-3}}$, The abundances of
$\CO$, $\OHx$, $\HCOplus$, and $\Heplus$ can change by a factor of $\sim 5$
around $A_V\sim 1$, but the chemical abundances at higher density 
$n=1000~\mr{cm^{-3}}$ are largely unaffected.

\subsection{Grain Surface Reactions\label{section:SR}}
\begin{figure*}[htbp]
     \begin{center}
        \subfigure[$n=100~\mr{cm^{-3}}$]{%
            \includegraphics[width=0.97\textwidth]{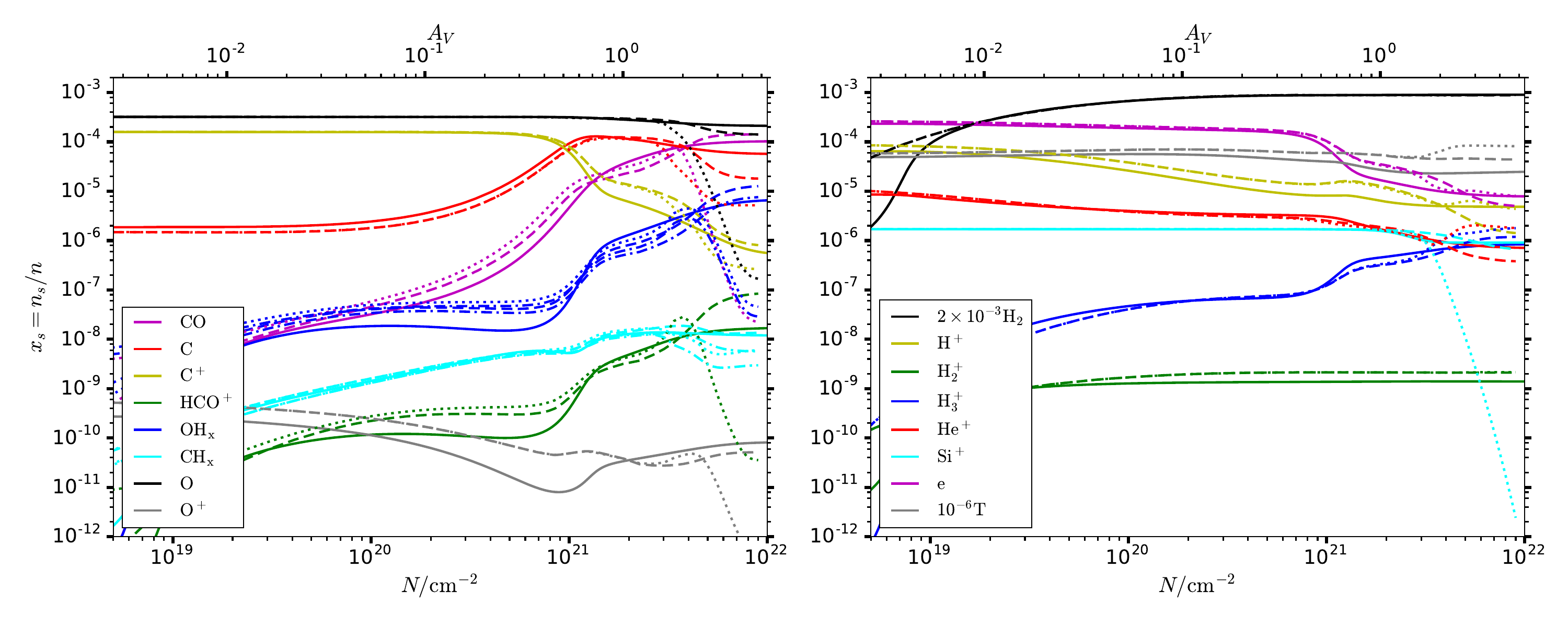}
        }
        \subfigure[$n=1000~\mr{cm^{-3}}$]{%
           \includegraphics[width=0.97\textwidth]{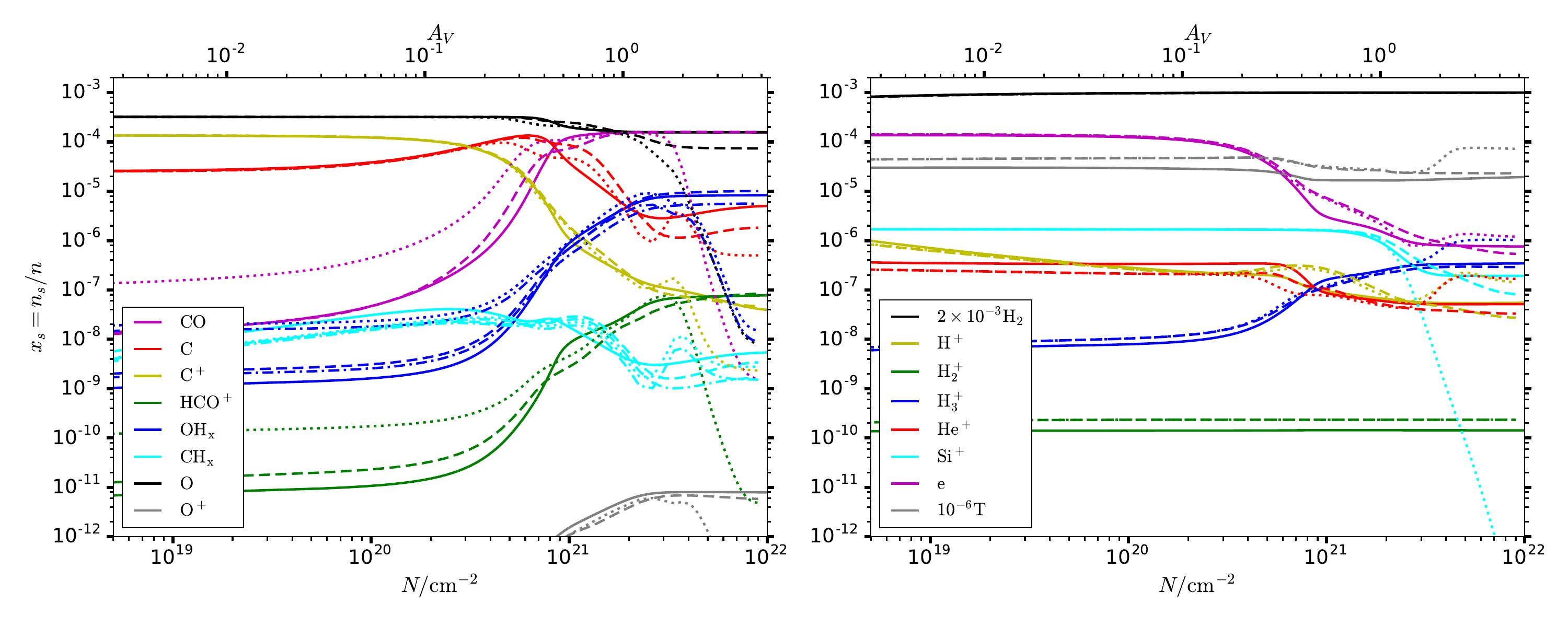}
        }

    \end{center}
    \caption{Comparison of chemical abundances in our network (solid), 
        the PDR code (dashed), and the PDR code with grain surface reactions
        in \citet{Hollenbach2009} (dotted).
        The temperature is in thermal equilibrium.  
\label{fig:species_all_SR}
}
\end{figure*}

\begin{figure*}[htbp]
     \begin{center}
        \subfigure[$n=100~\mr{cm^{-3}}$]{%
            \includegraphics[width=0.97\textwidth]{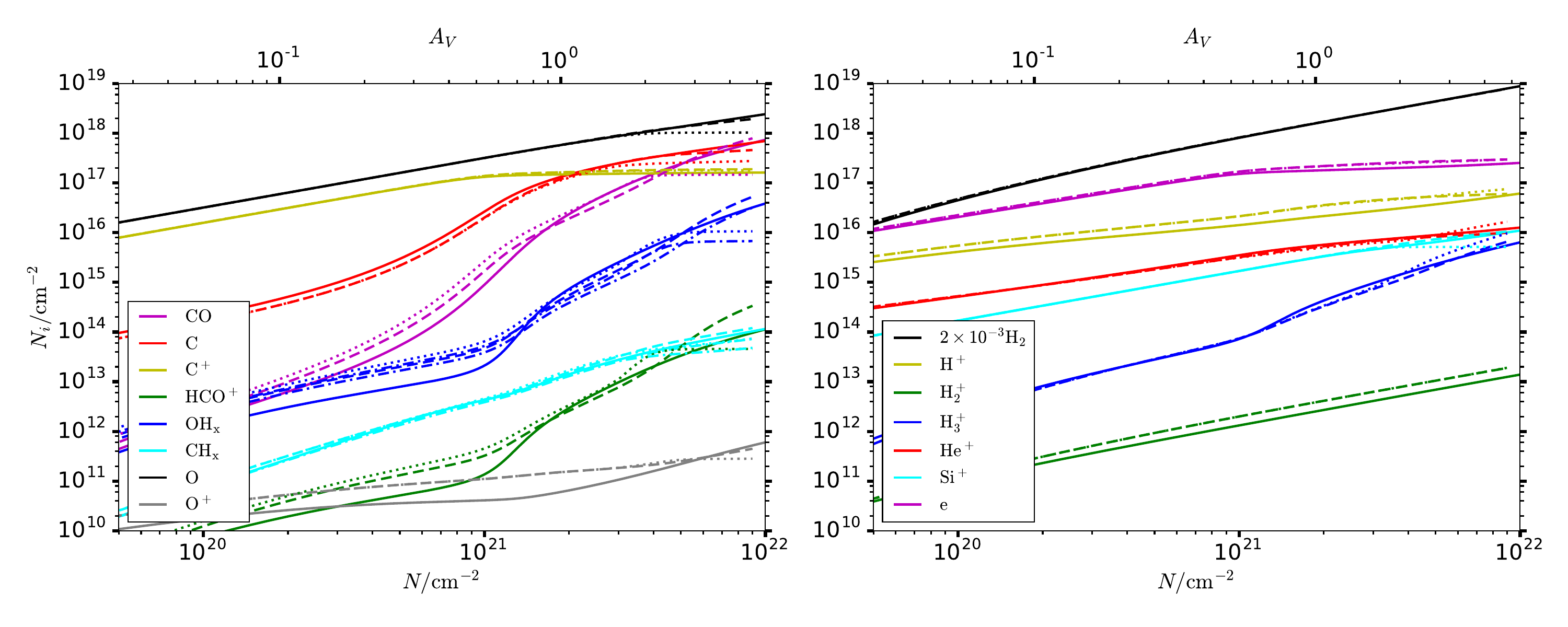}
        }
        \subfigure[$n=1000~\mr{cm^{-3}}$]{%
            \includegraphics[width=0.97\textwidth]{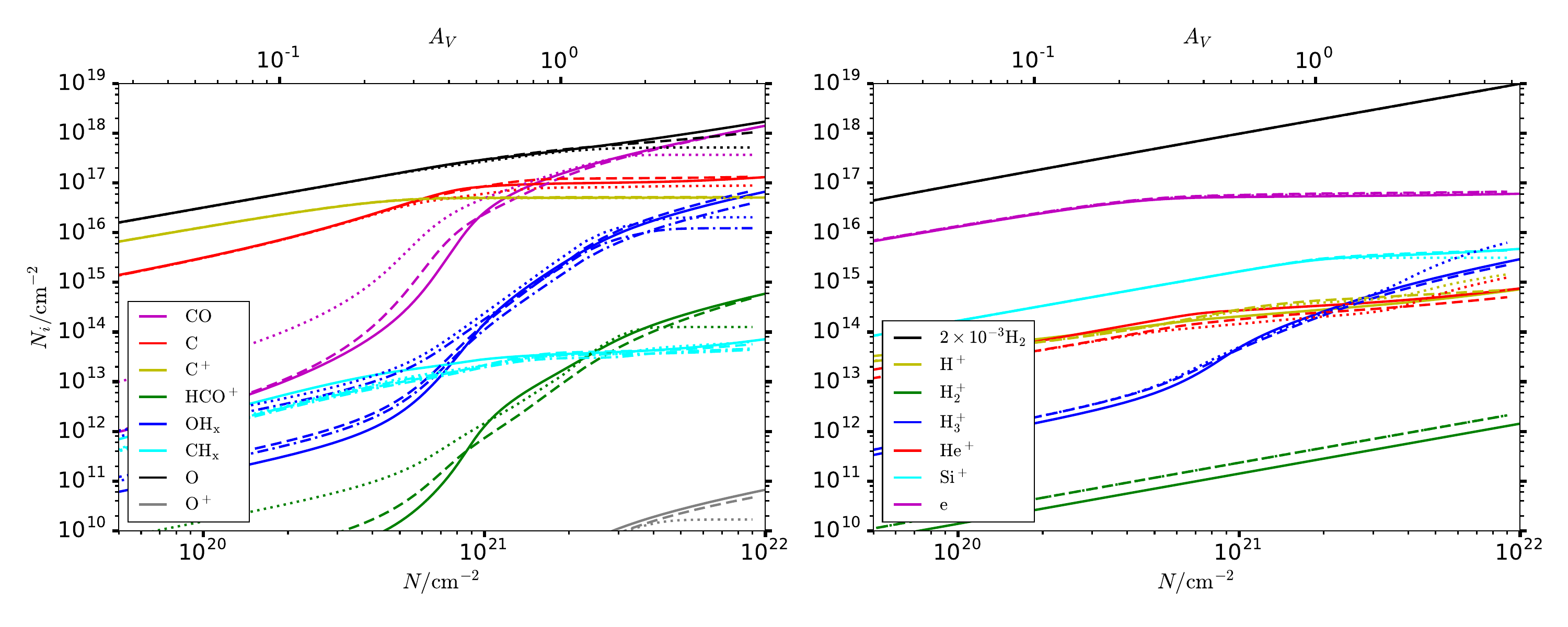}
        }

    \end{center}
    \caption{ Integrated column densities of species in Figure 
        \ref{fig:species_all_SR}.
\label{fig:Ni_all_SR}}
\end{figure*}

In addition to the grain-assisted recombinations, dust grains can also affect
the gas-phase chemistry by grain surface reactions \citep{Hollenbach2009}.
This includes the formation of $\mr{OH}$ and $\mr{H_2O}$ on
the surface of dust grains, and the freeze-out of molecules such as $\CO$ and 
$\mr{H_2O}$ on dust grains in cold and shielded regions. Figures
\ref{fig:species_all_SR} and \ref{fig:Ni_all_SR} show the comparisons of 
the chemical abundances with and without the
grain surface reactions. The formation of $\mr{OH}$ and $\mr{H_2O}$ on grains
results in a higher $\OHx$ and $\CO$ abundances at $n=1000~\mr{cm^{-3}}$
and $A_V < 1$. The freeze-out can greatly reduce the abundances of
$\CO$ and other species at $A_V\gtrsim 2$. Since the $\CO$ abundance is still
low in regions that are affected by the formation of $\mr{OH}$ and $\mr{H_2O}$ on
grains, and the freeze-out depends on the grain size distribution, which is very
uncertain in dense and shielded clouds, we do not include grain surface
reaction in our network. 
Ideally, the effect of freeze-out on carbon bearing species should
be calibrated with observations of diffuse and dense clouds as was done
with oxygen-bearing species in \citet{Hollenbach2009, Hollenbach2012} and
\citet{Sonnentrucker2015}.

\subsection{Comparison with the {\sl PyPDR} code \label{section:PyPDR} }
\begin{figure*}[htbp]
     \begin{center}
        \subfigure[Abundances]{%
            \includegraphics[width=0.97\textwidth]{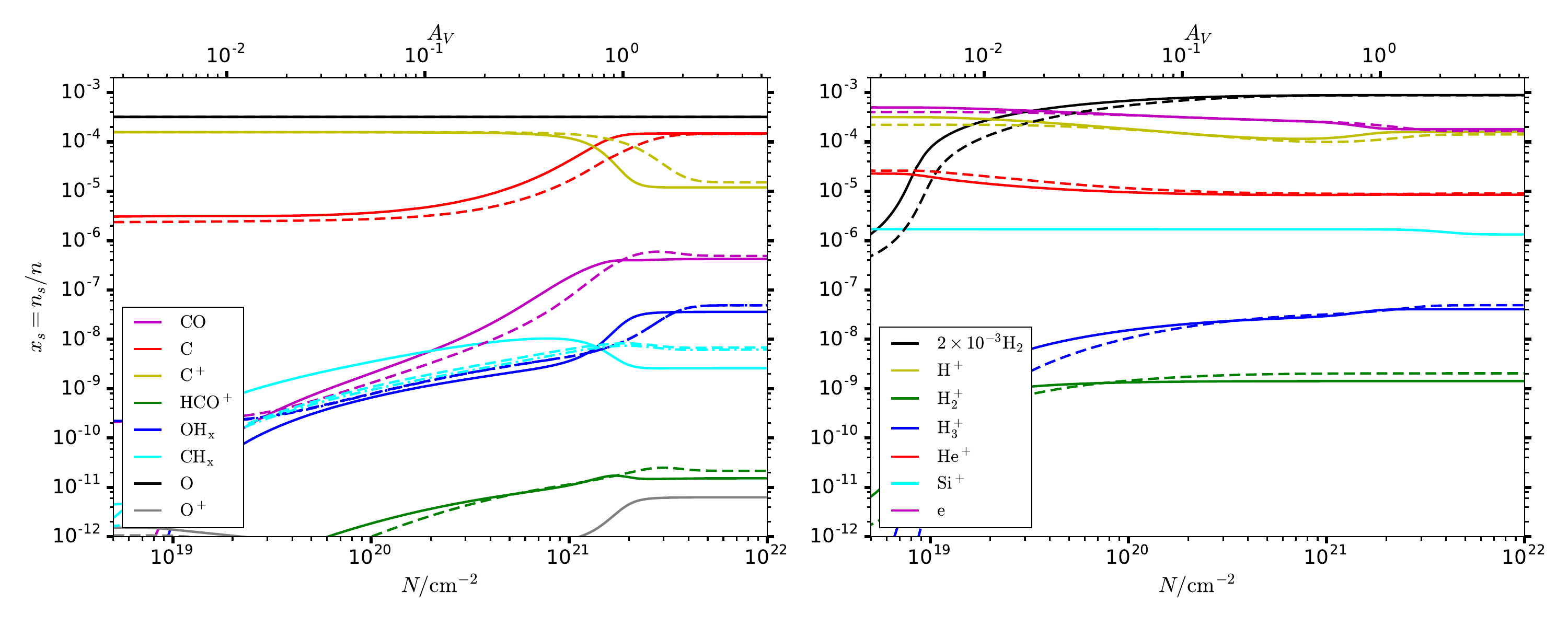}
        }
        \subfigure[Column densities]{%
            \includegraphics[width=0.97\textwidth]{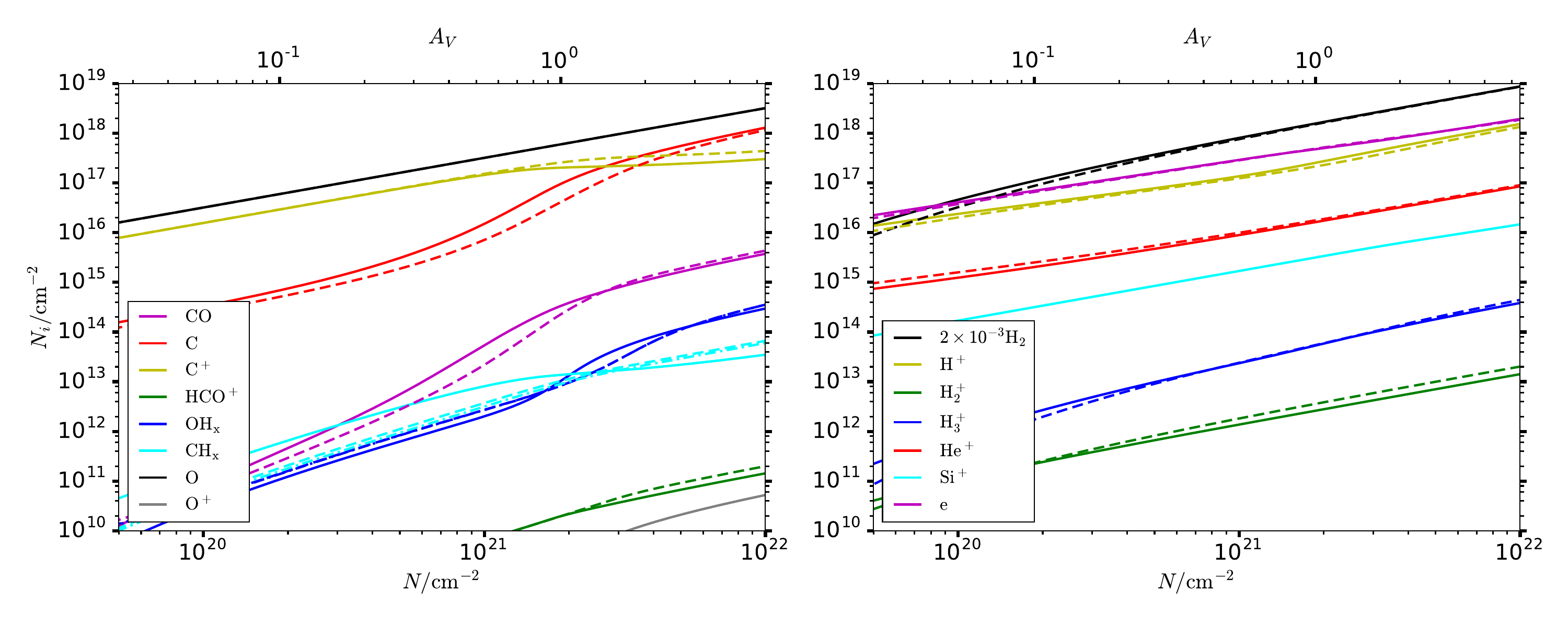}
        }

    \end{center}
    \caption{Abundances and column densities of species at density 
        $n=100~\mr{cm^{-3}}$ in our network (solid), and the {\sl PyPDR}
        code (dashed), as a function of $A_V$/$N$. This is similar to
        Figures \ref{fig:species_all_nH50-200} and \ref{fig:Ni_all_nH50-200},
        but with the dashed lines showing the 
        {\sl PyPDR} code instead of the Wolfire PDR code.
        We have also plotted the
        abundances of $\mr{CH}$ and $\mr{OH}$ species from the {\sl PyPDR} 
        code (cyan and blue dash-dotted lines), in addition to the sum of
        all $\CHx$ and $\OHx$ species.
        The cosmic-ray ionization rate $\xi_\mr{H}=2\times
        10^{-16}~\mr{s^{-1}H^{-1}}$, and gas temperature is fixed at $T=20~\mr{K}$.
        The reactions of 
        grain-assisted recombination of ions are excluded in our network to
        be consistent with {\sl PyPDR}.
\label{fig:PyPDR}}
\end{figure*}

Figure \ref{fig:PyPDR} shows a comparison between our chemical network and the
{\sl PyPDR} code. We disabled the grain-assisted recombination of ions in our
network to make a direct comparison with {\sl PyPDR}, which does not have these
reactions. We also updated the reaction rates in the {\sl PyPDR} code, to be
consistent with the reaction rates in this paper (Tables \ref{table:chem1} and
\ref{table:chem2}). Note that {\sl PyPDR} does not include metals such
as $\Si$, which is in our network. Figure \ref{fig:PyPDR} shows that our
chemical network (without grain-assisted recombination) agrees very well with
the results from the {\sl PyPDR} code.

\section{Characteristic Density in Slab and Spherical Turbulent
Clouds}\label{section:cloud_density}
Define $s\equiv \ln (\rho/\rho_0)$, where $\rho_0 = M/V = \langle \rho
\rangle_V$ is the
volume-weighted density. Both the volume- and the mass-weighted density obey
the log-normal distribution
\begin{equation}
    f_{V, M} (s) \di s = \frac{1}{\sigma_s\sqrt{2\pi}} \exp\left[-\frac{(s-\mu_{V,
    M})^2}{2\sigma_s^2}\right] \di s,
\end{equation}
where $\mu_V = -\sigma_s^2/2$ and $\mu_M = \sigma_s^2/2$ are the volume- and
mass- weighted means of the parameter $s$. Both observations of molecular clouds and
numerical simulations show that the variance $\sigma_s$ is related to the
turbulence Mach number ${\cal M}$ 
\citep[e.g.,][]{Padoan1997, LM2008, Brunt2010, KT2013}
\begin{equation}
    \sigma_s \approx \ln (1 + b^2 {\cal M}^2),
\end{equation}
with $b \approx 1/2 - 1/3$. 
Thus, the mass-weighted mean density is
\begin{equation}\label{eq:rho_M}
    \left\langle \frac{\rho}{\rho_0} \right\rangle_M 
    = \int \frac{\rho}{\rho_0} f_M(s) \di s =  \int e^s f(s) \di s =
    e^{\sigma_s^2} \approx 1 + b^2 {\cal M}^2.
\end{equation}

For slab clouds, the balance of the gas self-gravity with
the mid-plane pressure $P$ (assumed to be primarily from turbulence) gives
\begin{equation}
    P = \frac{\pi G \Sigma^2}{2},
\end{equation}
where $\Sigma$ is the gas surface density. The average
volume-weighted density is then $\rho_0 \equiv P/\sigma_\mr{1D}^2 = \pi G
\Sigma^2/(2\sigma_\mr{1D}^2)$, where $\sigma_\mr{1D}$ is the one-dimensional
velocity dispersion of the cloud-scale turbulence. The Mach number squared 
\begin{equation}\label{eq:Mach_slab}
    {\cal M}^2 = \frac{\sigma_\mr{3D}^2}{c_s^2} 
    = \frac{3\pi G \Sigma^2}{2\rho_0 c_s^2},
\end{equation}
where $c_s$ is the sound speed.
Therefore, according to Equation (\ref{eq:rho_M}),
the mass-weighted mean density of the slab cloud can be written as
\begin{equation}\label{eq:rho_M_slab}
    \langle \rho \rangle_{M, \mr{slab}} = \rho_0 + b^2 {\cal M}^2 \rho_0
    = \rho_0 + 3 b^2 \left( \frac{\pi G \Sigma^2}{2 c_s^2}\right).
\end{equation}

For uniform spherical clouds, the virial parameter 
\begin{equation}
    \alpha_\mr{vir} \equiv \frac{2E_k}{|E_G|} = \frac{5 \sigma_\mr{1D}^2 R}{GM},
\end{equation}
where $E_k$ is the kinetic energy of the cloud from turbulence and $E_G$ is
the gravitational energy. A virialized cloud has $\alpha_\mr{vir}=1$, and
marginally gravitationally bound cloud has $\alpha_\mr{vir}=2$. The Mach number 
squared can be written as 
\begin{equation}\label{eq:Mach_sphere}
    {\cal M}^2 = \frac{\sigma_\mr{3D}^2}{ c_s^2} 
    = \frac{3 \sigma_\mr{1D}^2 }{ c_s^2 }
= \frac{3\alpha_\mr{vir} G M}{5 R c_s^2} = \frac{1}{\rho_0} \alpha_\mr{vir}
\frac{9}{10} \left( \frac{\pi G \Sigma^2}{2 c_s^2}\right).
\end{equation}
Here we define the average cloud density $\rho_0 \equiv M/(\frac{4\pi}{3}R^3)$, and the
average cloud surface density $\Sigma \equiv M/(\pi R^2)$. Similar to Equation
(\ref{eq:rho_M_slab}), the mass-weighted mean density of the spherical cloud is
given by
\begin{equation}\label{eq:rho_M_sphere}
    \langle \rho \rangle_{M, \mr{sphere}} = \rho_0 + b^2 {\cal M}^2 \rho_0
    = \rho_0 + \left(\frac{9}{10}\alpha_\mr{vir}\right) b^2 \left( \frac{\pi G \Sigma^2}{2 c_s^2}\right).
\end{equation}
Equations (\ref{eq:rho_M_slab}) and (\ref{eq:rho_M_sphere}) can be written
together as

\begin{equation}\label{eq:rho_M0}
    \langle \rho \rangle_{M}
    = \rho_0 + a b^2 \left( \frac{\pi G \Sigma^2}{2 c_s^2}\right),
\end{equation}
where $a=3$ for slab clouds and $a=9\alpha_\mr{vir}/10$ for spherical clouds.
Using the relations $\rho = 1.4 m_\Ho n$ and $\Sigma = 1.4 m_\Ho N$,
where $m_\Ho = 1.67\times 10^{-24}~\mr{g}$ is the mass of a hydrogen atom, and
the factor $1.4$ comes from helium, Equation (\ref{eq:rho_M0}) gives
\begin{equation}\label{eq:n_M0}
    \langle n \rangle_{M}
    = \langle n \rangle_{V} + 1.4 m_\Ho a b^2 
    \left( \frac{\pi G N^2}{2 c_s^2}\right).
\end{equation}

\section{Additional Plots for Comparison with Observations of 
Diffuse and Translucent Clouds}
\begin{figure*}[htbp]
     \begin{center}
         \includegraphics[width=0.97\textwidth]{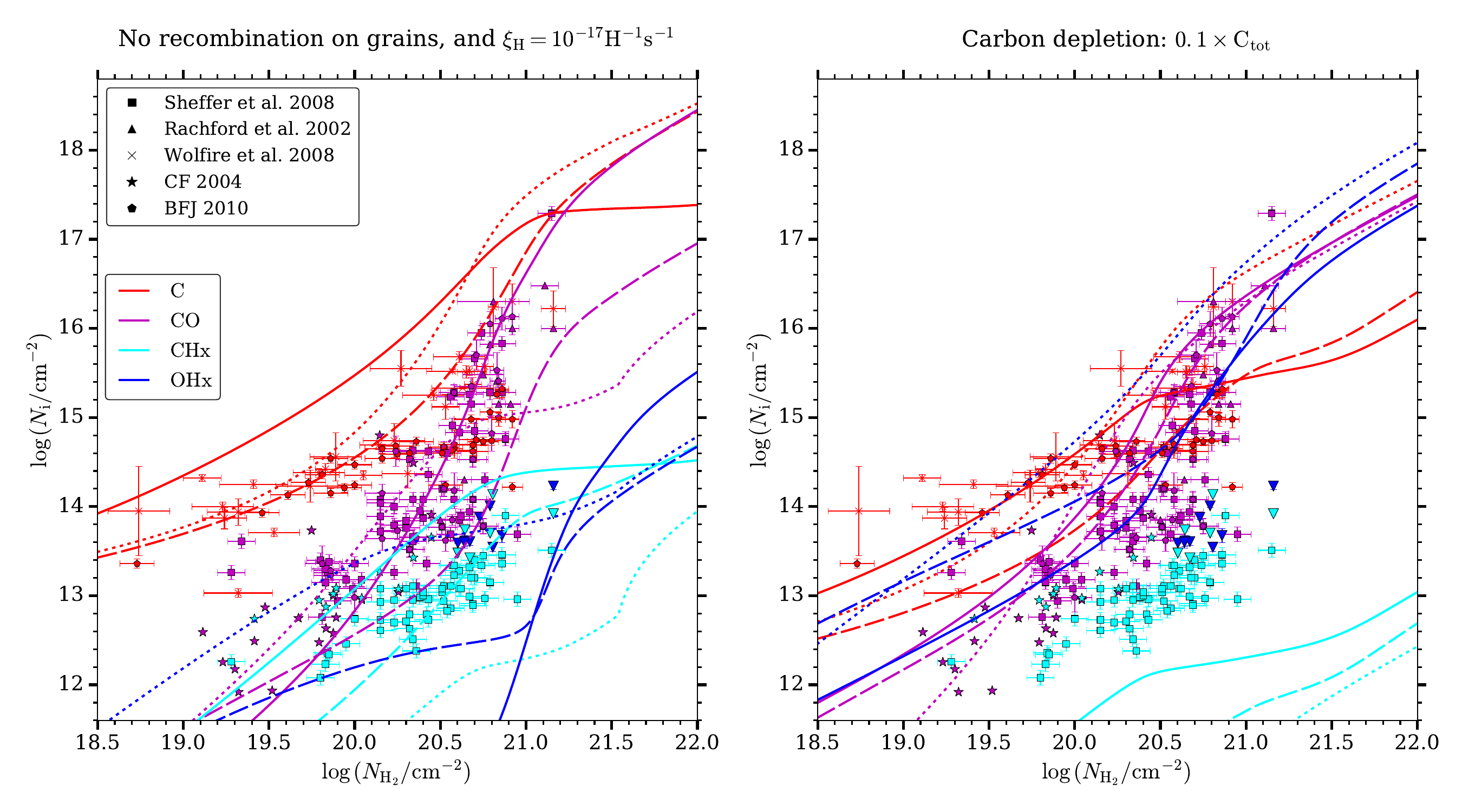}
    \end{center}
    \caption{Similar to the right panel of Figure \ref{fig:cloud_obs_NH_NH2},
        but with no grain-assisted recombination of ions ({\sl left}), or
        depletion of gas-phase carbon abundance by a factor of 10 ({\sl right}).
    \label{fig:cloud_obs_nogr_depCI}}
\end{figure*}

Figure \ref{fig:cloud_obs_nogr_depCI} shows a detailed comparison between
observations and our slab model with no grain-assisted recombination, and
depletion of carbon abundance. 

\clearpage
\newpage

\bibliographystyle{apj}
\bibliography{apj-jour,thesis}
\end{document}